\documentclass[useAMS,usenatbib]{mn2e}
\usepackage{hyperref}
\usepackage{times}
\usepackage{natbib}
\usepackage{aas_macros}
\usepackage{amsmath}
\usepackage{amssymb}
\usepackage{rotating}
\usepackage[labelfont=bf]{caption}
\usepackage{supertabular}
\usepackage{enumitem}

\newcommand{\chandra}{\textit{Chandra}}

\newcommand{\ergscm}{$\mathrm{erg\;s^{-1} cm^{-2}}$}
\newcommand{\cmsq}{cm$^{-2}$}
\newcommand{\ergs}{$\mathrm{erg\;s^{-1}}$}

\newcommand{\dd}{d}
  \newcommand{\NH}{$N_\mathrm{H}$}
\newcommand{\nh}{N_\mathrm{H}}
 \newcommand{\degrees}{$^{\circ}$}
\newcommand{\lx}{L_\mathrm{X}}
\newcommand{\LX}{$L_\mathrm{X}$}
\newcommand{\XI}{{\boldsymbol \xi}}
\newcommand{\xlfpars}{{\boldsymbol \theta}}
\newcommand{\nhfpars}{{\boldsymbol \Theta}}
\newcommand{\galpars}{{\boldsymbol \omega}}
\newcommand{\fabs}{$f_\mathrm{abs}$}

\newcommand{\giv}{\;|\;}
\newcommand{\gabs}{$\gamma_\mathrm{abs}$}
\newcommand{\gunabs}{$\gamma_\mathrm{unabs}$}

\newcommand{\secdraft}[1]{#1}\newcommand{\thirddraft}[1]{#1}\newcommand{\refone}[1]{{#1}}
\newcommand{\reftwo}[1]{{#1}}

\begin{document}

\title[The evolution of the XLFs of unabsorbed and absorbed AGNs]{The evolution of the X-ray luminosity functions of unabsorbed and absorbed AGNs out to $\mathbf{z\sim5}$}
\author[J. Aird et al.]{J. Aird$^{1,2}$\thanks{jaird@ast.cam.ac.uk}, A. L. Coil$^3$, A. Georgakakis$^{4,5}$, K. Nandra$^4$, 
G. Barro$^6$ and 
\newauthor
P. G. P\'{e}rez-Gonz\'{a}lez$^7$\\
$^1$Institute of Astronomy,
University of Cambridge,
Madingley Road,
Cambridge,
CB3 0HA\\
$^2$Department of Physics, Durham University, Durham, DH1 3LE\\
$^3$Center for Astrophysics and Space Sciences (CASS), Department of Physics, University of California, San Diego, CA 92093, USA\\
$^4$Max Planck Institute f\"{u}r Extraterrestrische Physik, Giessenbachstrasse, 85748 Garching, Germany\\
$^5$IAASARS, National Observatory of Athens, GR-15236 Penteli, Greece\\
$^6$University of California, Santa Cruz, 1156 High Street, Santa Cruz, CA 95064, USA\\
$^7$Departamento de Astrof\'{i}sica, Facultad de CC. F\'{i}sicas, Universidad Complutense de Madrid, E-28040 Madrid, Spain
}

\date{Accepted 2015 May 8. Received 2015 May 8; in original form 2015 February 24}

\pagerange{\pageref{firstpage}--\pageref{lastpage}} \pubyear{2015}

\maketitle
\label{firstpage}
\begin{abstract}
We present new measurements of the evolution of the X-ray luminosity functions (XLFs) of unabsorbed and absorbed Active Galactic Nuclei (AGNs) out to $z\sim5$. 
We construct samples containing 2957 sources detected at hard (2--7 keV) X-ray energies and 4351 sources detected at soft (0.5--2 keV) energies from a compilation of \textit{Chandra} surveys supplemented by wide-area surveys from \textit{ASCA} and \textit{ROSAT}.
We consider the hard and soft X-ray samples separately and find that the XLF based on either (initially neglecting absorption effects) is best described by a new flexible model parametrization where the break luminosity, normalization and faint-end slope all evolve with redshift. 
We then incorporate absorption effects, separately modelling the evolution of the XLFs of unabsorbed ($20<\log \nh<22$) and absorbed ($22<\log \nh<24$) AGNs, seeking a model that can reconcile both the hard- and soft-band samples. 
We find that the absorbed AGN XLF has a lower break luminosity, a higher normalization, and a steeper faint-end slope than the unabsorbed AGN XLF out to $z\sim2$.
Hence, absorbed AGNs dominate at low luminosities, with the absorbed fraction falling rapidly as luminosity increases.
Both XLFs undergo strong luminosity evolution which shifts the transition in the absorbed fraction to higher luminosities at higher redshifts. 
The evolution in the shape of the total XLF is primarily driven by the changing mix of unabsorbed and absorbed populations.
\end{abstract}
\begin{keywords}
galaxies: active --
galaxies: evolution --
galaxies: luminosity function, mass function -- 
X-rays: galaxies
\end{keywords}

\section{Introduction}
\label{sec:intro}

The luminosity function of Active Galactic Nuclei (AGNs) represents one of the crucial observational constraints on the growth of supermassive black hole (SMBHs) over the history of the Universe. 
The shape of the luminosity function reflects a combination of the underlying distribution of the SMBH masses and the distribution of their accretion rates or Eddington ratios \citep[e.g.][]{Aird13,Shankar13,Schulze15}.
Thus, accurate measurements of the shape and evolution of the luminosity function provide crucial insights into the physical processes that drive SMBH growth over cosmic time.

Many AGNs are surrounded by gas and dust that can obscure their emission at certain wavelengths.
Thus, it is vital to understand AGN obscuration in order to obtain accurate measurements of the luminosity function.
Quantifying AGN obscuration also reveals whether SMBHs undergo significant periods of obscured growth, when this takes place within the lifetimes of AGNs, and how it relates to the triggering and fuelling processes.

Obtaining accurate measurements of the luminosity function and revealing the extent of obscuration requires large, unbiased samples of AGNs selected over the widest possible range of redshifts and luminosities.
Optical surveys, combined with follow-up spectroscopy, can efficiently cover wide areas \citep[e.g. SDSS:][]{York00}
but are biased towards the most luminous, unobscured sources.
Alternatively, AGNs can be identified in the mid-infrared (mid-IR), which probes the reprocessed emission from the dusty, obscuring material.
Mid-IR selection should not be biased against obscured sources, but the contribution of the host galaxy is often significant at these wavelengths, which limits mid-IR selection to luminous sources where the galaxy light is overwhelmed by the AGN \citep[e.g.][]{Donley08,Mendez13}.

X-ray surveys can efficiently identify AGNs over a wide luminosity range, including low-luminosity sources where the host galaxy dominates at optical or infrared wavelengths \citep[e.g.][]{Barger03c}. 
Nevertheless, soft X-ray emission (at energies $\lesssim 2$ keV) will be absorbed by the same gas and dust that obscures the AGN at optical and UV wavelengths.
Thus, soft X-ray samples are generally dominated by unobscured AGNs. 
Absorption biases are reduced at hard X-ray energies ($\sim2-10$ keV), except in the most heavily-obscured, Compton-thick AGNs (equivalent line-of-sight hydrogen column densities $\nh\gtrsim 10^{24}$ cm$^{-2}$).
However, even Compton-thick sources may still be identified at soft or hard X-ray energies due to scattered emission, including the Compton-scattered emission (``reflection") from the obscuring material itself. 

A large number of deep and wide X-ray surveys have been carried out, taking advantage of the efficiency and power of X-ray selection
\citep[see a recent review by][]{Brandt15}.
A number of studies have measured the X-ray luminosity function (XLF) of AGNs out to high redshifts using these 
samples 
\citep[e.g.][]{Ueda03,Barger05,Miyaji15}.
These studies find that AGNs are a strongly evolving population, with a 
sharp decrease in their number density between $z\sim1-2$ and today.  
Bright X-ray-selected AGNs are found to peak in number density at $z\approx 2$, similar to optically-selected QSOs.  
Fainter AGNs peak later in the history of the Universe ($z\approx 1$) but with a much milder decline to the present day \citep[e.g.][]{Hasinger05}. 
These patterns have led several authors to propose a luminosity-dependent density evolution (LDDE) parametrization to describe the evolution of the XLF of AGNs \citep[e.g.][]{Miyaji00,Ueda03}.
In this model the XLF is modified by differing degrees of density
evolution that vary with luminosity and redshift.
This results in an XLF that changes shape over cosmic time.

In \citet[hereafter A10]{Aird10} we challenged this evolutionary model with a detailed study of the hard-band XLF that carefully accounted for numerous uncertainties and
biases that were generally not included in prior measurements.
These included
flux measurement errors, Eddington bias, incompleteness of optical
identifications, and the uncertainty in photometric redshift estimates. 
At high redshifts ($z\sim2-3$) we adopted a rest-frame UV colour pre-selection technique \citep{Aird08}.
By performing a robust model comparison based on Bayesian statistical techniques, we found that the evolution of the XLF could be described by a simpler model in which the XLF retains the same shape at all redshifts but evolves in both luminosity and density
\citep[see also][]{Assef11,Ross12}.

While A10 presented a number of important advances, absorption effects were not explored.
Other studies have attempted to measure the distribution of absorption column densities 
and present absorption-corrected measurements of the XLF. 
\citet{Ueda03} found that the fraction of absorbed AGNs (those with $\nh>10^{22}$ cm$^{-2}$) was strongly dependent on luminosity, decreasing at higher luminosities. 
Later studies found that the fraction of absorbed AGNs depends on both luminosity \emph{and} redshift, dropping at high luminosities but increasing (at a given luminosity) to higher redshifts \citep[e.g.][]{LaFranca05,Hasinger08}. 
The extent of any redshift evolution has been a matter of debate \citep[see][]{Akylas06,Dwelly06}, potentially due to 
difficulties regarding the selection functions for absorbed and unabsorbed sources. 
Recent work by \citet{Ueda14} re-examined the evolution of the XLF and the distribution of \NH\ (the ``\NH-function") using a large compilation of both soft and hard X-ray surveys and found that both a luminosity and redshift dependence of the absorbed fraction were required.
They also found that an LDDE parametrization was needed to describe the evolution of the XLF \citep[with some further modifications to describe the evolution at $z\gtrsim 3$, see also][]{Civano11,Hiroi12}.

Recently, the combination of extremely deep X-ray survey data and new approaches to X-ray spectral analysis have enabled improved measurements of \NH\ at $z\sim0.5-2$ and have been used to identify sizable samples of Compton-thick AGNs \citep[e.g.][]{Comastri11,Georgantopoulos13,Brightman14}. 
Building on this work, \citet{Buchner15} used a flexible, non-parametric method to estimate the space densities of AGNs as a function of redshift, luminosity and \NH, effectively measuring the XLF for different column densities. 
This work also recovered a luminosity and redshift dependence in the evolution of the fraction of absorbed AGNs (although the Compton-thick fraction was consistent with a constant value of $\sim35$ per cent).
However, a detailed comparison of parametric models for the evolution of the XLF of AGNs was not undertaken. 

In this paper we address some remaining issues in studies of the evolution of the XLF of AGNs:
the shape of the XLF and how it evolves with redshift, the extent of any luminosity and redshift dependence of the absorbed fraction, and the connection between the absorption properties and the evolution of the AGN population.
We combine samples selected at both hard and soft X-ray energies and determine the underlying XLF and distribution of \NH\ that adequately describes the observed fluxes in both samples (similar to the approach of \citealt{Ueda14}, cf. the X-ray spectral analysis used in \citealt{Buchner15}).

In Section \ref{sec:data} we describe our datasets
that we use to define large samples of X-ray sources selected in the hard (2--7 keV) and soft (0.5--2 keV) energy bands.
We also compile deep optical, near-infrared and mid-infrared imaging across our fields that we use to robustly identify counterparts to our X-ray sources and calculate photometric redshifts. 
In Section \ref{sec:method}, we describe our Bayesian statistical technique that allows us to incorporate a range of X-ray spectral shapes and account for the effects of absorption.
We also introduce an approach to account for the contribution from normal, X-ray detected galaxies on our measurements.
We then present measurements of the XLF based on our hard and soft samples individually (Section \ref{sec:xlf}), introducing a new flexible parametrization of the XLF. 
We show that significant discrepancies between the measurements at all redshifts warrant the further consideration of absorption effects.
In Section \ref{sec:absdist} we separately model the XLF of unabsorbed and absorbed AGNs (including a contribution from Compton-thick sources) and show how the combination of these populations can simulataneously account for both our hard- and soft-band samples.
Our results place constraints on the total XLF of AGNs and the absorbed fraction as a function of luminosity and redshift. 
In Section \ref{sec:discuss} we compare our results to prior work and discuss the wider implications of our findings. 
Section \ref{sec:summary} summarizes our paper and overall conclusions. 

Given the length of this paper, a casual reader may wish to skip to Section \ref{sec:xlfresults}, Section \ref{sec:absdist} and the discussion in Section \ref{sec:discuss} (and focus on Figures \ref{fig:xlfhardsoft}, \ref{fig:xlf_withabs}, and \ref{fig:xlf_abs_unabs}).
We adopt a flat cosmology with $\Omega_\Lambda=0.7$ and $h=0.7$ throughout this paper.

\section{Data}
\label{sec:data}

To constrain the evolution of the XLF we require large samples of X-ray selected AGNs.
By selecting samples in both the hard ($>2$ keV) and soft (0.5--2 keV) observed energy bands,
we can also constrain the distribution of \NH\ and correct for these effects on the XLF.
In this paper we combine a large number of \chandra\ X-ray surveys along with larger area surveys from \emph{ASCA} and \emph{ROSAT}.
We give further details of our datasets below. 		
Table \ref{tab:sourcenumbers} summarizes the different surveys and provides the number of hard and soft X-ray selected sources from each.

  \begin{table*}
\caption{Details of the X-ray surveys used in this paper.}
\setlength\tabcolsep{5pt}
\begin{tabular}{c c c c c c c c c c c c c c c}
\hline
{Field}  & {Survey}	& {RA}		& {Dec}		& {X-ray}  & {Survey} & \multicolumn{3}{l}{Soft band} & \multicolumn{3}{l}{Hard band} \\
       &              &               &                            & exposure               &  area           & {$N_\mathrm{X}$} & {$N_\mathrm{ctrprt}$} & {$N_\mathrm{spec-z}$} & {$N_\mathrm{X}$} & {$N_\mathrm{ctrprt}$} & {$N_\mathrm{spec-z}$} \\
          &                   & {\footnotesize (J2000)}		& {\footnotesize (J2000)}		& 		    & {\footnotesize (deg$^2$)} & & &    & & & \\
  (1) & (2) & (3) & (4) & (5) & (6) & (7) & (8) & (9) & (10) & (11) & (12) \\
  
\hline
      CDFS &   CDFS-4Ms &         03:32:27.2 & -27:47:55 &   4Ms &        0.075 &    413 &    397 (96.1\%) &   240 (58\%) &   283 &    273 (97\%) &   155 (55\%) \\
      CDFS &      ECDFS &         03:32:27.2 & -27:47:55 &   250ks &        0.181 &    334 &    328 (98.2\%) &   115 (34\%) &   273 &    268 (98\%) &   101 (37\%) \\
      CDFN &       CDFN &         12:36:49.3 & +62:13:19 &   2Ms &        0.112 &    384 &    363 (94.5\%) &   242 (63\%) &   286 &    273 (96\%) &   176 (62\%) \\
       EGS &   AEGIS-XD &         14:19:20.8 & +52:50:03 &   800ks &        0.260 &    698 &    673 (96.4\%) &   295 (42\%) &   552 &    539 (98\%) &   233 (42\%) \\
       EGS &   AEGIS-XW &         14:17:15.0 & +52:25:31 &   200ks &        0.204 &    334 &    332 (99.4\%) &   130 (39\%) &   274 &    274 (100\%) &   110 (40\%) \\
    COSMOS &   C-COSMOS &         10:00:20.3 & +02:11:20 &   160ks &        0.984 &   1213 &   1195 (98.5\%) &   694 (57\%) &   889 &    877 (99\%) &   530 (60\%) \\
    Bootes &    XBootes &         14:31:28.3 & +34:28:07 &   5ks &        7.124 &    754 &    744 (98.7\%) &   566 (75\%) &   257 &    255 (99\%) &   196 (76\%) \\
   ...     &    ALSS     & 13:14:00          & +31:30:00  &  ...		   &  5.800    & ...    & ...            & ...         &   34    &  34 (100\%) &   33 (97\%)  \\
   ...     &    AMSS     &  ...	       & ...	    &  ...		   &  81.77    & ...    & ...            & ...         &  109    & 109 (100\%) & 107 (98\%) \\
   ...     &    ROSAT    &  ...	       & ...	    &  ...	           & 20391    &   221  & 221 (100\%)    & 221 (100\%) &  ...    & ...         & ...          \\[3pt]
\hline\\
                 &             &                   &            &                      &  Total   &     4351 &     4253 &     2503 &     2957 &     2902 &     1641 \\
\hline
\end{tabular}
\\
 \label{tab:sourcenumbers}
 \secdraft{
 Column (1) name of the field; 
 (2) name of X-ray survey programme within this field; 
 (3,4) approximate center of the survey; 
 (5) nominal X-ray exposure time; 
 (6) total area covered by both the X-ray data and the optical or infrared imaging, excluding areas close to bright stars; 
 (7) number of X-ray sources detected in the soft (0.5--2 keV) energy band;
 (8) number (and fraction) of the soft X-ray sources that are associated with a robust multiwavelength counterpart;
 (9) number (and fraction) of the soft X-ray sources that have a spectroscopic redshift;
 (10) number of X-ray sources detected in the hard (2--7 keV) energy band;
 (11) number (and fraction) of the hard X-ray sources that are associated with a robust multiwavelength counterpart;
 (12) number (and fraction) of the hard X-ray sources that have a spectroscopic redshift.
 }
\end{table*}

\subsection{\chandra\ X-ray data}
\label{sec:xray}

In this paper we use \chandra\ X-ray observations from five distinct parts of the sky: the Chandra Deep Field-South (CDFS), Chandra Deep Field-North (CDFN), Extended Groth Strip (EGS), COSMOS, and Bootes fields.

In the CDFS field we identify two different ``surveys": 
1) the series of observations that have targeted the central $\sim0.07$ deg$^2$ of the field and reach a total combined exposure time of $\sim$4Ms \citep[the CDFS-4Ms survey,][]{Xue11};
and 2) the series of four 250ks observations that surround the central area \citep[the Extended-CDFS, or E-CDFS survey, ][]{Lehmer05}. 
The X-ray data reduction and source detection procedures were carried out independently for each of these surveys. 
To combine the two surveys,
we define a central region where the CDFS-4Ms survey takes precedence, roughly corresponding to the area within $\sim9$\arcmin\ of the center of the field. 
Outside this central region we adopt the sources detected in the E-CDFS survey only. 
This procedure ensures we have a well-defined sample where we can accurately determine the sensitivity.

In the EGS field we also identify two distinct surveys: 
1) the series of eight pointings that reach a nominal depth of $\sim$200ks and were presented in \citet{Laird09}, which we refer to as the AEGIS-XW(ide) survey; 
and 2) the AEGIS-XD(eep) survey \citep{Nandra15} which took three of the original 200ks to a depth of $\sim800$ks.
We adopt the deeper AEGIS-XD observations when available.

In each of the remaining fields we adopt data from a single \chandra\ survey:
the 2Ms survey in CDFN \citep{Alexander03}; the $\sim$160ks C-COSMOS observations \citep{Elvis09,Puccetti09}; and the 5ks XBootes survey \citep{Murray05}.

The X-ray data from all of our surveys were reduced with our own pipeline procedure, which is described in detail by \citet{Laird09} and \citet{Nandra15}. 
We performed point source detection using the procedure described by \citet{Laird09} and applied a false Poisson probability threshold of $<4\times 10^{-6}$ to generate catalogues of detected sources in the soft (0.5--2 keV), hard (2--7 keV), full (0.5--7 keV) or ultrahard (4--7 keV) observed energy bands. 
We combined the source lists in each band to create a merged catalogue, which is used in the counterpart identification procedure described in Section \ref{sec:counterparts} below.

For the 5ks XBootes survey we applied a stricter false probability cut ($<10^{-8}$) in addition to a requirement of $\ge 5$ total detected counts in each band. 
This cut reduces the sample size but applies an effective X-ray flux limit that helps raise the completeness of the spectroscopic follow-up in this field.
In this field, we also restrict our analysis to the $\sim7.1$ deg$^2$ of the Bootes field that corresponds to the 15 standard sub-fields of the AGES spectroscopic survey \citep{Kochanek12}. This cut also ensures we have a high spectroscopic completeness (see Section \ref{sec:specz} below).

\begin{figure}
\includegraphics[width=\columnwidth]{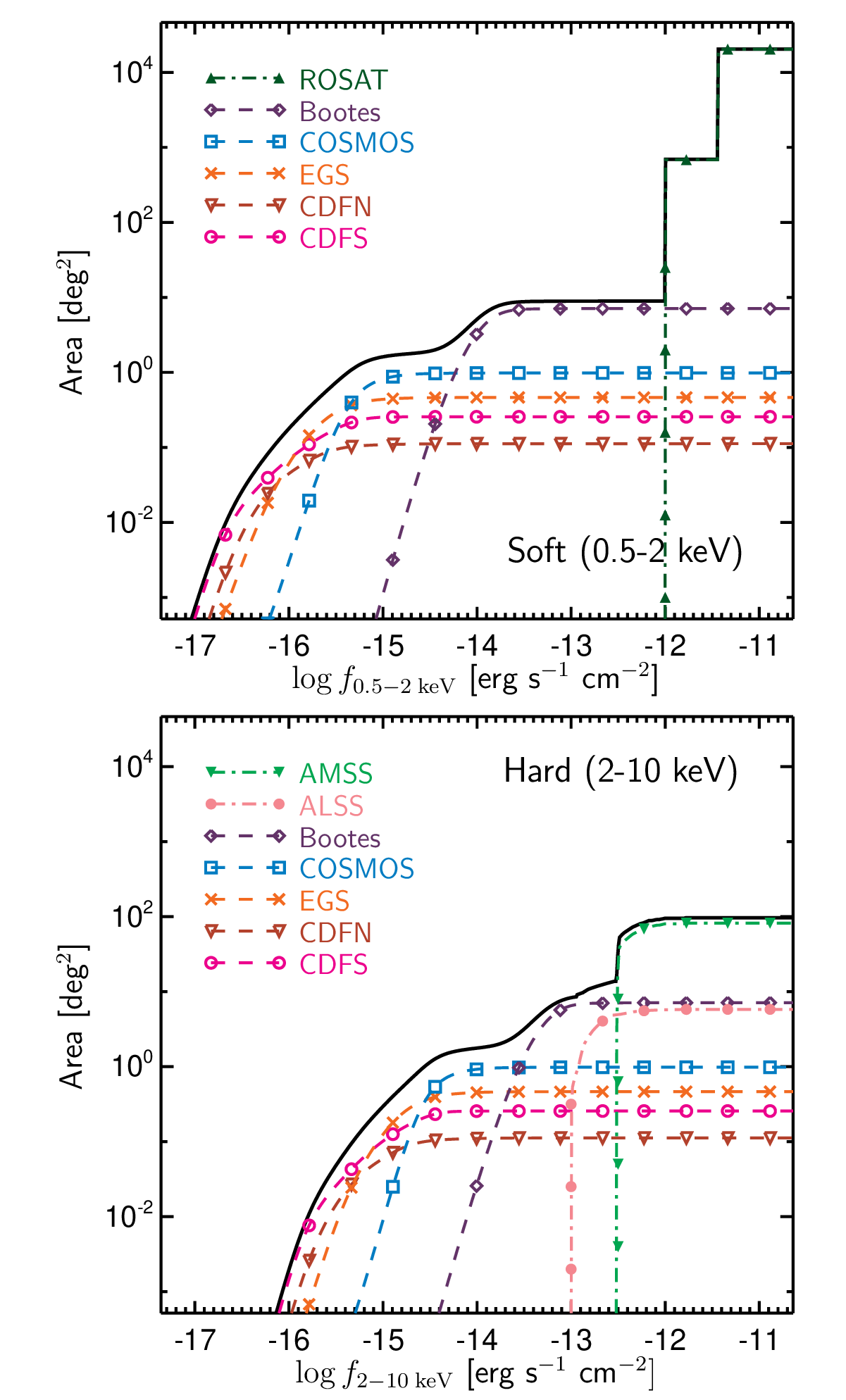}
\caption{X-ray area curves (sensitive area versus X-ray flux) for each of our surveys in the soft and hard bands. The black line indicates the total area curve for our study. 
We limit our sensitivity analysis to the area falling within the footprint of our multiwavelength photometry, excluding areas close to bright stars.
The ROSAT area curve includes the ROSAT bright survey and SA--N survey (see Section \ref{sec:largeareasurveys}).
The EGS area curve includes both the AEGIS-XD 800ks data and the additional area at 200ks from AEGIS-XW.
The CDFS area curve combines the CDFS 4Ms data with the flanking 250ks data from the E-CDFS.
These area curves assume a single, fixed conversion factor between the X-ray count rate and flux, corresponding to an X-ray spectrum with a photon index $\Gamma=1.4$ and Galactic absorption only; the effect of different X-ray spectral shapes---and the resulting uncertainties in the conversion factors---are accounted for in our Bayesian methodology described in Section \ref{sec:method}.
}
\label{fig:acurves}
\end{figure}

We determined X-ray sensitivity maps and area curves for each band as described in \citet{Georgakakis08}, 
accounting for the stricter false probability cut and minimum counts requirement for the XBootes survey.
We convert the sensitivity maps to area curves as a function of flux by assuming a fixed X-ray spectral slope of $\Gamma=1.4$.
We note that the flux calculated with this fixed conversion factor scales directly with the count rate; in Section \ref{sec:method} below we describe our procedure to convert between the count rate and intrinsic quantities (such as luminosity), which allows for a more complex X-ray spectrum and accounts for uncertainties in this conversion factor.
The sensitivity map calculation is limited to the footprint of the multiwavelength photometry for each field.
Figure \ref{fig:acurves} shows the corresponding area curves for each of our X-ray fields. 

We note that previously published X-ray source catalogues, often including mutiwavelength counterpart information, are available for the all of our \chandra\ fields \citep[e.g.][]{Alexander03, Goulding12, Brand06}.
Adopting our own X-ray reductions and source detection procedures ensures we can accurately determine the sensitivity in a consistent manner, which is essential for our Bayesian analysis of the XLF.
\refone{Our catalogues contain $\sim10-25$ per cent fewer sources than \citet{Xue11} and \citet{Puccetti09} in the CDFS-4Ms and C-COSMOS areas, mainly due to our stricter (i.e. lower) false probability threshold.
Thus, our catalogues are more conservative.}

\subsection{Multiwavelength counterparts and photometry}
\label{sec:counterparts}

In each of our \chandra\ fields, we identify multiwavelength counterparts to our X-ray sources using the likelihood ratio (LR) method 
\citep[e.g.][]{Ciliegi03,Brusa07,Civano12}, matching to multiple optical, near-IR and mid-IR bands to ensure a high completeness and reliability. 
We also compile multiwavelength photometry from a larger number of bands, which we use to calculate photometric redshifts (see Section \ref{sec:photz} below). 

\subsubsection{CDFS, CDFN and EGS}
\label{sec:rainbow}

In three of our fields---the CDFS, CDFN and EGS---we identify counterparts and extract the multiwavelength photometry using a custom version of the \textit{Rainbow} Cosmological Surveys Database\footnote{https://rainbowx.fis.ucm.es} \citep{Perez-Gonzlez05,Perez-Gonzlez08,Barro11,Barro11b}, which provides a compilation of the various photometric datasets.
Appendix \ref{sec:photoappendix} lists all the different photometric imaging datasets for each field that are used in this paper. 
We note that full coverage of the entire field is not always available in each photometric band.
All the images are registered to a common astrometric reference frame and photometry is performed in consistent apertures to produce spectral energy distributions that span from the UV to mid-IR. 
The X-ray source matching procedure is described in detail in \citet{Nandra15}, but we briefly summarise the method here.

First, we reran the \textit{Rainbow} photometric code, extracting all potential counterparts within 3.5\arcsec\ of the X-ray positions in any of the bands covered by \textit{Rainbow} (using initial SExtractor catalogues in each of the bands). 
The counterparts were then cross-matched using a 2\arcsec\ search radius to create a single multiband catalogue. 
We obtained consistent photometry by applying a single aperture across all optical and near-IR bands. 
We also extracted IRAC photometry, applying the procedure described in \citet{Perez-Gonzlez08} and \citet{Barro11} to deblend the IRAC photometry when a single IRAC source is associated with multiple optical/near-IR counterparts.

Next, we calculated the LR for all candidate counterparts detected in the IRAC 3.6\micron\ band and determined an LR threshold that maximises the sum of the completeness and reliability \citep[see ][]{Luo10}. 
A candidate counterpart that exceeded this LR threshold was deemed a secure counterpart (taking the counterpart with the highest LR value in cases of $>1$ secure candidate).
We then repeated the entire LR matching process for the bands indicated in the table in  Appendix \ref{sec:photoappendix}, retaining any additional secure counterparts identified in these bands.
This procedure enables us to identify secure counterparts for a high fraction ($>90\%$) of the X-ray sources in the CDFS, CDFN and EGS fields (see Table \ref{tab:sourcenumbers}).
The vast majority ($\sim92$\%) of the secure counterparts were identified in the (deblended) IRAC $3.6$\micron\ catalogue. 
Matching to the additional bands allows us to identify counterparts when the IRAC candidate is faint, blended, or non-existent.
No additional cross-matching is required as the full multiband photometry is provided through matched apertures for all sources in the Rainbow database.

\subsubsection{COSMOS}

In the COSMOS field, which is not currently included in the Rainbow surveys database, we matched directly between our X-ray source lists and two multiwavelength catalogues:
1) the COSMOS Intermediate and Broad Band Photometry Catalogue 2008\footnote{http://irsa.ipac.caltech.edu/data/COSMOS/datasets.html}, which is based on detection in the deep Subaru $i^+$ imaging of the entire COSMOS field \citep{Capak07};
and 
2) the S-COSMOS IRAC 3.6\micron\ based catalogue \citep{Sanders07}.
Unlike the Rainbow catalogues, the S-COSMOS IRAC 3.6\micron\ catalogue has not been deblended.
Thus, we first identified secure matches (using the LR method) from the higher-resolution Subaru $i^+$ catalogues. 
We found secure counterparts for 1348 of the 1621 X-ray sources (83 per cent) in our overall C-COSMOS catalogue.
Next, we applied the LR method to match between the X-ray and S-COSMOS catalogues. 
We identified secure counterparts for an additional 178 sources, taking our overall completeness to 94 per cent. 

Finally, we cross-matched between our secure positions and the original catalogues, again using the LR method to ensure only secure associations are considered.
For 1499 of the 1526 secure counterparts, we end up with both a Subaru $i^+$ and S-COSMOS counterpart.
For these sources we obtain photometry in up to 18 broad-band filters spanning from the UV to near-IR, as well as 15 intermediate- or narrow-band optical filters, from the Subaru $i+$ catalogue \citep[see ][]{Ilbert09, McCracken10}. 
The photometry was extracted in 3\arcsec\ diameter apertures from PSF-matched images.
We also adopt IRAC photometry in the 3.6, 4.5, 5.8 and 8.0\micron\ bands from the S-COSMOS catalogue, applying the aperture corrections given in \citet{Ilbert09}. 

In 27 cases we have an $i^+$ counterpart but do not find an S-COSMOS counterpart, either due to the limited depth of the IRAC imaging or because the $i^+$ source is blended at the IRAC resolution. 
For these sources we simply ignore the IRAC bands in our photometric redshift estimates (see Section \ref{sec:photz} below).

In 23 cases, we identify an S-COSMOS counterpart, but no Subaru $i^+$ source.
For these sources we extracted photometry in 3\arcsec\ diameter apertures at the S-COSMOS position 
in the PSF-matched images for all the UV to near-IR bands, generally obtaining only upper limits for these optically faint sources.

\subsubsection{Bootes}

For the Bootes field, we compiled multiwavelength catalogues from the NOAO Deep-Wide Field Survey \citep{Jannuzi99} DR3, SDSS DR9 \citep{Ahn12}, GALEX GR7 \footnote{http://galex.stsci.edu/GR6/}, FLAMINGOS Extragalactic Survey \citep{Elston06}, and the Spitzer Deep-Wide Field Survey \citep{Ashby09}.
We use the LR method to match our X-ray catalogues to the appropriate selection band for each of these surveys, assigning secure matches from the surveys in the order of priority indicated in the table in Appendix \ref{sec:photoappendix}. 
The vast majority of our secure matches are identified in the NDWFS $I$ band.
We identify secure counterparts for 95.8 per cent of our X-ray sources. 
Finally, we cross-matched between the original catalogues and our secure counterpart positions, again applying the LR method, to identify common sources.
The combined surveys provide photometry in up to 17 different bands (see Appendix \ref{sec:photoappendix}).

\subsection{Large-area surveys}
\label{sec:largeareasurveys}

Combining our five \chandra\ fields provides a sample of over 4000 soft band detections and over 2800 hard band detections from a total area of $\sim 9$ deg$^2$. 
However, to accurately constrain the bright end of the XLF requires samples of higher luminosity X-ray sources identified from larger-area surveys. 
We thus supplement our sample with sources from large-area surveys carried out with \textit{ASCA} (in the hard band) and \textit{ROSAT} (in the soft band). 

For our large-area hard-band sample, we include the 34 sources from the \textit{ASCA} Large Sky Survey \citep[ALSS: ][]{Ueda99}, which covers a contiguous area of 5.8 deg$^2$ near the north Galactic pole. 
We adopt the optical identifications from \citet{Akiyama00}: 2 sources are optically identified as galaxy clusters, 1 is a star, 1 source remains unidentified and the remaining 30 are associated with AGNs, all of which have spectroscopic redshifts. 
We also select sources from the \textit{ASCA} Medium Sensitivity Survey \citep[AMSS,][]{Ueda01}, which combines data from a large number of \textit{ASCA} observations at high Galactic latitudes over an area $\sim82$ deg$^2$.
We include sources from the AMSSn sub-sample, selected in the hard (2--10 keV) band, with optical identifications presented by \citet{Akiyama03}. 
The sample includes 87 X-ray sources and has 100 per cent spectroscopic completeness.
We include additional sources with from the AMSSs sub-sample (Ueda \& Akiyama, private communication), which includes 20 AGN; 2 sources in this sample remain unidentified. 
We adopt area curves from the ALSS and AMSS from \citet{Akiyama03} and \citet{Ueda03} respectively. 

For our soft band sample we combine samples from
the \textit{ROSAT} bright survey \citep{Fischer98,Schwope00} and the Selected-Area--North survey \citep[SA--N:][]{Appenzeller98}, removing duplicate sources. 
We include sources with significant detections in the 0.5--2 keV band and adopt the unabsorbed flux estimates (corrected for Galactic absorption).
We cut our sample at flux limits of $f_\mathrm{0.5-2keV}> 3.6 \times 10^{-12}$ \ergscm\ for the RBS sample and $f_\mathrm{0.5-2keV}> 1\times 10^{-12}$ \ergscm\ for the SA--N sample.
These high flux limits ensure our sources all lie well above the sensitivity limits of the surveys, allowing us to adopt simple sensitivity curves that correspond to the entire area of each survey and cut off sharply at each of the flux limits (see Figure \ref{fig:acurves}). 
All the sources above these flux limits have spectroscopic classifications and we identify a total of 221 AGNs with spectroscopic redshifts (excluding BLLac type objects).

We note that, in contrast to our \chandra\ fields, our area curves for our large-area surveys do not account for the Poisson nature of the detection. 
As we restrict the samples from the large-area surveys to highly significant detections, this simplification will have a minimal effect on our XLF measurements. 
However, differences in the assumed spectral shape can have a significant impact on the estimate of a flux and thus the assumed sensitivity. 
Uncertainties in the spectral shape and the resulting differences in sensitivity are accounted for by our Bayesian methodology described below.

\subsection{Identification and masking of stars}
\label{sec:stars}

Bright stars can contaminate our photometry, leading to issues with counterpart identification and photometric redshift estimates in these regions. 
We have therefore masked out areas close to bright stars from all of our \chandra\ fields in a consistent manner. 
We searched for stars brighter than $V=15$ in the \textit{HST} Guide Star Catalog 2.3 \citep{Lasker08}. 
We masked all areas within a radius, $r$, given by 
\begin{equation}
r = (16-V) \times 6\arcsec
\end{equation}
where $V$ is the $V$-band magnitude from the Guide Star Catalog. 
We set a maximum masking radius of 40\arcsec. 
We have removed any X-ray sources within this radius from our samples (which can include the star itself or a nearby source).
We also excluded these regions when calculating the X-ray sensitivity and area curves.

We have also identified stars with fainter magnitudes that are detected at X-ray wavelengths and removed them from our samples. 
In the fields with Rainbow coverage (CDFS, CDFN, EGS) stars were identified by a range of colour and morphology criteria, as described in \citet{Barro11}. 
In the COSMOS and Bootes fields we applied a single colour criterion based on the region of colour-colour space occupied by stars in \citet{Ilbert09},
\begin{equation}
R-[3.6] < 3.0\times (R-I) -1.2
\end{equation}
where $[3.6]$ is the magnitude in the IRAC 3.6\micron\ imaging, $R$ is the magnitude in the Subaru $r+$ or NDWFS $R$ filter, and $I$ is the magnitude in the Subaru $i+$ or NDWFS $I$ filter.
For all fields, we also required that the X-ray sources exhibit a low X-ray-to-optical flux ratio, $\log f_\mathrm{X}/{f_\mathrm{opt}}<-1$, where the ratio is calculated as
\begin{equation}
\log \frac{f_\mathrm{X}}{f_\mathrm{opt}} = \log f_\mathrm{0.5-2keV} +5.4+ \frac{I}{2.5}.
\end{equation}
This cut ensures that we do not exclude bright QSOs from our sample that may satisfy the other stellar criteria.
When a spectroscopic classification is available (see Section \ref{sec:specz}) this over-rides our photometric classification. 
Of the 50 spectroscopically classified stars in our five \chandra\ fields, 40 (80 per cent) were also identified by our photometric procedure.

\subsection{Spectroscopic redshifts}
\label{sec:specz}

All of our \chandra\ surveys have been the subject of intense spectroscopic campaigns.These campaigns include large-scale follow-up of the general galaxy population, in addition to those directly targeting X-ray sources.

In the CDF-S we first searched for spectroscopic redshifts in the catalogue of \citet{Xue11}.
The majority of the spectroscopic redshifts for our CDFS-4Ms survey are taken from this catalogue, as well as some of our ECDFS survey (in the area that overlaps with the 4Ms data). 
We also searched for spectroscopic redshifts from the Arizona CDFS Environment Survey \citep{Cooper12b}, the spectroscopic sub-sample of sources from the MUSYC sample \citep{Cardamone10}, and PRIMUS \citep{Coil11}.

In the CDF-N we used spectroscopic redshifts from DEEP3 \citep{Cooper11} as well as the surveys of 
\citet{Trouille08,Barger08,Reddy06,Wirth04,Cowie04} and \citet{Steidel03}.

In the EGS we compiled spectroscopic redshifts from a number of surveys including DEEP2, DEEP3, the Canada-France Redshift Survey and \textit{MMT} follow-up of X-ray sources. See \citet{Nandra15} and references therein for full details.

In COSMOS we initially searched for spectroscopic redshifts of X-ray sources in the C-COSMOS catalogue of \citet{Civano12}. We also search for additional spectroscopic redshifts from the bright zCOSMOS survey catalogue \citep{Lilly09} and PRIMUS \citep{Coil11}.

In Bootes we adopt spectroscopic redshifts from the AGN and Galaxy Evolution Survey \citep{Kochanek12}.
For all fields we also searched for spectroscopic redshifts from the Sloan Digital Sky Survey \citep[SDSS:][]{York00}.

We matched the spectroscopic catalogues to the secure counterparts of our X-ray sources using a 2\arcsec\ search radius, corrected for any overall astrometric offset, and repeated the matching with a 1\arcsec\ radius.
We only adopt those spectroscopic redshifts that are flagged as high-quality, reliable redshifts in the original catalogues.
Table \ref{tab:sourcenumbers} gives the number of X-ray sources in our hard- and soft-band selected samples with reliable spectroscopic redshifts.

\subsection{Photometric redshifts}
\label{sec:photz}

The levels of spectroscopic completeness vary over our \chandra\ surveys from $\sim35$ per cent (in our ECDFS area) to $\sim75$ per cent (in the Bootes field). 
For the remaining X-ray sources we must resort to photometric redshift estimates, which are determined by fitting a set of template spectra to the observed spectral energy distributions (SEDs) of our sources.
Such redshifts can be highly uncertain, particularly when considering faint X-ray sources.
A key advantage of our Bayesian analysis (see Section \ref{sec:method}) is that we are able to account for the uncertainties in photometric redshifts by adopting probability distributions for the redshift, $p(z)$, rather than a single redshift estimate.
We thus require that our photo-$z$ approach recovers a $p(z)$ distribution that accurately reflects the uncertainties in our redshift estimates.

\begin{figure*}
\begin{center}
\includegraphics[width=0.99\textwidth, trim=40 20 0 0]{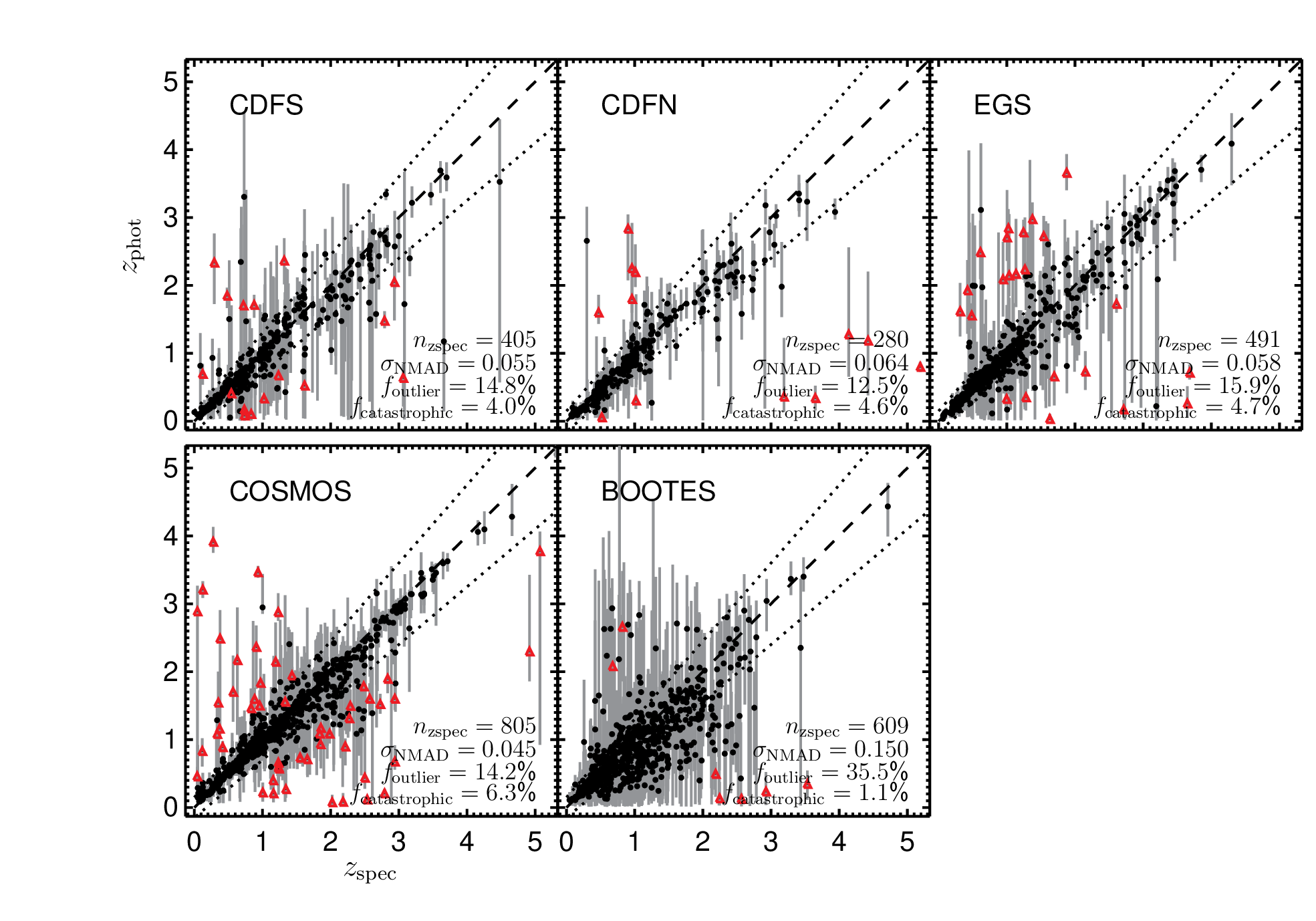}
\end{center}
\caption{Comparison of photometric redshifts ($z_\mathrm{phot}$) and spectroscopic redshifts ($z_\mathrm{spec}$) for our five \chandra\ fields. 
We plot the mean of the $p(z)$ distribution as the best estimate of the photometric redshifts (black circles); error bars indicate the 95 per cent central confidence interval. 
The dashed line indicates a 1:1 relation, whereas the dotted lines correspond to $\Delta z/(1+z_\mathrm{zspec})=\pm 0.15$ (sources where the best estimate of the photo-$z$ lies outside this range are flagged as outliers).
In the legend for each panel we give 
the number of X-ray sources with reliable spectroscopic redshifts ($n_\mathrm{zspec}$), 
the accuracy based on the normalized median absolute deviation ($\sigma_\mathrm{NMAD}$), 
the fraction of outliers ($f_\mathrm{outlier}$)
and the fraction of catastrophic failures ($f_\mathrm{catastrophic}$); see text for details. 
We highlight catastrophic failures with a red triangle. 
Over four of our fields (CDFS, CDFN, EGS and COSMOS) we obtain a consistent accuracy of $\sigma_\mathrm{NMAD}\approx 0.05$, with $\sim 15$ per cent outliers and $\sim 5$ per cent catastrophic failures.
In the Bootes field---our largest area, shallowest field---we have poorer accuracy and a higher outlier fraction, reflecting the more limited photometric imaging in this field. 
\thirddraft{While our photo-$z$ have a poorer accuracy ($\sigma_\mathrm{NMAD}$) and a higher outlier rate than in some prior works, we choose to adopt our estimates as they have been calculated in a consistent manner across all five of our \chandra\ fields and have representative errors that we can fully track via the $p(z)$ in our Bayesian methodology (see text for further discussion).}
}
\label{fig:zvsz}
\end{figure*}

We calculate photometric redshifts using the \textit{EaZY} photo-$z$ code \citep{Brammer08}.  
We use \textit{EaZY} in two-template mode, allowing for combinations of a galaxy and AGN template.
For the galaxy templates we adopt the ``pegase13" template set provided with \textit{EaZY}.
This template set consists of 259 synthetic galaxy templates drawn from the range of parameters described by \citet{Grazian06}. 
Additional ``star-forming and dusty" templates are included by applying the \citet{Calzetti00} reddening law for a range of different extinctions to a subset of the galaxy templates. 
We adopt seven AGN templates from the \citet{Salvato09} template set\footnote{\href{http://www.mpe.mpg.de/~mara/PHOTOZ_XCOSMOS/}{http://www.mpe.mpg.de/$\sim$mara/PHOTOZ\_XCOSMOS/}}, 
namely the Sey 1.8, Sey 2, Mrk231, pl\_TQSO1,  
pl\_QSOH, pl\_QSO and the S0-10\_QSO2-90 hybrid template.
We also include the Type-2 ``Torus" template from the \citet{Polletta07} library\footnote{\href{http://www.iasf-milano.inaf.it/~polletta/templates/swire_templates.html}{http://www.iasf-milano.inaf.it/$\sim$polletta/templates/swire\_templates.html}}, which is not included in the final \citet{Salvato09} set.
We note that some of these templates will include host galaxy contributions (particularly the Sey 1.8, Sey 2, Mrk231 and S0-10\_QSO2-90 templates).
This is not a major issue as we want to determine the distribution of possible redshifts, rather than perform an accurate host-AGN decomposition. 
The two-template mode in \textit{EaZY} allows any possible combination of one of our AGN templates and one of our galaxy templates and thus allows for a large amount of flexibility in the fitted template SEDs.

Another advantage of the \textit{EaZY} photo-$z$ code is that it allows for the inclusion of a ``template error function". 
This feature accounts for additional uncertainty in the template SED as a function of (rest-frame) wavelength and thus allows for the fact that our template set may not accurately represent the true diversity of SED shapes.
This uncertainty in the true range of template SEDs is particularly useful as our observed SEDs extend into the UV and mid-IR, where the templates are poorly calibrated, especially for AGNs.
In addition, as the optical emission from an (unobscured) AGN can vary on timescales of months-to-years, our observed SEDs may not be well-matched by a single underlying template.
The template error function can also allow for any overall calibration uncertainties in the diverse sets of photometric observations used to construct our observed SEDs.
We derive a template error function on a field-by-field basis---to ensure that it represents the calibration uncertainties in a given data set---using the basic procedure laid out in \citet{Brammer08}.
First, we attempt to fit the observed SEDs with our templates, fixing the redshift at the spectroscopic value, where available.
We then calculate
\begin{equation}
\Delta f_j = \frac{F_j - T_j}{F_j}
\end{equation}
where $F_j$ indicates the observed flux in a filter for source $j$, and $T_j$ is the flux from the best-fitting template.
We calculate $\Delta f_j$ as function of \emph{rest-frame} wavelength for all sources and filters and take the median of every 400 individual measurements across the rest-frame wavelength range. 
We subtract the median photometric error, in quadrature, to estimate the contribution from ``template error" to the uncertainty as a function of rest-frame wavelength. 
The value of the template error is typically around 10 per cent (in flux) but varies between $\sim4$ per cent and $\sim 20$ per cent depending on the wavelength and the data in a given field.

In Figure \ref{fig:zvsz} we compare our photo-$z$ estimates for X-ray sources to secure spectroscopic redshifts across our 5 \chandra\ fields. 
\thirddraft{The panels for the EGS and CDFS fields combine the deep survey areas (AEGIS-XD, CDFS-4Ms), where the best photometry is available, with the larger-area shallow surveys (AEGIS-XW, ECDFS).
We include all X-ray sources with a high-quality spectroscopic redshift in these plots and do not apply any cuts based on the estimated quality of the photo-$z$.
Thus, we include sources with extremely broad $p(z)$ distributions, which are often flagged as unreliable and excluded when assessing the success of photo-$z$ techniques.}

In each panel of  Figure \ref{fig:zvsz} we present a number of summary statistics: 
\begin{enumerate}[leftmargin=20pt]
\item $\sigma_\mathrm{NMAD}$: the accuracy based on the normalized median absolute deviation between the best photo-$z$ and the spectroscopic value, defined as $\sigma_\mathrm{NMAD}= 1.48 \times \mathrm{median}\left(| z_\mathrm{phot} - z_\mathrm{spec}| / (1 + z_\mathrm{spec})\right)$.
\item $f_\mathrm{outlier}$: the fraction of outliers, defined as the fraction of sources where
$| z_\mathrm{phot}-z_\mathrm{spec} |/(1+z_\mathrm{spec}) > 0.15$; 
\item $f_\mathrm{catastrophic}$: the catastrophic outlier rate, calculated as the fraction of sources where less than 5 per cent of the integrated $p(z)$ lies within $-0.15<(z-z_\mathrm{spec})/z_\mathrm{spec}< +0.15$.
\end{enumerate}

Over four of our fields (CDFS, CDFN, EGS and COSMOS) we obtain a consistent accuracy of $\sigma_\mathrm{NMAD}\approx 0.06$, with $\sim 15$ per cent outliers.
Our approach ensures we assign an appropriate uncertainty, traced by the $p(z)$, to the bulk of our sources and only $\sim5$ per cent of sources are thus flagged as catastrophic failures. 
In the Bootes field---our largest area, shallowest field---we have poorer accuracy and a much higher outlier fraction, reflecting the more limited photometric imaging in this field. 
However, the catastrophic outlier fraction is only $\sim1$ per cent, indicating that this additional uncertainty is represented by our $p(z)$ distributions.
We also note that in this field we have the highest spectroscopic completeness ($\sim75$ per cent) and so only use photometric redshifts for a relatively small fraction of the sources in our eventual analysis of the XLF; when we do resort to a photometric redshift we account for the large uncertainty in the redshift.

\begin{figure}
\includegraphics[width=\columnwidth]{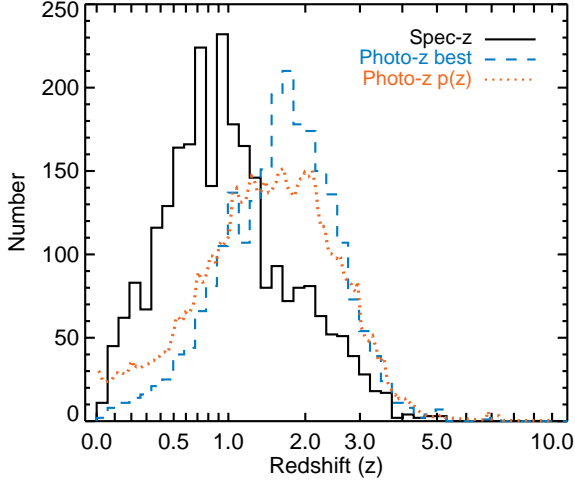}
\caption{Redshift distribution of X-ray sources in our five \chandra\ fields with high-quality  spectroscopic redshifts (black solid line) and those where we adopt photometric redshifts. 
The blue dashed line indicates the distribution of the ``best" photo-$z$ estimates (mean of the $p(z)$ distribution), whereas 
the orange dotted line shows the distribution obtained by combining the individual $p(z)$ distributions. 
}
\label{fig:zdist}
\end{figure}

In Figure \ref{fig:zdist} we plot the distribution of redshift estimates for sources in our \chandra\ fields where we have a spectroscopic redshift and those where we adopt the photometric redshift information.
Generally the sources with photometric redshifts lie at higher redshifts, a consequence of spectroscopic follow-up programmes being biased towards optically bright sources.
We also show the integrated contribution from the $p(z)$ of all the photo-$z$ sources. 
\refone{The integrated $p(z)$ is skewed towards lower redshifts.
This skew is due to the possibility that many potential high-redshift sources could actually lie at lower redshifts, which is reflected by their $p(z)$ and must be accounted for in measurements of the XLF.}

A small fraction of our X-ray sources ($<2$ per cent) lack a multiwavelength counterpart, precluding a photometric redshift estimate. 
We retain these sources in our analysis, ensuring completeness of our sample, but adopt a $p(z)$ that is constant in $\log(1+z)$ over our allowed redshift range ($0<z<10$). 
This reflects our lack of \emph{a priori} knowledge of the redshift; the X-ray flux information is retained and thus \emph{a posteriori}
(after folding the constant $p(z)$ through the final XLF) there may be a preferred redshift solution. 
The lack of a multiwavelength counterpart could imply that a high redshift solution should be given higher a priori preference for such sources, thus our approach is conservative. 
There is also a possibility that these X-ray sources lack counterparts as they are spurious detections, corresponding to positive fluctuations in the background count rate. Our analysis accounts for the Poisson nature of the X-ray detection and thus allows for this possibility.

Other estimates of photometric redshifts are available for X-ray sources in many of our fields \citep[e.g.][]{Barger03c, Cardamone10b, Hsu14}.
Many of these studies take additional steps to improve the quality of the photo-$z$ estimates.
These steps can include optimizing the template set \citep[e.g.][]{Luo10}, attempting to correct the observed photometry for variability \citep[e.g.][]{Salvato09}, or applying priors based on the source morphology and X-ray flux \citep[e.g.][]{Salvato11}. 
These studies often achieve a higher accuracy and lower outlier rate than our own photo-$z$ analysis. 
However, these additional steps can lead to underestimates of the true uncertainties in the photo-$z$.
Conversely, we retain a large set of possible templates 
to ensure we produce $p(z)$ distributions that account for the large uncertainties in the redshift and template degeneracies.
While we have higher outlier rates, our fraction of catastrophic failures remains $\lesssim5$ per cent, indicating that our $p(z)$ distributions are representing the uncertainties.
\thirddraft{
The nominal errors (i.e. the 68 per cent confidence intervals) on our photo-$z$ are also comparable to the residuals between our best photo-$z$ estimate and the available spectroscopic redshifts, in contrast to most previous work where errors may be underestimated by a factor $\sim 2-6$ \citep[e.g.][]{Luo10,Hsu14}.
Furthermore, we require the full $p(z)$ distribution, which most prior studies do not provide.}
\refone{
We thus choose to use our own photometric redshifts:
our poorer accuracy and higher outlier rates are accounted for and compensated by our Bayesian analysis that incorporates the full $p(z)$ information.}

\section{Bayesian methodology}
\label{sec:method}

In this paper we expand on the Bayesian methodology developed by A10, incorporating the distribution of X-ray absorption properties and accounting for the effects on the inferred shape and evolution of the XLF. 
Our method also accounts for the uncertainty in the measured X-ray flux (due to photon counting statistics), uncertainties in the redshift (for sources with photometric redshifts or no counterparts), uncertainties in the X-ray spectral shape, and the resulting uncertainty in the X-ray luminosity for an individual source.
Our methodology is described below.

\subsection{Probability distribution function for a single source}
\label{sec:singlesource}

For a single source in either our hard- or soft-band sample we can derive the probability distribution function for $z$, \LX\ and \NH\ based on our observed data for that source alone, which is given by $p(z, \lx, \nh \giv D_i)$ where $D_i$ indicates the observed data from source $i$ in our sample. 
This function is normalized such that
\begin{equation}
\int \dd z \int \dd \log \lx \int \dd \log \nh\;  p(z, \lx, \nh \giv D_i) = 1.
\end{equation}
We can re-write the probability distribution function as
\begin{equation}
p(z, \lx, \nh \giv D_i) = p(z \giv d_i)\; p(\lx, \nh \giv z, T_i, b_i)
\end{equation}
where $d_i$ indicates the multiwavelength data used to estimate the redshift and $T_i$ and $b_i$ correspond to the X-ray data for this source: the total observed X-ray counts in the given band and the estimated background respectively.

Our knowledge of $z$ is based on either a spectroscopic redshift, in which case we assume $p(z \giv d_i)$ is described by a $\delta$-function at the spectroscopic value, or a photometric redshift, when $p(z \giv d_i)$ is given by the $p(z)$ from our photometric redshift fitting described in Section \ref{sec:photz} above.
For the small fraction of sources where we were unable to identify a multiwavelength counterpart---and thus have no redshift information---we adopt a $p(z)$ distribution with a constant density in $\log(1+z)$ over $0<z<10$. 

The observed X-ray data can be described by a Poisson process. 
Thus, the likelihood of observing $T_i$ counts from a source is given by 
\begin{equation}
\mathcal{L}(T_i \giv c_i, b_i) = \frac{(c_i+b_i)}{T_i!} e^{-(c_i+b_i)}
\label{eq:poisslik}
\end{equation}
where $c_i$ is the X-ray count rate from source $i$ in the observed energy band and we have assumed that the expected background count rate, $b_i$, is well determined.
The uncertainty in $c_i$ is fully described by the Poisson likelihood given in Equation \ref{eq:poisslik}. 
However, to convert from a count rate, $c_i$, to an estimate of \LX\ and \NH\ (given $z$) we must assume a model for the X-ray spectral shape and fold this model through the appropriate instrumental response. 

\begin{figure}
\includegraphics[width=\columnwidth]{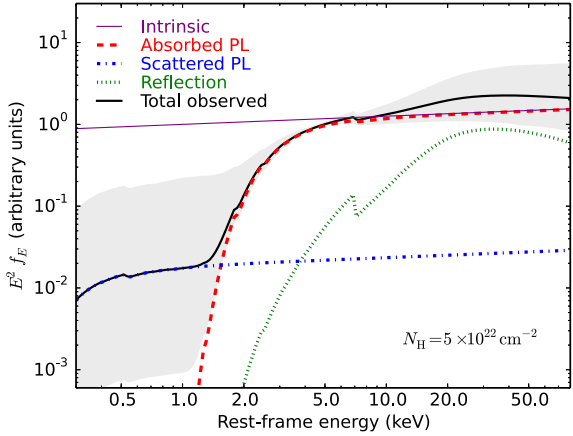}
\caption{
Our assumed X-ray spectral model for an AGN. We assume the intrinsic X-ray spectrum is a power law with photon index $\Gamma=1.9\pm0.2$ (purple line). 
The observed spectrum is absorbed by the intervening column density, fixed at $\nh=5\times10^{22}$ cm$^{-2}$ for this example (red dashed line). 
We also allow for a fraction ($f_\mathrm{scatt}\approx2$ per cent) of the intrinsic power-law to be scattered, unabsorbed, into the line-of-sight (blue dot-dashed line).
In addition, we allow for a component from Compton-reflection from cold, optically thick matter (such as a torus or accretion disc) that leads to the characteristic hump at high energies (green dotted line).
The black line indicates the total observed spectrum, while the grey region indicates the 95 per cent confidence interval on the spectrum, allowing for the range of possible spectral parameters (see Section \ref{sec:singlesource}, Table \ref{tab:xspecpriors}).
}
\label{fig:xspec}
\end{figure}

Figure \ref{fig:xspec} shows an example of our X-ray spectral model.
We assume the intrinsic X-ray continuum is described by a power law with photon index $\Gamma$ and a high energy cut-off (we fix the folding energy at 300 keV, although this has a negligible impact on the lower energies we observe).
The observed spectrum is attenuated along the line-of-sight by the intervening column density, \NH. We include both photoelectric absorption (using the \textit{wabs} model in {\sc XSpec}) and Compton-scattering (via \textit{cabs}), which suppresses the continuum further for Compton-thick column densities.
A fraction of this power-law, $f_\mathrm{scatt}$, is allowed to emerge as an unabsorbed component that is thought to be scattered into the line-of-sight by ionized gas in the vicinity of the AGN. 
We also include a contribution due to Compton reflection from cold, optically thick matter, which gives rise to a characteristic ``hump" in the X-ray spectrum at $\sim 30$ keV. 
Such a reflection component is expected due to reprocessing of the primary X-ray emission by a surrounding, dusty torus or an accretion disc. 
We adopt the \emph{pexrav} model \citep{Magdziarz95}, which is based on Monte-Carlo simulations of Compton reflection from a cold, optically thick slab of material. 
We assume the incident power-law has the same shape as the directly transmitted component.
We fix the inclination angle to 30\degrees\ as a representative value and allow the
intensity to be set by the normalization, $R$, relative to that expected from a slab subtending a solid angle of $2\pi$
\refone{(which is allowed to vary between 0 and 2, spanning the extreme cases of no reflection up to an effective $4\pi$ solid angle coverage).}
We absorb the reflection component by the same column density seen by the primary emission. This is a good assumption when the reflection arises from the accretion disc and also provides reasonable agreement with the shape and intensity of the reflection component based on sophisticated models of toroidal obscurers \citep[e.g][]{Brightman11}.
All components are subjected to Galactic absorption, with column densities determined from \citet{Dickey90} via the \textsc{Heasoft} \textsc{Nh} tool.
In \textsc{XSpec} terminology, our model is described by
\begin{eqnarray}
\mathrm{wabs*\big[ } 	&& \mathrm{ (1-constant)*zwabs*cabs*zpowerlw*zhighect }  \nonumber\\
	                   	&& +\; \mathrm{ constant*zpowerlw } \nonumber\\
		     		&&  +\; \mathrm{zwabs}*\mathrm{pexrav \big] }
\end{eqnarray}
where the constant corresponds to the scattered fraction, $f_\mathrm{scatt}$.
\secdraft{We note that more physically motivated models could be adopted to describe the X-ray spectrum, self-consistently modeling the emission, reflection and absorption due to the accretion disc or torus \citep[e.g.][]{Ross05,Brightman11}.
However, \citet{Buchner14} found that more simplistic models such as ours are generally sufficient to reproduce the observed spectral shape of individual, distant AGNs, especially considering our analysis uses broad-band fluxes rather than performing a detailed X-ray spectral analysis.}

To describe our X-ray spectral model requires three additional parameters---the photon index, $\Gamma$, the scattered fraction, $f_\mathrm{scatt}$, and the relative normalization of the reflection component, $R$---which we introduce as ``nuisance" parameters (collectively designated by $\XI$) in our Bayesian analysis.
For a given set of spectral parameters ($\XI$), $z$, \LX\ and \NH, we can determine the expected count rate, $c_i$, and thus link the probability distribution function for \LX, \NH\ and $\XI$ to the Poisson likelihood given in Equation \ref{eq:poisslik} above. Thus,
\begin{equation}
p(\lx, \nh, \XI \giv z, T_i, b_i) \propto \mathcal{L}\big(T_i \giv c_i(z, \lx, \nh, \XI ), b_i \big) \pi\big(\XI\big)
\end{equation}
where $c_i(z, \lx, \nh, \XI )$ is the expected count rate for a source with redshift $z$, luminosity \LX, absorption column \NH, and additional spectral parameters $\XI$ based on our X-ray spectral model, folded through the appropriate instrumental response for source $i$.

A priori, the spectral parameters $\XI$ for a given source are not well known; 
$\pi\big(\XI\big)$ denotes the prior distribution that we adopt for our spectral parameters, which describes the range of possible values and thus encapsulates the uncertainty in X-ray spectral shape.
We assume the photon index, $\Gamma$, is drawn from a Gaussian distribution with a mean of 1.9 and standard deviation of 0.2 
\citep[corresponding to the observed distribution of intrinsic photon indices in X-ray spectral studies of nearby AGNs, e.g.][]{Nandra07}. 
We assume the scattered fraction, $f_\mathrm{scatt}$, is drawn from a lognormal distribution with mean of 2 per cent and scatter of 0.8 dex based on the observed distribution of partial covering factors of sources in the \textit{Swift}/BAT sample from \citet{Winter09}.
For the reflection strength, $R$, we assume a uniform distribution in the range $0<R<2$.  
We thus allow for a large uncertainty in the strength of the reflection component, which is reasonable to encapsulate uncertainties in the geometry of the accretion disc and/or torus with our simplified modeling of the reflection.
Table \ref{tab:xspecpriors} summarizes this prior information.
The coloured lines in Figure \ref{fig:xspec} show each component evaluated at the prior mean of the spectral parameters and the black line corresponds to the total observed spectrum for these values.
The grey region indicates the 95 per cent confidence interval on the total observed spectrum, adopting our prior distributions for the spectral parameters.
We note that the large uncertainties in the scattered fraction, $f_\mathrm{scatt}$, lead to large uncertainties in the spectrum at softer energies for moderately and heavily absorbed sources and thus leads to large uncertainties in the intrinsic \LX.

\begin{table}
\caption{Priors on the spectral parameters for an individual X-ray AGN, $\XI$.}
\begin{center}
\begin{minipage}[c]{0.8\columnwidth}
\begin{center}
\begin{tabular}{c c c}
\hline
{Parameter} & {Prior type} & {Prior specification} \\
\hline
$\Gamma$         		     & Gaussian     & $1.9\pm0.2^a$\\
$\log f_\mathrm{scatt}$ & lognormal & $-1.73\pm0.8^b$\\
$R$ & constant    &  0 -- 2\\
\hline
\end{tabular}\\
\end{center}
$^{a}${Based on observed distribution from \citet{Nandra07}.}\\
$^{b}${Based on observed distribution from \citet{Winter09}.}
\end{minipage}
\end{center}
\label{tab:xspecpriors}
\end{table}

\begin{figure*}
\includegraphics[width=\textwidth]{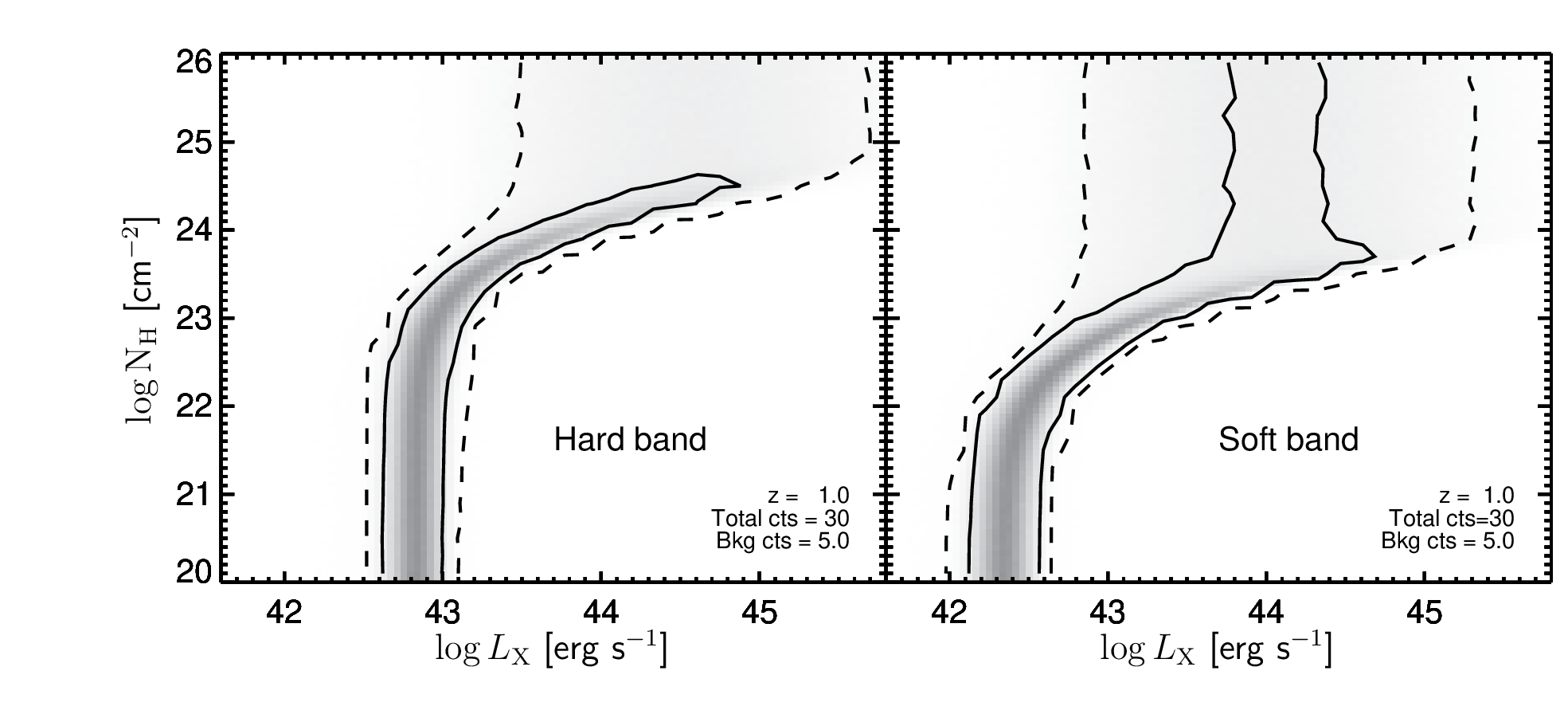}
\caption{
Example constraints on the rest-frame 2--10keV luminosity, \LX, and the line-of-sight absorption column density, \NH, for sources detected in the observed hard band (2--7keV, left) and soft band (0.5--2keV, right) in a typical deep \chandra\ observation.
In both cases, we assume 30 total counts detected in the observed energy band with an expected background of 5 counts, typical of our sample in the AEGIS-XD survey region.
We fix the redshift at $z=1.0$.
The grey shading indicates the probability distribution function for \LX\ and \NH, allowing for the Poisson uncertainty from the observed counts and assuming the X-ray spectral model shown in Figure \ref{fig:xspec}, marginalized over the uncertainty in the ``nuisance" spectral parameters ($\Gamma$, $f_\mathrm{scatt}$, and $R$).
The solid and dashed lines indicate the 68.3 per cent and 95.4 per cent (i.e. 1 and 2 $\sigma$ equivalent) confidence intervals on the joint parameter space. 
Detection in a single band does not constrain the value of \NH\ but does place constraints on the range of possible values of \LX.
At low values of \NH, the X-ray luminosity is constrained to within $\sim 0.25$ dex. 
However, for higher \NH\ values the X-ray luminosity must be correspondingly higher to produce the same number of observed counts.
Absorption affects the X-ray spectrum (and thus the constraints on \LX) for lower column densities in the soft band ($\nh\gtrsim 10^{22}$ cm$^{-2}$) than the hard band ($\nh\gtrsim 10^{23}$ cm$^{-2}$) at $z=1.0$.
}
\label{fig:lx_vs_nh}
\end{figure*}

To obtain the probability distribution function for \LX\ and \NH\ only (for a given $z$ and our observed X-ray data), we must marginalize over the nuisance spectral parameters. Thus,
\begin{equation}
p(\lx, \nh \giv z, T_i, b_i) = \int \dd \XI \; p(\lx, \nh, \XI \giv z, T_i, b_i).
\end{equation}
In Figure \ref{fig:lx_vs_nh} we show an example of $p(\lx, \nh \giv z, T_i, b_i)$ for a source detected in the hard band (left) and a source detected in the soft band (right) with \chandra.
In both cases we fix $z=1.0$ and assume $T_i=30$ total observed counts with an expected background of $b_i=5.0$ counts. 
We restrict the possible column densities to $20<\log \nh (\mathrm{cm}^{-2})<26$.
The shading and contours indicate the range of possible values of \LX\ and \NH\ for each source.
Detection in a single, broad energy band does not place constraints on the value of \NH, but we are able to place constraints on the range of possible values of \LX\ and how this depends on \NH. 
For the hard band example, we can constrain \LX\ to within $\sim0.25$ dex, provided the absorption column is $\nh\lesssim 10^{23}$ cm$^{-2}$.
If the column density is higher, then the same observed counts must correspond to a higher value of \LX. 
For a source detected in the soft band, absorption effects are apparent for lower column densities ($\nh\gtrsim10^{22}$ cm$^{-2}$).
At column densities $\nh\gtrsim 2 \times 10^{23}$ cm$^{-2}$ (at $z=1.0$) the observed flux in the 0.5--2keV band is dominated by the scattered component only; the intrinsic luminosity is poorly constrained but must be a factor $\sim 30$ greater than if the source had a lower \NH.

We note that the effect of absorption on the observed counts in the hard or soft energy bands will vary significantly with redshift. 
For example, at $z\sim3$, where the observed 0.5--2 keV energy band probes rest-frame energies $\sim2-8$keV, the inferred \LX\ is only affected for column densities $\gtrsim10^{23}$ cm$^{-2}$.
These redshift-dependent effects are fully accounted for in our calculation of the expected counts, $c_i(z, \lx, \nh, \XI )$, using our X-ray spectral model.

For the analysis in this paper, we treat each source in our hard- or soft-band samples as independent detections. 
Thus, the only X-ray data we use for an individual source is the detected counts (and the expected background), which as described above does not allow us to constrain the value of \NH\ for an individual source.
Instead, our approach (see Sections \ref{sec:likelihood} below) involves determining the overall distribution of X-ray luminosities and absorption column densities (given by the XLAF) that correctly describes both our hard- and soft-band samples.
Using this method, certain values of \LX\ and \NH\ may be disfavored \emph{a posteriori} (after performing our full analysis of the overall sample).
For example, higher values of \LX\ may be disfavored \emph{a posteriori} based on the shape of the XLF (that generally shows higher \LX\ sources should be rarer). 
Conversely, lower values of \NH\ may be disfavored if our overall XLAF requires a large fraction of absorbed sources.

Nevertheless, it is worth noting that a more sophisticated analysis could place constraints on the value of \NH\ (or indeed any of the other spectral parameters) for an individual source \emph{before} performing fits to the overall samples. 
We could combine the information on the counts in both the hard and soft bands for an individual source to place constraints on the underlying X-ray spectrum, based on this hardness ratio information \citep[e.g.][]{Hasinger08, Xue10}.
Alternatively, the full X-ray spectrum could be extracted for a source and fitted with our spectral model or any other well-motivated model \citep[e.g.][]{Tozzi06,Buchner14}.
However, accurately incorporating this information into our Bayesian analysis---and fully propagating the uncertainties in the estimated spectral parameters, \LX\ and \NH---is beyond the scope of this paper.

\subsection{The likelihood function for the overall sample}
\label{sec:likelihood}

Here, we describe how we combine the individual probability distribution functions for each source in our sample with a given model of the XLAF to construct the likelihood function for our overall sample.

Following A10 \citep[see also][]{Loredo04}, we assume our individual sources are effectively Poisson points drawn from a distribution described by the XLAF and the overall sample selection function.
The expected number of sources in our hard-band sample is given by
\begin{eqnarray}
\mathcal{N}_\mathrm{hard} (\nhfpars) & = &
\int \dd \log \lx 
\int \dd \log \nh 
\int \frac{\dd V}{\dd z} \dd z 
\int \dd \XI  \\
&& \bigg[ \psi(\lx, z, \nh \giv \nhfpars) \;
 \pi(\XI) \sum_{j=1}^{N_\mathrm{fields}} A_j ( \lx, z, \nh, \XI ) \bigg]\nonumber
\label{eq:nexpected}
\end{eqnarray}
where $\psi(\lx, z, \nh \giv \nhfpars)$ is the XLAF---the differential co-moving number density of AGNs per logarithmic interval in \LX\ and \NH---and is described by a model with a given set of parameters, $\nhfpars$. 
We note that the XLF, $\phi(\lx,z\giv\nhfpars)\equiv \frac{d\Phi(\lx,z)}{d\log\lx}$ (i.e. the differential co-moving number density of AGNs per logarithmic interval in \LX), can be recovered from the full XLAF by integrating over \NH. Thus,
\begin{equation}
\phi(\lx,z\giv \nhfpars) = \int d\log\nh \; \psi(\lx,z,\nh \giv \nhfpars).
\end{equation}
We investigate and compare various different parametrizations of the XLAF (or the XLF directly), which are described in Sections \ref{sec:xlf} and \ref{sec:absdist} below. 

The distribution of possible spectral parameters is given by $\pi(\XI)$, the prior on $\Gamma$, $f_\mathrm{scatt}$ and $R$ that we describe in Section \ref{sec:singlesource} above.
$\frac{\dd V}{\dd z}$ is the differential comoving volume per unit area, as a function of $z$. 
$A_j(\lx, z, \nh, \XI)$ is the area for field $j$ that is sensitive to a source of luminosity \LX, redshift $z$, absorption column \NH, and spectral parameters $\XI$, effectively giving the probability of a source entering our sample.
To calculate $A_j(\lx, z, \nh, \XI)$ we use our X-ray spectral model described in Section \ref{sec:singlesource} above to convert from a given \LX, $z$, \NH\ and $\XI$ to an expected count rate, $c_i(z, \lx, \nh, \XI ).$\footnote{
We assume an appropriate Galactic hydrogen column density, $\nh^\mathrm{Gal}$, for each survey. For the \textit{ROSAT} surveys and AMSS we assume the median $\nh^\mathrm{Gal}$ of the sample.}
We then convert the count rate to an ``effective flux" using our standard conversion factors and determine the area sensitive to this flux from our sensitivity curves (see Section \ref{sec:xray} and Figure \ref{fig:acurves}). 
This calculation is performed for each of our $N_\mathrm{fields}$ fields and the total area is summed. 

The likelihood function may now be constructed from the product of the probability distribution functions for the individual sources in the sample and the probability that no other sources were detected \citep[A10, ][]{Loredo04}.
For a Poisson process, the probability of no detected sources (in the hard-band sample) for a given model XLAF is $e^{-\mathcal{N}_\mathrm{hard}(\nhfpars)}$.
Thus, for the hard-band sample the likelihood is
\begin{eqnarray}
 \mathcal{L} &( {\boldsymbol D}_\mathrm{hard}&  \giv \nhfpars) =
e^{-\mathcal{N}_\mathrm{hard}(\nhfpars)} \times \nonumber\\
&
\displaystyle \prod_{i=1}^{n_\mathrm{hard}} \Bigg( & 
\int \dd \log \lx 
\int \dd \log \nh 
\int \frac{\dd V}{\dd z} \dd z \label{eq:likelihood}\\
&& \bigg[ p(z, \lx, \nh \giv D_i) \;
\psi(\lx, z, \nh \giv \xlfpars) \bigg] \Bigg) \nonumber
\end{eqnarray}
where $\mathcal{L} ( {\boldsymbol D}_\mathrm{hard}  \giv \nhfpars)$
is the likelihood of obtaining the observed data from the entire hard-band sample (${\boldsymbol D}_\mathrm{hard}$) for
a given set of parameters ($\nhfpars$) for the XLAF, 
and $p(z, \lx, \nh \giv D_i)$ is the probability distribution function for $z, \lx$ and \NH\ for an individual hard-band source with observed data ($D_i$), marginalized over the nuisance spectral parameters ($\XI$).
The product is taken over all $n_\mathrm{hard}$ sources in our hard-band sample.

The likelihood function for the soft-band sample can be constructed in a completely analogous manner to that described for the hard-band sample above.
As we treat the hard- and soft-band samples as independent samples, the overall likelihood function for both samples is simply given by the product of the two likelihood functions. Thus,
\begin{equation}
\mathcal{L} ( {\boldsymbol D}_\mathrm{hard}, {\boldsymbol D}_\mathrm{soft}  \giv    \nhfpars) 
= \mathcal{L} ( {\boldsymbol D}_\mathrm{hard}  \giv    \nhfpars) \;
    \mathcal{L} ( {\boldsymbol D}_\mathrm{soft}  \giv  \nhfpars) 
\end{equation}
where we assume the same underlying XLAF (with parameters $\nhfpars$) can describe the observed data from both samples. 
This approach is taken in Section \ref{sec:absdist} where we combine the two samples to 
determine the form of the XLAF that adequately describes our two samples.

\subsection{Including the contribution of normal galaxies}
\label{sec:gallfunc}

So far we have implicitly assumed that all of the X-ray point sources in our two samples are actually due to the emission from AGNs, rather than an alternative process.
We have already excluded stars from our sample, either by masking known bright stars or according to the criteria given in Section \ref{sec:stars}.
However, we have so far made no attempt to exclude normal galaxies from our sample, where the X-ray emission may be predominantly due to the combined emission from X-ray binaries within the galaxy---with the total luminosity tracing the star-formation rate of the galaxy \citep[e.g.][]{Ranalli03, Mineo14}---rather than due to an AGN.
The total luminosity produced by a galaxy is usually much lower than could be produced by an AGN, even in galaxies with extremely high star-formation rates. 
Thus, a common solution is to simply cut any source with an X-ray luminosity below $\lx\lesssim10^{42}$ \ergs, and not consider sources below this luminosity any further in measurements of the XLF of AGNs \citep[e.g.][]{Barger05,Silverman08}.
Such a cut is fairly conservative as a high fraction of the sources below this \LX\ cut will be dominated by AGN emission rather than star formation.
Including such sources can also improve constraints on the faint end of the XLF.

In our Bayesian analysis we do not assign a single \LX\ to any individual source; we have a probability distribution describing the range of possible \LX.
Thus, we are unable to apply a strict luminosity cut to identify and remove normal galaxies from our sample.
Instead, we account for the potential contamination of our AGN sample by including the X-ray luminosity function of normal galaxies (hereafter, the ``galaxy luminosity function" or GLF) directly in our analysis.
We assume the GLF is described by a Schechter function, 
\begin{eqnarray}
\phi_\mathrm{gal}(\lx, z &&\giv \galpars) \; \dd \log \lx = \label{eq:gallfunc}\\
&& A \left(\frac{\lx}{L_*(z)}\right)^{-\alpha} \exp\left(-\frac{\lx}{L_*(z)}\right) \dd \log \lx
\nonumber
\end{eqnarray}
where $\phi_\mathrm{gal}(\lx, z \giv \galpars)$ is the GLF for a given set of parameters, $\galpars$.
We allow for pure luminosity evolution of the GLF by allowing the characteristic luminosity, $L_*$, to vary with redshift as
\begin{equation}
\log L_*(z) = \begin{cases}  \log L_{0} + \beta \log(1+z) & \text{if } z < z_{c} \\
 							 	 \log L_{0} + \beta \log(1+z_c)       & \text{if } z \ge z_{c}
					  \end{cases}
\end{equation}
where $L_0$ is the characteristic luminosity at $z=0$ and $z_c$ is a cutoff redshift. 
The GLF function is thus described by five parameters, $\galpars = [A, \alpha, L_0, \beta, z_c]$.
We adopt fairly strong (Gaussian or lognormal) priors on the principal parameters $A$, $\alpha$, and $L_0$ that describe the GLF at $z=0$, based on a previous study using a completely independent dataset \citep[see Table \ref{tab:galpriors}]{Georgakakis06b}. 
As the redshift evolution is less well constrained, we apply a weak, Gaussian prior on $\beta$ \citep[based on][]{Georgakakis07} and adopt a constant prior on $z_c$, allowing any value in the range $0.5<z_c<2.5$.
Ultimately, we allow our own data to constrain these parameters through our exploration of the posterior parameter space (see Section \ref{sec:nestedsamp} below).

\begin{table}
\caption{Priors on the galaxy luminosity function parameters, $\galpars$.}
\begin{center}
\begin{tabular}{c c c}
\hline
{Parameter} & {Prior type} & {Prior specification}\\
\hline
$\log A$ (Mpc$^{-3}$ dex$^{-1}$) & lognormal & $-3.96 \pm 0.25^{a}$\\
$\alpha$					& Gaussian & $0.76\pm 0.10^{a}$ \\
$\log L_0$ (\ergs)			& lognormal & $41.15\pm0.14^{a}$\\
$\beta$					& Gaussian  & $2.7\pm0.5^{b}$ \\
$z_c$					& constant    & 0.5 -- 2.5\\
\hline
\end{tabular}
\end{center}
$^{a}${Based on measurements in \citet{Georgakakis06b}, converted to rest-frame 2--10keV luminosities and our parametrization of the galaxy luminosity function (Equation \ref{eq:gallfunc}).}\\
$^{b}${Weak prior, based on \citet{Georgakakis07}.}
\label{tab:galpriors}
\end{table}

We must now modify the likelihood function for our overall sample to include the contribution of normal galaxies.
The expected number of galaxies in our hard-band X-ray sample is given by
\begin{eqnarray}
\mathcal{N}_\mathrm{gal,hard} (\galpars) & = &
\int \dd \log \lx 
\int \frac{\dd V}{\dd z} \dd z 
\int \dd \Gamma_\mathrm{gal} \\
&&  \bigg[ \phi_\mathrm{gal}(\lx, z \giv \galpars) \;
 \pi(\Gamma_\mathrm{gal})
\sum_{j=1}^{N_\mathrm{fields}} A_j ( \lx, z, \Gamma_\mathrm{gal} ) \bigg]. \nonumber
\label{eq:ngal}
\end{eqnarray}
An analogous expression gives the predicted number of galaxies in the soft-band sample, $\mathcal{N}_\mathrm{gal,soft}(\galpars)$.

We assume the spectrum for a galaxy's X-ray emission can be described by a single power law (subjected only to Galactic absorption), and thus is described by a single parameter, the photon index $\Gamma_\mathrm{gal}$. 
We adopt a Gaussian distribution with mean 1.9 and standard deviation 0.2 \citep[e.g.][]{Young12} as the prior, $\pi(\Gamma_\mathrm{gal})$, on the range of possible photon indices.

Our likelihood function (for the hard-band sample) must then be modified to include the expected number of normal galaxies and allow for the possibility that an individual detection is associated with a galaxy.
Thus, Equation \ref{eq:likelihood} is modified to
\begin{eqnarray}
 \mathcal{L} ( {\boldsymbol D}_\mathrm{hard}  \giv  \nhfpars, \galpars) &=&
e^{-\big(\mathcal{N}_\mathrm{hard}(\nhfpars) + \mathcal{N}_\mathrm{gal,hard}(\galpars)\big)} \times \nonumber\\
&&
\displaystyle \prod_{i=1}^{n_\mathrm{hard}} \Big(  
P^\mathrm{[AGN]}_i + P^\mathrm{[gal]}_i \Big)
\end{eqnarray}
where
\begin{eqnarray}
P^\mathrm{[AGN]}_i & = &    \label{eq:pagn}
\int \dd \log \lx 
\int \dd \log \nh 
\int \frac{\dd V}{\dd z} \dd z \\ 
&& \; \bigg[ (1-\eta_i)
   p(z, \lx, \nh \giv D_i) \;
\psi(\lx, z, \nh \giv \nhfpars) \bigg] \nonumber
\end{eqnarray}
corresponds to the probability that source $i$ is an AGN, for a given realization of the XLF and \NH\ function,
and
\begin{eqnarray}
P^\mathrm{[gal]}_i & = &    \label{eq:pgal}
\int \dd \log \lx 
\int \frac{\dd V}{\dd z} \dd z \\
&& \;\; \bigg[ \eta_i \; p_\mathrm{gal}(z, \lx \giv D_i) \;
\phi_\mathrm{gal}(\lx, z \giv \galpars) \;
\bigg] \nonumber
\end{eqnarray}
corresponds to the probability that source $i$ is a galaxy, for a given realization of the GLF.
$p_\mathrm{gal}(z, \lx \giv D_i)$ describes the probability distribution function for \LX\ and $z$ under the assumption that source $i$ is a galaxy and is marginalized over the single nuisance spectral parameter, $\Gamma_\mathrm{gal}$. 
Thus,
\begin{eqnarray}
p_\mathrm{gal}(z, \lx \giv D_i) &\propto& \; p(z \giv d_i) \times \\
&&	\int \dd \Gamma_\mathrm{gal}\;
						\mathcal{L}\Big(T_i \giv c'_i(z, \lx, \Gamma_\mathrm{gal}), b_i\Big) \; \pi(\Gamma_\mathrm{gal})\nonumber
\end{eqnarray}
where $c'_i(z, \lx, \Gamma_\mathrm{gal})$ is the expected counts based on our single power-law X-ray spectral model (for a galaxy with redshift $z$, rest-frame 2--10 keV X-ray luminosity \LX, and photon index $\Gamma_\mathrm{gal}$)
and $\mathcal{L}\big(T_i \giv c'_i(z, \lx, \Gamma_\mathrm{gal}), b_i\big)$ is the Poisson likelihood (cf. Equation \ref{eq:poisslik}).
We retain the same redshift probability distribution function, $p(z \giv d_i)$, for the galaxy as was assumed in the case of an AGN.

The parameter $\eta_i$ in Equations \ref{eq:pagn} and \ref{eq:pgal} represents any prior knowledge as to whether source $i$ is a galaxy or an AGN.
This prior knowledge could be based on some other available data; for example an optical spectrum, morphological information, or the multiwavelength SED.
However, for this work we simply assume no \emph{a priori} preference for a galaxy or AGN; thus we set $\eta_i=0.5$ for all sources. 
Instead, we allow our prior knowledge of the GLF itself (the relatively strong priors that we set on the parameters, $\galpars$) to drive our \emph{a posteriori} inferences on the probability that a source is an AGN or a galaxy.
As the GLF drops rapidly above $\lx\approx10^{41-42}$\ergs, any X-ray source with a higher luminosity is very unlikely to be associated with a galaxy.
Due to their unabsorbed spectrum, galaxies will just as easily be seen in the soft band as the hard band; indeed, our approach naturally accounts for the rise in the number of sources detected in the soft-band at the very faintest fluxes \citep[e.g.][]{Georgakakis08, Lehmer12}.

As in Section \ref{sec:likelihood} above, the final likelihood function for the full data from both the hard- and soft-band samples, 
$\mathcal{L} ( {\boldsymbol D}_\mathrm{hard}, {\boldsymbol D}_\mathrm{soft}  \giv     \nhfpars, \galpars)$,
is given by the product of the hard-band and soft-band likelihood functions.

\subsection{Parameter estimation and model comparison}
\label{sec:nestedsamp} 

Having determined the likelihood function for our overall sample, our knowledge of the XLAF and GLF can be fully described by the posterior probability distribution function,
\begin{equation}
p(\nhfpars, \galpars \giv {\boldsymbol D}, M) = 
\frac{ \mathcal{L}({\boldsymbol D} \giv \nhfpars, \galpars) \; 
          \pi(\nhfpars, \galpars \giv M) }
       { p({\boldsymbol D} \giv M) }
\label{eq:posterior}       
\end{equation}
where ${\boldsymbol D}$ indicates all of our observed data and we have introduced $M$ to indicate the hypothesis that a particular model parametrization for the XLAF and GLF describes our observed data. 
The model itself is described by the parameters $\nhfpars, \galpars$; our prior knowledge of these parameters for a given model is described by 
\begin{equation}
\pi(\nhfpars, \galpars \giv M) = 
\pi(\nhfpars \giv M_\mathrm{XLAF}) \; \pi(\galpars \giv M_\mathrm{gal})
\end{equation}
where $M_\mathrm{XLAF}$ and $M_\mathrm{gal}$ indicate our models for the XLAF and GLF respectively.

The denominator in Equation \ref{eq:posterior}, $p({\boldsymbol D} \giv M)$, is known as the Bayesian evidence, $\mathcal{Z}$. 
This factor is calculated by integrating the likelihood function over the prior parameter space:
\begin{eqnarray}
\mathcal{Z} &=& p({\boldsymbol D} \giv M)  \label{eq:evidence} \\
	&=&  \int\dd\nhfpars \int\dd\galpars \;
				\mathcal{L}({\boldsymbol D} \giv \nhfpars, \galpars) \;           
				\pi( \nhfpars, \galpars \giv M).\nonumber
\end{eqnarray}
The posterior probability for a particular model (given the data) can be calculated from the evidence and any prior knowledge of the probability that the model is correct:
\begin{equation}
p(M \giv {\boldsymbol D}) = p({\boldsymbol D} \giv M) \; \pi(M)
\end{equation}
The ratio of the posterior probabilities for different models, known as the Bayes factor, can be used for model comparison; if neither model is favored \emph{a priori}, then the Bayes factor is simply given by the ratio of the evidences,
$\mathcal{Z}$, of each model.
We adopt this approach to compare between different models for the XLAF.
We report the difference in logaritmic evidence between two models, 
\begin{eqnarray}
\Delta \ln \mathcal{Z} & = & \ln \mathcal{Z}_1 - \ln \mathcal{Z}_2 \\
					   & = & \ln \left( \frac{p(M_1 \giv {\boldsymbol D})}{p(M_2 \giv {\boldsymbol D})}\right)\nonumber
\end{eqnarray}
where $\mathcal{Z}_1$ is the Bayesian evidence for model 1 ($M_1$) and $\mathcal{Z}_2$ is the Bayesian evidence for model 2 ($M_2$).
A difference in the logarithmic evidence of $\Delta \ln \mathcal{Z}>4.6$ (corresponding to posterior odds of 100:1) indicates very strong evidence in favour of the model 1, whereas $\Delta \ln \mathcal{Z}<-4.6$ indicates very strong evidence in favor of model 2 \citep{Jeffreys61}. 
If $-4.6<\Delta \ln  \mathcal{Z} < 4.6$ then, while we may still favour the model with the higher evidence, we cannot decisively rule out one model in favour of the other. 

This Bayesian model comparison approach fully incorporates uncertainties in the underlying parameters and naturally applies Occam's Razor, favouring a simpler model with fewer free parameters over a more complex one, unless the latter is required by the data \citep[e.g.][]{Kass95}.
Our different models for the XLF or the full XLAF are described in Sections \ref{sec:xlf} and \ref{sec:absdist} below (along with a description of the appropriate priors on the different parameters).
Our model for the GLF, described in Section \ref{sec:gallfunc} above, is kept the same throughout the analysis of this paper.

To perform the integration in Equation \ref{eq:evidence} and thus calculate the evidence for a given model of the XLAF, we adopt the \textsc{MultiNest} algorithm \citep{Feroz09}, which extends the ``nested sampling" approach of \citet{Skilling04} to allow for efficient exploration of large, multimodal parameter spaces.
We use the \textsc{MultiNest} code v3.6 with 400 ``live points" and an efficiency factor of 0.3.
The algorithm also provides estimates of the posterior probability distribution for the parameters, 
$p(\nhfpars, \galpars \giv {\boldsymbol D}, M)$,
which we use to obtain our ``best estimates" of the parameters for a given model. 
We choose to report the posterior mean for a parameter as our best estimate.
To characterize the uncertainty in a parameter, we report the 68.3 per cent (i.e. 1$\sigma$ equivalent) central confidence interval, marginalized over all other parameters.

\subsection{Binned estimates}
\label{sec:binned}

To aid in the visualization of our results and the comparison between different models, it is useful to produce binned estimates of the XLF (including the GLF at low luminosities) in fine bins of \LX\ and $z$ for both the hard- and soft-band samples.
To achieve this, we adopt the $N_\mathrm{obs}/N_\mathrm{mdl}$ method \citep{Miyaji01}, as expanded on in A10. 
This method compares the observed number of sources in a given \LX-$z$ bin to that predicted based on a model fit.
The binned estimate is then given by 
\begin{equation}
\phi_b \approx 
\frac{N_\mathrm{obs}}{N_\mathrm{mdl}}
\bigg[\phi(L_b, z_b \giv \hat{\nhfpars}) + \phi_\mathrm{gal}(L_b, z_b \giv \hat{\galpars}) \bigg] 
\end{equation}
where $L_b$ and $z_b$ are the luminosity and redshift of the center of the bin, $\phi_b$ is the binned estimate, and $\hat{\nhfpars}$ and $\hat{\galpars}$ are our best (\emph{a posteriori}) estimates of the parameters for the XLAF and GLF.
 
The predicted number of sources in a bin, $N_\mathrm{mdl}$, for a given model XLAF and GLF, can be calculated from Equations \ref{eq:nexpected} and \ref{eq:ngal}, restricting the integration to the appropriate range in \LX\ and $z$. 
To calculate the observed number of sources, $N_\mathrm{obs}$, we must account for the distribution of possible values of \LX\ (and in many cases $z$) for an individual source.
A single source can make a partial contribution to the \emph{effective} observed number in various \LX-$z$ bins.
To calculate $N_\mathrm{obs}$, we sum the partial contributions of the individual sources to each \LX-$z$ bin:
\begin{equation}
N_\mathrm{obs} = \displaystyle \sum_{i=0}^n \int_{\log L_\mathrm{lo}}^{\log L_\mathrm{hi}} \dd \log \lx
	\int_{z_\mathrm{lo}}^{z_\mathrm{hi}}	\dd z \;
		p(\lx, z \giv D_i, \hat{\nhfpars}, \hat{\galpars}).
\end{equation}
where $\log L_\mathrm{lo}$, $\log L_\mathrm{hi}$, $z_\mathrm{lo}$ and $z_\mathrm{hi}$ indicates the limits of the bin 
and 
$p(\lx, z \giv D_i,  \hat{\nhfpars}, \hat{\galpars})$ is the \emph{a posteriori} probability distribution for \LX\ and $z$ for source $i$, given the data for that source, $D_i$, and our best estimate of the overall XLAF and GLF (described by our best estimates of the parameters, $\hat{\nhfpars}, \hat{\galpars}$).
Thus,
\begin{eqnarray}
&&p(\lx, z   \giv D_i, \hat{\nhfpars}, \hat{\galpars})  \propto \\
&& \;\;\;\;\bigg[ p_\mathrm{gal}(z, \lx \giv  D_i) \phi_\mathrm{gal}(\lx,  z \giv \hat{\galpars}) \frac{\dd V}{\dd z} \bigg] \nonumber\\
&&\;\;+ \displaystyle \int \dd \log \nh  \bigg(
  p(z, \lx, \nh \giv D_i) 
\psi(\lx, z, \nh \giv \hat{\nhfpars}) \frac{\dd V}{\dd z} \bigg) \nonumber
\end{eqnarray}
where we have marginalized over the probability distribution for \NH\ for source $i$.
We assign an error to each binned estimate based on the approximate Poisson error in the effective number of observed sources in each bin. 
We only plot points for bins with $N_\mathrm{obs}\ge 1$.

Our binned estimates of the XLF calculated in this manner will depend on the underlying XLAF model. 
For example, an XLAF that requires a high fraction of heavily absorbed sources would increase the binned estimate of the XLF based on the soft-band sample, to account for absorbed sources that would by missing from our sample.
Furthermore, the individual binned estimates are not independent as a single source can enter multiple bins.
Nevertheless, the binned estimates serve to visualize and compare how well the data from our soft- and hard-band samples are described by a particular model.

\section{X-ray luminosity functions from the hard-band and soft-band samples}
\label{sec:xlf}

In this section we investigate and compare different models to describe the evolution of the XLF, neglecting the effects of absorption. 
For this initial study, we fit the hard-band and soft-band samples separately and determine the model parameters for the XLF from each sample. 
We adopt a log-constant distribution for \NH\ over a limited range ($20<\log \nh<21$) but otherwise use our Bayesian methodology described in Section \ref{sec:method} above (accounting for the distributions of spectral parameters and including the GLF).
Given the lack of absorption corrections, the measurements of the XLF presented in this section can be seen as tracing the ``observed'' luminosity, corrected to 2--10keV rest-frame values.
In Sections \ref{sec:ldde} to \ref{sec:fdpl} below we describe our various parametrizations of the XLF.
Section \ref{sec:xlfresults} summarizes our findings.

\subsection{Luminosity-dependent density evolution (LDDE)}
\label{sec:ldde}

Following a number of studies of the XLF of AGNs based on both soft and hard X-ray selected samples
\citep[e.g.][]{Miyaji00,Ebrero09,Ueda14},
we first investigated luminosity-dependent density evolution (LDDE) model parametrizations.
These models start with the assumption that the XLF at $z=0$ can be described by a smoothly 
connected double power-law, 
\begin{eqnarray}
\phi(\lx, z=0) & =  & \frac{\dd \Phi(\lx, z=0)}{\dd \log \lx} \nonumber\\
	& = & 
K\left[ \left(\frac{\lx}{L_*}\right)^{\gamma_1} + \left(\frac{\lx}{L_*}\right)^{\gamma_2} \right]^{-1}
\label{eq:dpl}
\end{eqnarray}
where $\gamma_1$ is the faint-end slope, $\gamma_2$ is the bright-end slope, $L_*$ is the characteristic break luminosity,
and $K$ is the overall normalization.
The XLF is then modified by an evolution term, $e(z,\lx)$, which is a function of both redshift and luminosity:
\begin{equation}
\phi(\lx, z) = \phi(\lx, z=0) \; e(z,\lx).
\end{equation}
We adopt two different models for the evolution term.
The first, hereafter referred to as the LDDE1 model, is taken from \citet{Ueda03}, and assumes that $e(z,\lx)$ is a power-law function of $(1+z)$, with different indices above and below a cutoff redshift, $z_{c1}$, which is itself a function of $z$. Thus,
\begin{equation}
\resizebox{\hsize}{!}{$
e(z,\lx) =  \begin{cases} (1+z)^{e_1} & \left[z \le z_{c1}(\lx)\right] \\
	 			  (1+z_{c1}(\lx))^{e_1} \left( \frac{1+z}{1+z_{c1}(\lx)}\right)^{e_2}   & \left[z > z_{c1}(\lx)\right]
		\end{cases}
$}
\end{equation}
where
\begin{equation}
z_{c1}(\lx)= \begin{cases} z_{c1}^*   & \left[\lx \ge L_{a1}\right] \\
				z_{c1}^* \left(\frac{\lx}{L_{a1}}\right)^{\alpha_1}      &     \left[\lx < L_{a1}\right].
	       \end{cases}	       
\end{equation}

For our second paraterization (hereafter LDDE2), we adopt the further refinements of LDDE1 proposed by \citet{Ueda14}.
The evolutionary term is modified to allow for a stronger density evolution at the highest redshifts, above a second cutoff redshift, $z_{c2}$. Thus,
\begin{equation}
\resizebox{\hsize}{!}{$
e(z,\lx) =  \begin{cases} (1+z)^{e_1} & \left[z < z_{c1}(\lx)\right] \\
			             (1+z_{c1}(\lx))^{e_1}
						\left( \frac{1+z}{1+z_{c1}(\lx)}\right)^{e_2}   & 
							\left[z_{c1}(\lx) < z < z_{c2}(\lx)\right]\\
			             (1+z_{c1}(\lx))^{e_1}
						\left( \frac{1+z_{c2}(\lx)}{1+z_{c1}(\lx)}\right)^{e_2}  
						\left( \frac{1+z}{1+z_{c2}(\lx)}\right)^{e_3}   & 
							\left[z > z_{c2}(\lx)\right].
		\end{cases}
$}		
\end{equation}
where the cutoff redshifts are both functions of \LX, given by
\begin{equation}
	z_{c1}(\lx)= \begin{cases} z_{c1}^*   & \left[\lx \ge L_{a1}\right] \\
				z_{c1}^* \left(\frac{\lx}{L_{a1}}\right)^{\alpha_1}      &     \left[\lx < L_{a1}\right]
	\end{cases}	      
\end{equation}
and 
\begin{equation}
	z_{c2}(\lx)= \begin{cases} z_{c2}^*   & \left[\lx \ge L_{a2}\right] \\
				z_{c2}^* \left(\frac{\lx}{L_{a2}}\right)^{\alpha_2}      &     \left[\lx < L_{a2}\right].
	\end{cases}	      
\end{equation}
Furthermore, in the LDDE2 model the $e_1$ parameter is allowed to depend on luminosity, with the form
\begin{equation}
e_1(\lx) = e_1^* + \beta_1(\log \lx - \log L_\mathrm{p})
\end{equation}
as proposed in \citet{Hasinger05}.

Both of these LDDE models have a large number of parameters: 9 parameters are needed for the LDDE1 model, whereas the LDDE2 model has 15 parameters. 
Given the large parameter space, earlier studies often fix some of these parameters \citep[e.g.][ fix 6 of the parameters in the LDDE2 model]{Ueda14}.
In our Bayesian analysis, we instead apply priors for all parameters but otherwise allow them to vary, thus allowing the increased parameter space to be accounted for and appropriately penalized when calculating the Bayesian evidence. 
The physical meaning of the parameters is somewhat obscure, thus we apply constant (or log-constant) priors over reasonable ranges given the form of the parametrization and the range of our data.
Table \ref{tab:ldde} gives our best \emph{a posteriori} estimates of the parameters based on either the hard-band sample or the soft-band sample, along with our prior limits and the Bayesian evidence for LDDE2 relative to the simpler LDDE1 parametrization.
We note that the GLF is also included in each of the fits, introducing additional parameters, although these are partly constrained by the priors given in Table \ref{tab:galpriors}.

As can be seen from Table \ref{tab:ldde}, we have very strong evidence in favor of the LDDE2 model over the LDDE1 model for both the hard-band and soft-band samples, indicating that the introduction of the additional parameters in LDDE2 is justified   by the data. 
 
\begin{table*}
\caption{Prior limits and best \emph{a posteriori} estimates of parameters for the LDDE1 and LDDE2 models for separate fits to the hard-band and soft-band samples (neglecting absorption effects).}
\begin{tabular}{l r r r r r r}
\hline
{Parameter}         &  {Lower limit} &  {Upper limit} &  \multicolumn{2}{c}{Hard band}              & \multicolumn{2}{c}{Soft band} \\
                            &                        &                        &  {LDDE1}   &  {LDDE2}         &  {LDDE1}         &  {LDDE2}\\
\hline
$\log K$  (Mpc$^{-3}$)      &                    -7.0 &                  -3.0 &  $ -5.63 \pm 0.07 $ &  $ -5.72 \pm 0.07 $     &   $ -5.87  \pm 0.05 $    &  $ -5.97 \pm 0.05 $\\
$\log L_*$ (\ergs)          &                    43.0 &                  46.0 &  $ 44.10 \pm 0.05 $ &  $ 44.09 \pm 0.05 $     &   $ 44.17  \pm 0.04 $    &  $ 44.18 \pm 0.03 $\\
$\gamma_1$                  &                    -1.0 &                   1.5 &  $\;0.72 \pm 0.02 $ &  $\;0.73 \pm 0.02 $     &   $\;0.67  \pm 0.02 $    &  $\;0.79 \pm 0.01 $\\
$\gamma_2$                  &                     1.5 &                   4.0 &  $\;2.26 \pm 0.07 $ &  $\;2.22 \pm 0.06 $     &   $\;2.37  \pm 0.06 $    &  $\;2.55 \pm 0.05 $\\
$e_1$ (or $e_1^*$)          &                     2.0 &                   6.0 &  $\;3.97 \pm 0.17 $ &  $\;4.34 \pm 0.18 $     &   $\;3.67  \pm 0.09 $    &  $\;4.35 \pm 0.35 $\\
$\beta_1$                   &                    -2.0 &                   2.0 &  ...            &  $ -0.19 \pm 0.09 $     &    ...               &  $\;0.65 \pm 0.06 $\\
$\log L_\mathrm{p}$ (\ergs) &                    43.0 &                  46.0 &  ...            &  $ 44.48 \pm 0.44 $     &    ...               &  $ 44.45 \pm 0.53 $\\
$e_2$                       &                    -5.0 &                   0.0 &  $ -2.08 \pm 0.17 $ &  $ -0.30 \pm 0.13 $     &   $ -2.92  \pm 0.14 $    &  $ -0.96 \pm 0.12 $\\
$e_3$                       &                   -10.0 &                  -5.0 &  ...            &  $ -7.33 \pm 0.62 $     &    ...               &  $ -7.84 \pm 0.51 $\\
$z_{c1}^*$                  &                     0.4 &                   2.5 &  $\;2.02 \pm 0.09 $ &  $\;1.85 \pm 0.08 $     &   $\;2.27  \pm 0.09 $    &  $\;1.80 \pm 0.08 $\\
$z_{c2}^*$                  &                     2.5 &                   3.5 &  ...            &  $\;3.16 \pm 0.10 $     &    ...               &  $\;3.09 \pm 0.11 $\\
$\log L_{a1}$ (\ergs)       &                    43.0 &                  46.0 &  $ 44.71 \pm 0.09 $ &  $ 44.78 \pm 0.07 $     &   $\;0.92  \pm 0.11 $    &  $ 44.92 \pm 0.12 $\\
$\log L_{a2}$ (\ergs)       &                    43.0 &                  46.0 &  ...            &  $ 44.46 \pm 0.17 $     &    ...               &  $ 44.27 \pm 0.30 $\\
$\alpha_1$                  &                    -1.0 &                  1.0  &  $\;0.20 \pm 0.01$  &  $\;0.23 \pm 0.01 $     &   $\;0.18  \pm 0.01 $    &  $\;0.16 \pm 0.01 $\\
$\alpha_2$                  &                    -1.0 &                  1.0  &  ...            &  $\;0.12 \pm 0.02 $     &    ...               &  $\;0.06 \pm 0.02 $\\[5pt]
\hline\\
$\Delta \ln \mathcal{Z}$    &                         &                       &  0.0               &   $+36.1$                & 0.0                      &  $+74.8$    \\
\hline
\end{tabular}
\label{tab:ldde}
\end{table*}

\subsection{Luminosity and density evolution (LADE)}
\label{sec:lade}

The LDDE model substantially warps the shape of the XLF with redshift, in particular introducing a flattening of the faint-end slope at $z\sim1$ (as well as a further turn up at the lowest luminosities). 
Whether the observational evidence supports such changes in the shape of the XLF has been a matter of debate \citep[e.g. A10,][]{Aird08,Fiore12,Ueda14,Buchner15}.
Furthermore, a physical interpretation of this complex parametrization is difficult. 
Therefore, in A10 we proposed a model where the shape of the XLF (the smoothly-connected double power-law given in Equation \ref{eq:dpl} above) is kept the same at all redshifts, but undergoes a shift in luminosity with redshift as well as an overall decrease in density.
We refer to this model as Luminosity And Density Evolution \citep[LADE, see also e.g.][ for similar parametrizations]{Yencho09,Ross12}. 
The luminosity evolution is achieved by allowing $L_*$ to change with redshift,
\begin{equation}
\log L_*(z) = \log L_0 - \log\left[ \left(\frac{1+z_c}{1+z}\right)^{p_1} + \left(\frac{1+z_c}{1+z}\right)^{p_2} \right]
\end{equation}
where $p_1$ and $p_2$ allow for a different evolution of $L_*$ above and below a transition redshift, $z_c$. 
The additional density evolution was introduced by allowing the normalization, $K$, to evolve as
\begin{equation}
\log K(z) = \log K_0 + d(1+z).
\label{eq:densevol}
\end{equation}
In A10, we found that the Bayesian evidence for the LADE model was comparable to the evidence for the LDDE1 model\footnote{The evidence for LDDE1 was higher but not by enough to decisively favour LDDE1 over LADE.} when considering a hard-band selected sample of X-ray sources at $z<1.2$ and colour pre-selected samples at higher redshifts. 

Here, we have repeated the fitting of the LADE model with our updated samples and improved analysis (allowing for a distribution of spectral parameters and the contribution from the GLF, but assuming all sources are unabsorbed). 
Our best a posteriori estimates of the parameters for the LADE model (along with the prior constraints) are reported in Table \ref{tab:lade}, along with 
the Bayesian evidence, relative to the LDDE1 model fit.
For both our samples, we find strong evidence in favour of the LADE model compared to the LDDE1 model, indicating that it provides a better description of the XLF despite having (slightly) fewer free parameters. 
The luminosity evolution of the best-fit LADE model in this work is somewhat different to the A10 findings, with a slightly weaker luminosity evolution at low redshifts ($p_1\approx4$, compared to $p_1\approx6$ in A10) that continues to higher redshifts ($z_c\approx2$, compared to $z_c\approx0.8$ in A10) and has a much stronger, negative luminosity evolution at higher redshifts ($p_2\approx -2$, compared to $p_2\approx -0.2$ in A10). 
These differences indicate that the LADE provides a better description than LDDE1 for the high-redshift evolution, which is probed by our updated samples.
Nevertheless, the evidence for the LDDE2 model---which introduces an additional density evolution at the highest redshfits---is significantly stronger than for the LADE model, despite the larger number of free parameters and additional complexity of LDDE2.

\begin{table*}
\caption{Prior limits and best \emph{a posteriori} estimates of parameters for the LADE model for separate fits to the hard-band and soft-band samples (neglecting absorption effects).}
\begin{tabular}{l r r r r r}
\hline
{Parameter}   &  {Lower limit} &  {Upper limit} & {Hard band}         & {Soft band} \\
                      &                        &                        &  {LADE}             &  {LADE}     \\
        
\hline   
$\log K$  (Mpc$^{-3}$)&                    -7.0 &                  -3.0 &     $ -4.03  \pm  0.08 $    &      $ -4.28  \pm   0.05 $     \\
$\log L_*$ (\ergs)    &                    43.0 &                  46.0 &     $ 44.84  \pm  0.05 $    &      $ 44.93  \pm   0.03 $     \\
$\gamma_1$            &                    -1.0 &                   1.5 &     $\;0.48  \pm  0.03 $    &      $\;0.44  \pm   0.02 $     \\
$\gamma_2$            &                     1.5 &                   4.0 &     $\;2.27  \pm  0.07 $    &      $\;2.18  \pm   0.04 $     \\
$p_1$                 &                     3.0 &                  10.0 &     $\;3.87  \pm  0.17 $    &      $\;3.39  \pm   0.08 $     \\
$p_2$                 &                    -4.0 &                   3.0 &     $ -2.12  \pm  0.39 $    &      $ -3.58  \pm   0.26 $     \\
$z_{c}$               &                     0.4 &                   3.0 &     $\;2.00  \pm  0.13 $    &      $\;2.31  \pm   0.07 $     \\
$d$                   &                    -1.5 &                   0.5 &     $ -0.19  \pm  0.02 $    &      $ -0.22  \pm   0.01 $     \\
\hline\\
$\Delta \ln \mathcal{Z}$ &                      &                       &     $+9.6$             &           $+38.4$          \\
\hline
\end{tabular}
\label{tab:lade}
\end{table*}

\subsection{Flexible double-power law (FDPL)}
\label{sec:fdpl}

The LADE model has a number of issues that may limits its ability to accurately described the observed behaviour of the XLF.
Firstly, the forms of the luminosity evolution (in $L_*$) and the density evolution (in $K$) are restricted to the specified functional form. 
Thus, the model is unable to reproduce any more complex evolutionary behaviour across certain redshift ranges, even if required by the data (e.g. at the highest redshifts).
Secondly, the LADE model strictly requires that the shape of the XLF, described by the faint-end and bright-end slopes ($
\gamma_1$, $\gamma_2$), remains exactly the same at all redshifts.

We therefore propose a new set of models to parametrize the evolution of the XLF, which we refer to as the Flexible Double Power-Law (FDPL) models.
These models assume that the XLF can be described by a double power-law form (Equation \ref{eq:dpl}) at any redshift with four parameters: $K$, $L_*$, $\gamma_1$ and $\gamma_2$. 
We then allow any of these parameters to evolve with redshift, parametrizing the evolution by a polynomial function 
of $\log(1+z)$.
By comparing the Bayesian evidence for models with different order polynomials,
we can determine whether our data require a more complicated density or luminosity evolution, or whether there is any evidence for changes in the faint-end or bright-end slopes with redshift. 

With the simplest form of the polynomial,
it is difficult to place meaningful priors on the free parameters: the various polynomial coefficients. 
Instead, we use Chebyshev polynomials over the interval $0 < \log(1+z) < \zeta_\mathrm{max}$, 
where $\zeta_\mathrm{max}=\log(1+z_\mathrm{max})$ and we set $z_\mathrm{max}=7$. 
For a given order, $n$, the polynomial form is defined by the value of the parameter (e.g. $\log K$) at each of the $n+1$ Chebyshev nodes, i.e. at redshifts, $z_k$, where
\begin{equation}
\log (1+z_k) = \frac{1}{2}\zeta_\mathrm{max} + \frac{1}{2}\zeta_\mathrm{max} \cos\left(\frac{2k+1}{2(n+1)}\pi\right).
\end{equation}
The value of e.g. $\log K(z)$ is then given by a sum of Chebyshev polynomials,
\begin{equation}
\log K(z) = \sum_{j=0}^n c_j T_j(x)
\end{equation}
where $T_j(x)$ are the Chebyshev polynomials of the first kind 
and $x$ is a re-scaled version of $\zeta=\log(1+z)$ over the interval $[-1,1]$,
\begin{equation}
x = 2 \frac{\log(1+z)}{\zeta_\mathrm{max}} -1.
\end{equation}
The coefficients, $c_j$, are given by
\begin{equation}
c_j = \frac{2}{n+1}  \sum_{k=1}^{n+1} \log K(z=z_k) T_k(x=x_k)
\end{equation}
where $\log K(z=z_k)$ is the value of $\log K$ at $z_k$. 

This scheme may appear complicated but it can easily be understood as choosing values for a parameter (e.g. $\log K$) at a number of redshifts, $z_k$, and then finding a polynomial function to interpolate between them for all $z$. 
We can thus set a prior on the value of the parameter at each of the Chebyshev nodes ($z_k$).
The choice of Chebyshev polynomials minimizes Runge's phenomenon and thus prevents the adoption of highly oscillatory solutions for the redshift dependence.

The FDPL models could take a large number of possible forms, with a polynomial of arbitrary order describing the redshift dependence of each of the four double power-law parameters ($K$, $L_*$, $\gamma_1$, $\gamma_2$).
Fully exploring all possible combinations is computationally prohibitive.
To find an appropriate description of the XLF we proceed as follows.
First, we try to find the best parametrization for the overall density evolution and luminosity evolution by allowing $\log K(z)$ and $\log L_*(z)$ to each be described by a polynomial of up to fourth order, but with no redshift-dependence for $\gamma_1$ and $\gamma_2$.
We evaluate the Bayesian evidence for all possible combinations of polynomials of up to fourth order for $\log K(z)$ and $\log L_*(z)$ , corresponding to 25 different models.
We find that the model where $\log K(z)$ is described by a second-order polynomial and $\log L_*(z)$ is described by a third-order polynomial has the highest Bayesian evidence for the hard-band sample (although we note that alternative combinations cannot be ruled out).
We then fix the polynomial orders for $\log K(z)$ and $\log L_*(z)$\footnote{The form of the redshift-dependence is thus fixed for $\log K(z)$ and $\log L_*(z)$ but the parameters that describe it are still allowed to vary in subsequent fits.}
and evaluate the evidence for models where $\log \gamma_1(z)$ (the faint-end slope) is described by a polynomial of up to fourth order. 
We choose to describe $\log \gamma_1(z)$ as a polynomial function, rather than $\gamma_1$ directly, so that extrapolation of our model to high redshifts does not result in negative, likely unphysical slopes \citep[see][]{Hopkins07b}.
We find strong evidence that $\gamma_1(z)$ changes with redshift, with a dependence that is described by a first-order polynomial (i.e. a linear relation) in $\log(1+z)$; the higher-order polynomial dependences have lower Bayesian evidence but cannot be definitively ruled out.
Finally, we fix the form for $\log \gamma_1(z)$ to a first-order polynomial and test models where $\log \gamma_2(z)$ (the bright-end slope) is allowed to vary with redshift. 
We find strong Bayesian evidence favouring the model with no redshift-dependence for $\gamma_2$.

Our final ``best-fit" model thus consists of $\log K(z)$ being described by a second-order polynomial, $\log L_*(z)$ being described by a third-order polynomial, $\log \gamma_1(z)$ being described by a first-order polynomial and $\gamma_2$ being constant with redshift. 
Table \ref{tab:fdplhard} gives the best \emph{a posteriori} estimates of the parameters, corresponding to the values of the parameters at each of the Chebyshev nodes. 
We also provide a simplified polynomial expression for the redshift dependence of each of the double power-law parameters.
In addition, we give the Bayesian evidence for this FDPL model, relative to the evidence for the LDDE1 model (from Table \ref{tab:ldde}). 
We find strong evidence in favour of our final FDPL model compared to any of the previously considered models, including LDDE2.

We then repeat the entire process with the soft-band sample. 
We find that the soft-band sample requires the same model form as the hard-band sample, consisting of a second-order polynomial for $\log K(z)$, a third-order polynomial for $\log L_*(z)$, a linear function for $\log \gamma_1(z)$ and a constant $\gamma_2$ (although the values of these parameters are different). 
The results, including the Bayesian evidence relative to the LDDE1 model fit to the soft-band sample, are given in Table \ref{tab:fdplsoft}.
Again, we find strong evidence favouring the FDPL model over LDDE1, LDDE2, or the LADE model.

\begin{table*}
\caption{Prior limits and best \emph{a posteriori} estimates of parameters for the FDPL model for the hard-band sample (neglecting absorption effects).}
\begin{tabular}{l r r r r r r}
\hline
{Parameter}          &  {Lower limit} &  {Upper limit} & \multicolumn{4}{c}{Parameter value at node $k$}   \\
                             &                        &                        & $k=0$            & $k=1$            & $k=2$            & $k=3$      \\
\hline
$\log K(z=z_k)$  (Mpc$^{-3}$)&                   -7.0 &                  -3.0 &  $-6.19 \pm 0.14$ & $-4.44 \pm 0.05$ & $-4.87 \pm 0.09$ & ...          \\
$\log L_*(z=z_k)$ (\ergs)    &                   43.0 &                  46.0 &  $44.45 \pm 0.17$ & $44.61 \pm 0.05$ & $44.04 \pm 0.04$ & $43.57 \pm 0.08$ \\
$\gamma_1(z=z_k)$            &                   0.01 &                   1.5 &  $ 0.27 \pm 0.04$ & $ 0.58 \pm 0.02$ & ...          & ...          \\
$\gamma_2(z=z_k)$            &                    1.5 &                   4.0 &  $ 2.31 \pm 0.07$ & ...          & ...          & ...          \\
\hline\\
$\Delta \ln \mathcal{Z}$    &                        &                       &     $+45.0$        &                  &                    &               \\
\\
\multicolumn{7}{l}{$\log K(z) =   -5.13 + 4.73\zeta    -7.10\zeta^2$} \\
\multicolumn{7}{l}{$\log L_*(z) =   43.53 +      1.23\zeta + 3.35\zeta^2 -4.08\zeta^3$}\\
\multicolumn{7}{l}{$\log \gamma_1(z) = -0.17   -0.51\zeta$}\\
\multicolumn{3}{l}{$\gamma_2(z) =  2.31$} & \multicolumn{4}{l}{where $\zeta=\log(1+z)$}\\
\hline
\end{tabular}
\label{tab:fdplhard}
\end{table*}

\begin{table*}
\caption{Prior limits and best \emph{a posteriori} estimates of parameters for the FDPL model for the soft-band sample (neglecting absorption effects).}
\begin{tabular}{l r r r r r r}
\hline
{Parameter}          &  {Lower limit} &  {Upper limit} & \multicolumn{4}{c}{Parameter value at node $k$}   \\
                             &                        &                        & $k=0$            & $k=1$          & $k=2$            & $k=3$      \\
        
\hline
$\log K(z=z_k)$  (Mpc$^{-3}$)&                   -7.0 &                  -3.0 &  $-6.47 \pm 0.09$ & $-4.73 \pm 0.03$ & $-5.20 \pm 0.07$ & ...          \\
$\log L_*(z=z_k)$ (\ergs)    &                   43.0 &                  46.0 &  $44.18 \pm 0.15$ & $44.61 \pm 0.04$ & $44.12 \pm 0.03$ & $43.81 \pm 0.05$ \\
$\gamma_1(z=z_k)$            &                   0.01 &                   1.5 &  $ 0.25 \pm 0.02$ & $ 0.55 \pm 0.02$ & ...          & ...          \\
$\gamma_2(z=z_k)$            &                    1.5 &                   4.0 &  $ 2.34 \pm 0.05$ & ...          & ...          & ...          \\
\hline\\
$\Delta \ln \mathcal{Z}$    &                        &                       &   $+93.1$        &                  &                    &               \\
\\
\multicolumn{7}{l}{$\log K(z) =    -5.47 + 4.88\zeta   -7.20\zeta^2$}\\
\multicolumn{7}{l}{$\log L_*(z) =   43.81 -0.27\zeta + 6.82\zeta^2 -6.94\zeta^3$}\\
\multicolumn{7}{l}{$\log \gamma_1(z) = -0.18 -0.55\zeta$}\\
\multicolumn{3}{l}{$\gamma_2(z) = 2.34$}              & \multicolumn{4}{l}{where $\zeta=\log(1+z)$}\\
\hline
\end{tabular}
\label{tab:fdplsoft}
\end{table*}

\subsection{Summary of results}
\label{sec:xlfresults}

In this section we have explored a number of different model parametrizations that can be used to describe the XLF, measured using either our hard-band or soft-band sample (neglecting the effects of absorption).
The analysis of both samples leads to similar conclusions. 
We find that there is strong evidence for the updated LDDE2 model of \citet{Ueda14} compared to the simpler LDDE1 parametrization or the LADE model of A10 (where the shape of the XLF remains the same at all redshifts).
We also introduce a more flexible parametrization (FDPL) that models the XLF as a double power-law but
allows for arbitrary evolution in the overall luminosity or density, as well as allowing for changes in the overall shape. 
The final FDPL model requires 10 free parameters to describe the XLF. 
Our FDPL provides a simpler parametrization that is nevertheless able to reproduce the overall evolution of the XLF. 
The Bayesian evidence strongly favours our FDPL model over LDDE1, LDDE2 and LADE.

\begin{figure*}
\includegraphics[width=\textwidth,trim=0 0 0 0]{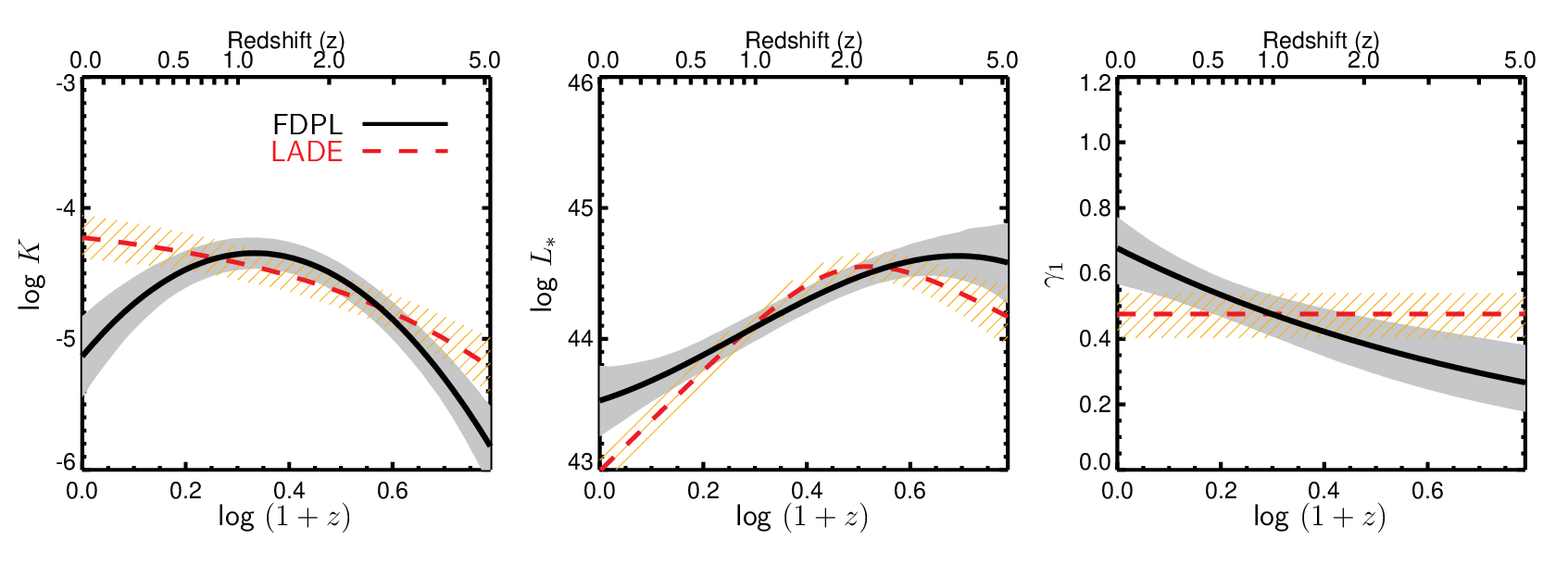}
\caption{
Dependence of the parameters that specify the double power-law form of the XLF on redshift based on best fits to the hard-band sample (neglecting absorption effects) with our FDPL model (black solid line) and LADE model (red dashed line). The solid grey and hatched orange regions indicate the 99 per cent confidence interval on the parameters for FDPL and LADE respectively based on the posterior distributions of the model parameters from our Bayesian analysis. 
The LADE model as proposed by A10 is restricted to a monotonic overall density evolution (traced by the $\log K$ parameter) and does not allow for any flattening of the faint-end slope ($\gamma_1$).
Thus LADE is unable to produce the evolutionary behaviour described by the more flexible FDPL model, which provides a better description of our data and is thus favoured according to the Bayesian evidence.
}
\label{fig:pars_vs_z_hard}
\end{figure*}

\begin{figure*}
\includegraphics[width=\textwidth,trim=30 0 20 0]{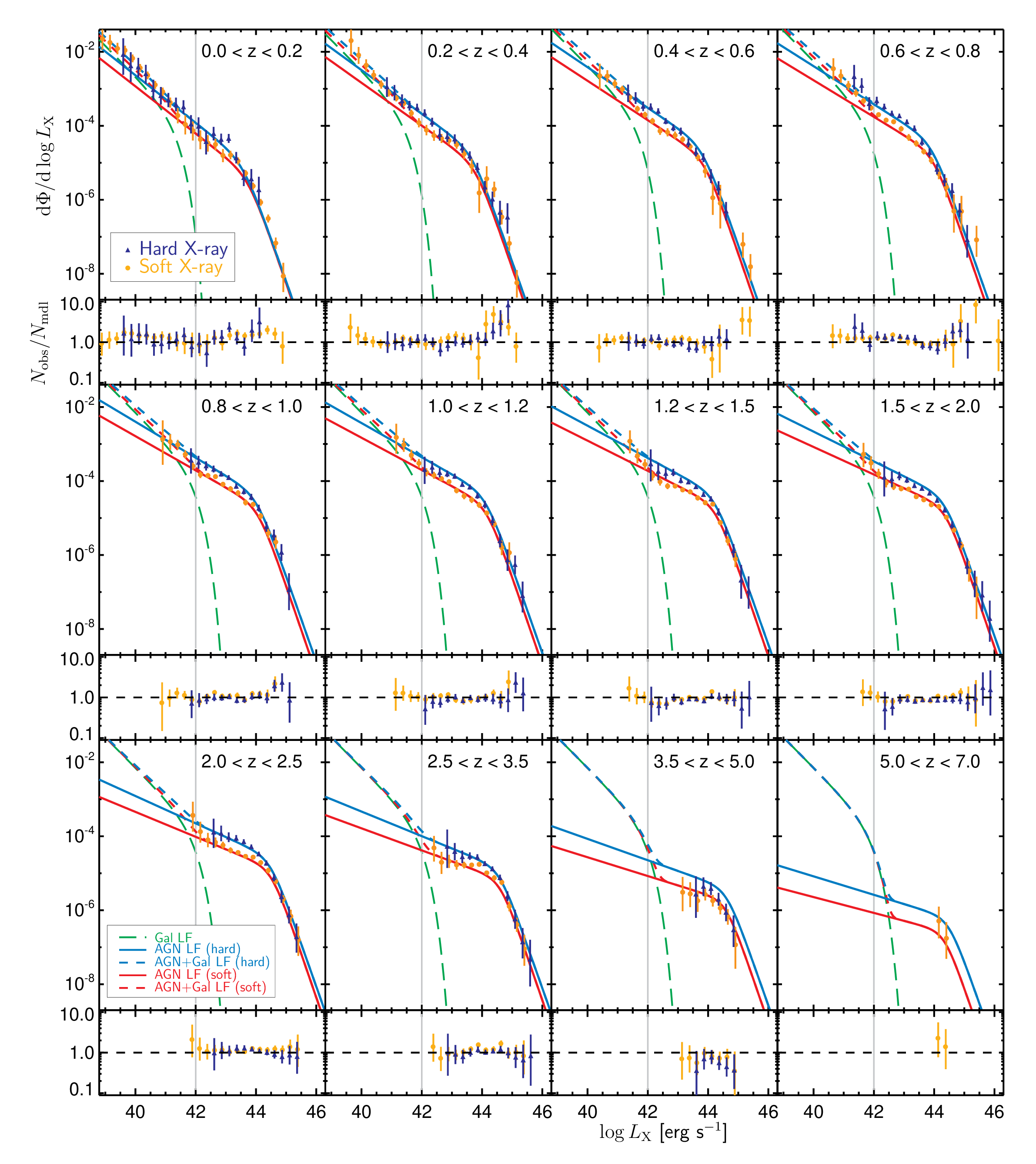}
\caption{
XLFs calculated using the hard-band and soft-band X-ray samples, neglecting absorption effects and fitting each band individually.
The solid blue and solid red lines show our best-fit FDPL models for the AGN XLF based on the hard-band and soft-band samples respectively. 
The green dashed line shows the GLF (soft-band fit only for clarity) and the dashed blue and dashed red lines show the total luminosity function, including both AGNs and galaxies, which should be compared to the binned estimates (dark blue triangles and orange circles for the hard-band and soft-band samples respectively).
\refone{The sub-panels show the residuals between the data and model in terms of the $N_\mathrm{obs}/N_\mathrm{mdl}$ ratio.}
While the form of the evolution based on the two samples is very similar, there are significant discrepancies in the space densities below the break in the XLF across all redshifts.
To reconcile these discrepancies we must include the distribution of absorption column densities in our modeling and the effects on the inferred luminosities and space densities for both samples.
}
\label{fig:xlfhardsoft}
\end{figure*}

In Figure \ref{fig:pars_vs_z_hard} we show how the double power-law parameters change with redshift based on our final FDPL model for the hard-band sample (black line).
We find that the evolution of the XLF is driven by a combination of 
a) an overall density evolution (traced by the $\log K$ parameter) that peaks at $z\sim2$,
b) a relatively mild luminosity evolution that shifts the overall XLF towards higher luminosities at higher redshifts (which continues out to high redshifts),
and c) a mild flattening of the faint-end slope of the XLF with increasing redshift.
For comparison, we also show the redshift dependence for the parameters based on our LADE model fit to the hard-band sample (dashed red line, along with uncertainties shown by the orange hatched region).
The LADE model is unable to re-produce the form of density evolution required by our FDPL model and does not allow for any flattening of the faint-end slope.

Figure \ref{fig:xlfhardsoft} presents our best-fit FDPL model at a range of redshifts based on the hard-band (blue line) and soft-band (red line) samples, along with binned estimates using the method described in Section \ref{sec:binned}. 
While we find that the form of the evolution of the XLF (described by the FDPL model) is very similar for both the hard-band and soft-band samples, Figure \ref{fig:xlfhardsoft} shows that there are significant differences between the XLFs from the two samples.
At high luminosities ($\lx \gtrsim L_*$) the space densities based on the hard-band and soft-band samples are broadly in agreement but at lower luminosities the best-fit model and binned estimates from the soft-band sample fall significantly below the hard-band estimates. 
This discrepancy is most likely due to our neglect of absorption effects
and motivates further investigation in Section \ref{sec:absdist} below.

\section{Joint constraints on the X-ray luminosity and absorption distribution function}
\label{sec:absdist}

In this section we attempt to find a parametrization for the full X-ray luminosity and absorption-distribution function (XLAF) that---after accounting for sensitivity effects and selection biases---can simultaneously describe both our hard-band and soft-band samples. 
For each individual detection, we adopt a log-constant prior distribution for \NH\ in the range $20<\log\nh<26$ (reflecting our lack of \emph{a priori} knowledge of the absorption) and the corresponding constraints on \LX\ (e.g. Figure \ref{fig:lx_vs_nh}). 
Below, we describe our model for the XLAF (Section \ref{sec:xlafmodel}) and present our results (Section \ref{sec:xlafresults}).

\subsection{Model for the XLAF}
\label{sec:xlafmodel}

Most previous studies adopt a parametrization for the overall XLF of all AGNs (or, in many cases, all Compton-thin AGNs) and separately model the distribution of \NH\ at a given \LX\ and $z$ \citep[often described as the ``\NH\ function" e.g.][]{Ueda03,LaFranca05}. 
The overall shape of the \NH\ function and any luminosity or redshift dependence is  defined by \fabs, the fraction of absorbed AGNs\footnote{Defined here as the fraction of AGNs with column densities $22<\log \nh <24$ relative to all AGNs with $\log \nh<24$}.
Previous studies have shown that \fabs\ is strongly dependent on luminosity
and may also increase at higher redshifts (e.g. \citealt{Hasinger08,Ueda14}, but see also \citealt{Dwelly06}). 
These dependencies may be modeled directly using specific parametrizations to describe \fabs\ as a function of \LX\ and $z$ (e.g. power-law functions). 
Further assumptions, such as setting limits on \fabs\ at certain luminosity or redshift thresholds, are often required. 
Additional parameters are also required to fully describe the overall distribution of \NH\ for a given value of \fabs. 
Such models can have a very large number of parameters and thus many of the parameters may be fixed based on measurements in the local Universe or prior expectations \citep[e.g.][]{Ueda14}.
The resulting model parametrizations often have sharp breaks and discontinuities at certain luminosities or redshifts that may not be physical.

In this work, we therefore choose to take a different approach that builds on the flexibility of our FDPL model described above. 
We allow the unabsorbed ($20<\log \nh<22$) and absorbed ($22<\log \nh<24$) AGN populations to be described by independent luminosity functions---$\phi_\mathrm{unabs}(\lx,z \giv \xlfpars_\mathrm{unabs})$ and $\phi_\mathrm{abs}(\lx,z \giv \xlfpars_\mathrm{abs})$---with independent sets of parameters describing their double power-law shape that can evolve differently with redshift.
We assume a fixed fraction of unabsorbed AGNs, $f_\mathrm{21-22}$, have column densities in the range $21<\log \nh<22$.
Likewise, we assume that a fixed fraction of absorbed AGNs, $f_\mathrm{23-24}$, have column densities in the range $23<\log \nh<24$.
We assume a log-constant distribution of \NH\ within each of these 1 dex wide ranges (at a given \LX\ and $z$).
The addition of $f_{21-22}$ and $f_{23-24}$ ensures we have sufficient freedom to describe the overall distribution of \NH\ within the unabsorbed or absorbed populations.
We also allow for a population of Compton-thick AGNs with $24<\log \nh<26$, which for this work are assumed to completely track the evolution of absorbed AGNs. Thus, the XLF of the Compton-thick AGNs is assumed to be simply a factor, $\beta_\mathrm{Cthick}$, times the XLF of absorbed AGNs.
Our overall model of the XLAF, $\psi(\lx,z,\nh \giv \nhfpars)$, is thus given by

\begin{equation}
\resizebox{\hsize}{!}{$
\psi(\lx, z, \nh \giv \nhfpars) =  
\begin{cases} 
	(1-f_\mathrm{21-22})\phi_\mathrm{unabs}(\lx,z \giv \xlfpars_\mathrm{unabs})
										 & \left[20 \le \log \nh < 21 \right] \\
	 f_\mathrm{21-22}\; \phi_\mathrm{unabs}(\lx,z \giv \xlfpars_\mathrm{unabs})
									 & \left[21 \le \log \nh < 22 \right] \\
	(1-f_\mathrm{23-24})\phi_\mathrm{abs}(\lx,z \giv \xlfpars_\mathrm{abs})
									 & \left[22 \le \log \nh < 23 \right] \\
	f_\mathrm{23-24} \;   \phi_\mathrm{abs}(\lx,z \giv \xlfpars_\mathrm{abs})
									 & \left[23 \le \log \nh < 24 \right] \\
	\frac{\beta_\mathrm{Cthick}}{2} 	\phi_\mathrm{abs}(\lx,z \giv \xlfpars_\mathrm{abs})
							 		 & \left[24 \le \log \nh < 26 \right] \\				\end{cases}
$}
\label{eq:xlaf_nh}
\end{equation}
where $\nhfpars=\{\xlfpars_\mathrm{unabs}, \xlfpars_\mathrm{abs}, f_{21-22}, f_{23-24}, \beta_\mathrm{Cthick}\}$ and thus encapsulates all parameters required to describe the XLAF. 
The factor $\frac{1}{2}$ in Equation \ref{eq:xlaf_nh} for Compton-thick AGNs ($24 \le \log \nh < 26$) represents our assumption that the column densities are evenly distributed over the 2 dex wide bin in $\log \nh$.

It is worth noting that in our analysis both the hard- and soft-band samples will contain AGN from the unabsorbed, absorbed, and Compton-thick AGN populations, although the soft-band sample is likely to contain a lower fraction of absorbed sources than the hard-band sample (due to the detection biases against absorbed sources at soft energies).  
We do not know if an individual hard- or soft-band detection corresponds to an unabsorbed, absorbed, or Compton-thick source; we are only able to make statements about the population as a whole, based on our model for the XLAF.

Marginalizing over $\log \nh$, we can recover the total XLF,
\begin{eqnarray}
\phi(\lx, z \giv \nhfpars) =& &\int_{20}^{26} d \log \nh\;  \psi(\lx, z, \nh \giv \nhfpars)\nonumber\\
						   =& &\phi_\mathrm{unabs}(\lx,z \giv \xlfpars_\mathrm{unabs})     \\
						   	& + & \phi_\mathrm{abs}(\lx,z \giv \xlfpars_\mathrm{abs}) \nonumber\\
						   	& + & \beta_\mathrm{Cthick}\phi_\mathrm{abs}(\lx,z \giv \xlfpars_\mathrm{abs})  \nonumber
						   	\label{eq:xlf_tot}
\end{eqnarray}

We assume that $\phi_\mathrm{unabs}$ and  $\phi_\mathrm{abs}$ are each individually described by an FDPL model paramaterization where the four XLF parameters---$K$, $L_*$, $\gamma_1$, $\gamma_2$---can all vary with redshift according to polynomial functions of $\log (1+z)$.
Fully exploring all possible combinations of polynomials of different orders for the redshift-dependence of all eight of these shape parameters is computationally prohibitive. 
Instead, we initially assume that $\phi_\mathrm{unabs}$ and  $\phi_\mathrm{abs}$ are both described by the form found in Section \ref{sec:fdpl} when analyzing the hard- and soft-band samples separately. 
Thus, we adopt a second-order polynomial for $\log K(z)$, a third-order polynomial for $\log L_*(z)$, a first-order polynomial for $\log \gamma_1(z)$, and a constant $\gamma_2$ as our baseline model. 
We then increase or decrease (if possible) the polynomial order for each of the eight shape parameters in turn and evaluate the Bayesian evidence relative to our baseline model. 

\subsection{Results}
\label{sec:xlafresults}

In Table \ref{tab:xlaf_comp} we give the change in the logarithmic Bayesian evidence, $\Delta \ln \mathcal{Z}$, found when increasing or decreasing the polynomial order for the redshift dependence of each of the FDPL parameters in turn, relative to the evidence for our baseline model.
In the majority of cases, the altered model has a \emph{lower} Bayesian evidence, thus favouring our baseline model.
For the normalization ($K$) of both the unabsorbed and absorbed AGN XLF and the $L_*$ of the absorbed AGN XLF, we can decisively rule out the simpler models (with lower order polynomials used to describe the redshift dependence) as $\Delta \ln \mathcal{Z}<-4.6$.
For the $L_*$ of the absorbed AGN, a simpler (second-order polynomial) model is weakly favoured by our data ($\Delta \ln \mathcal{Z}>0$), but the evidence for this simplification is not strong.
In all other cases, the baseline model is very weakly favoured ($\Delta \ln \mathcal{Z} <0$).

We also tested a model with no change in the faint-end slope for \emph{both} the unabsorbed and absorbed AGN XLFs (but otherwise retaining the baseline model). 
While such a model is weakly favored over the baseline ($\Delta \ln \mathcal{Z}=0.71$), we are unable to choose decisively between these models based purely on our data.
We therefore retain the baseline model, which has greater flexibility to describe the faint end of the XLF.
Nevertheless, as we are unable to fully explore all possible combinations of different polynomials, our final model must be seen as one possible, but adequate, way of describing our observational data.
The evolution of the faint-end slope of the XLF is discussed in more detail in Section \ref{sec:faintendslope} below. 

In Table \ref{tab:XLAFrestable} we give our best estimates of the XLAF parameters: the shape parameters for the unabsorbed and absorbed XLFs and their redshift dependence in terms of Chebyshev polynomials and a simplified form, as well as $f_{21-22}$, $f_{23-24}$ and $\beta_\mathrm{Cthick}$.
\secdraft{Table \ref{tab:galresults} provides our best \emph{a posteriori} estimates of the parameters of the GLF, which are constrained along with the AGN XLAF.}

\begin{table*}
\caption{Bayesian evidence (relative to our baseline XLAF model) for increasing or decreasing the polynomial order for the redshift dependence of each paramter in turn in the independent FDPL models (for the unabsorbed and absorbed populations) adopted for our modeling of the XLAF.}
\begin{tabular}{l l c r c r}
\hline
{AGN population} & {Parameter}  & {Order}   & {$\Delta \ln \mathcal{Z}$} & {Order}   & {$\Delta \ln \mathcal{Z}$}  \\
& 	& {(increased)} & & {(decreased)} & \\
\hline
Unabsorbed & $\log K$  (Mpc$^{-3}$)&  3 &    -0.86  &  1 &     -22.74             \\
           & $\log L_*$ (\ergs)   &  4 &    -3.12  &  2 &      -10.26             \\
           & $\gamma_1$           &  2 &    -3.24  &  0 &       -0.25             \\
           & $\gamma_2$           &  1 &    -2.45  &  ... & ...             \vspace{10pt}\\
Absorbed   & $\log K$  (Mpc$^{-3}$)&  3 &    -4.18  &  1 &      -27.23             \\
           & $\log L_*$ (\ergs)   &   4 &   -3.54  &  2 &       +2.85             \\
           & $\gamma_1$           &   2 &   -4.23  &  0 &      -2.15             \\
           & $\gamma_2$           &   1 &    -2.96  &  ... & ...             \\
\hline
\end{tabular}
\label{tab:xlaf_comp}
\end{table*}

\begin{table*}
\caption{Prior limits and best \emph{a posteriori} estimates of parameters for our full XLAF model, consisting of independent XLFs (described by the FDPL model) for the unabsorbed and absorbed AGN populations, and a scalar times the absorbed AGN XLF for the Compton-thick population.}
\begin{tabular}{l r r r r r r}
\hline
{Parameter}          &  {Lower limit} &  {Upper limit} & \multicolumn{4}{c}{Parameter value at node $k$}   \\
                             &                        &                        & $k=0$            & $k=1$            & $k=2$            & $k=3$      \\
        
\hline
\multicolumn{7}{c}{Unabsorbed AGNs ($20<\log \nh<22$)}\vspace{5pt}\\
$\log K(z=z_k)$  (Mpc$^{-3}$)&                   -7.0 &                  -3.0 &  $-6.17 \pm 0.11$ & $-4.81 \pm 0.04$ & $-5.03 \pm 0.06$ & ...          \\
$\log L_*(z=z_k)$ (\ergs)    &                   43.0 &                  46.0 &  $43.26 \pm 0.14$ & $44.49 \pm 0.03$ & $44.15 \pm 0.03$ & $43.79 \pm 0.05$ \\
$\gamma_1(z=z_k)$            &                   0.01 &                   1.5 &  $ 0.04 \pm 0.02$ & $ 0.25 \pm 0.04$ & ...          & ...          \\
$\gamma_2(z=z_k)$            &                    1.5 &                   4.0 &  $ 2.32 \pm 0.04$ & ...          & ...          & ...          \\
$f_{21-22}$                   &                    0.1 &                   0.9 &  $ 0.43 \pm  0.04$ &  ...          & ...          & ...          \\
\\
\multicolumn{7}{l}{$\log K(z) =        -5.21 + 0.72\zeta -0.26\zeta^2$} \\
\multicolumn{7}{l}{$\log L_*(z) =      43.56 + 0.32\zeta +0.32\zeta^2 -0.12\zeta^3$} \\
\multicolumn{7}{l}{$\log \gamma_1(z) = -0.44 - 0.52\zeta$}\\
\multicolumn{3}{l}{$\gamma_2(z) =  2.32$} & \multicolumn{4}{l}{where $\zeta=\log(1+z)$}
\vspace{10pt}\\
\hline\vspace{-2pt}\\
\multicolumn{7}{c}{Absorbed AGNs ($22<\log \nh<24$)}\vspace{5pt}\\
$\log K(z=z_k)$  (Mpc$^{-3}$) &                   -7.0 &                  -3.0 &   $  -6.81 \pm  0.09 $ & $  -4.44 \pm  0.05 $ & $  -4.30 \pm  0.09 $ & ...              \\
$\log L_*(z=z_k)$ (\ergs)    &                   43.0 &                  46.0 &    $  44.96 \pm  0.13 $ & $  44.57 \pm  0.06 $ & $  43.85 \pm  0.06 $ & $  43.17 \pm  0.08 $ \\
$\gamma_1(z=z_k)$            &                   0.01 &                   1.5 &    $   0.16 \pm  0.04 $ & $   0.43 \pm  0.04 $ & ...              & ...              \\
$\gamma_2(z=z_k)$            &                    1.5 &                   4.0 &    $   2.33 \pm  0.18 $ & ...              & ...              & ...              \\   
$f_{23-24}$                   &                    0.1 &                   0.9 &   $   0.70 \pm  0.02 $ & ...              & ...              & ...              \\
\\
\multicolumn{7}{l}{$\log K(z) =        -4.48 + 0.76\zeta -0.37\zeta^2$} \\
\multicolumn{7}{l}{$\log L_*(z) =      43.06 + 0.98\zeta -0.14\zeta^2 +0.01\zeta^3$} \\
\multicolumn{7}{l}{$\log \gamma_1(z) =  -0.28 - 0.28\zeta$}\\
\multicolumn{3}{l}{$\gamma_2(z) =  2.33$} & \multicolumn{4}{l}{where $\zeta=\log(1+z)$}
\vspace{10pt}\\
\hline\vspace{-2pt}\\
\multicolumn{7}{c}{Compton-thick AGNs ($24<\log \nh<26$)}\vspace{5pt}\\
$\beta_{Cthick}$           &                     0.2  &                    2.0 &  $   0.34 \pm 0.08 $  &  ...          & ...          & ...              \\
\hline
\end{tabular}
\label{tab:XLAFrestable}
\end{table*}

\begin{figure*}
\includegraphics[width=\textwidth,trim=30 0 20 0]{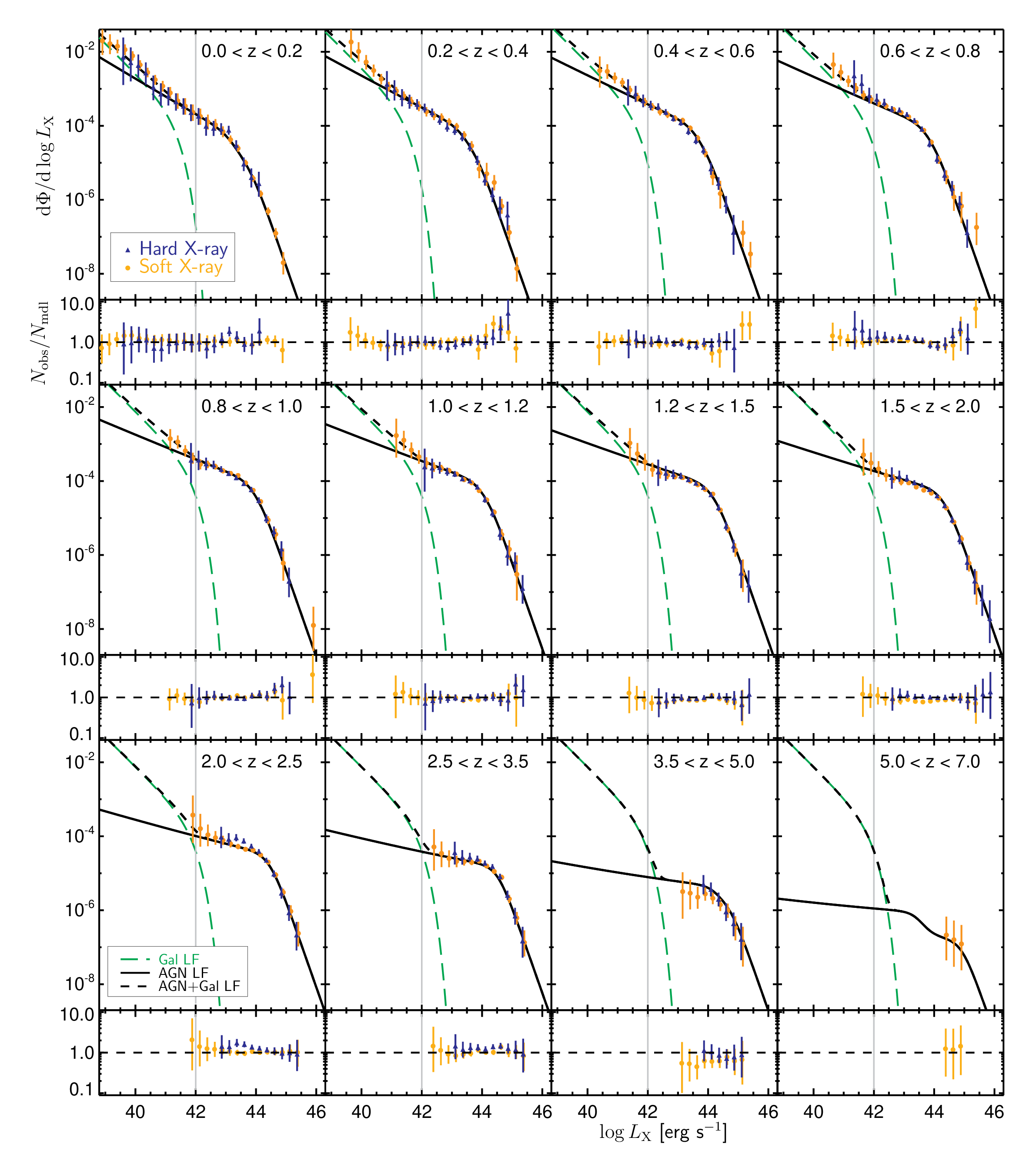}
\caption{
XLF of AGNs across the full range of absorption ($20<\log \nh<26$) calculated by fitting both the hard- and soft-band X-ray samples to our full X-ray luminosity and absorption-distribution function (XLAF) model described in Section \ref{sec:xlafmodel} (solid black line).
We also show the GLF (green dashed line) which is fitted simultaneously with the AGN XLAF. 
Binned estimates are based on either the hard-band (dark blue triangles) or soft-band (orange circles) X-ray samples and should be compared to the total XLF, including both AGNs and galaxies (black dashed line).
\refone{The sub-panels show the residuals between the data and model in terms of the $N_\mathrm{obs}/N_\mathrm{mdl}$ ratio.}
Our final model of the full XLAF of AGNs allows us to reconcile measurements based on the hard- and soft-band X-ray samples as ultimately coming from the same AGN population (but subject to different selection biases and incompleteness as a function of \NH, $z$ and \LX). 
Hence, both the hard- and soft-band estimates are consistent with the total XLF (cf. Figure \ref{fig:xlfhardsoft}).
}
\label{fig:xlf_withabs}
\end{figure*}

Figure \ref{fig:xlf_withabs} shows our model for the total AGN XLF (found by marginalizing over \NH, see Equation \ref{eq:xlf_tot}) compared to binned estimates based on our hard- and soft-band samples. 
The binned estimates from both the hard- and soft-band samples are consistent with the total XLF (with an additional contribution from the GLF at low luminosities), showing that our overall model of the XLAF is able to describe both of our samples (cf. Figure \ref{fig:xlfhardsoft}, where absorption effects were neglected).

\begin{figure*}
\includegraphics[width=\textwidth]{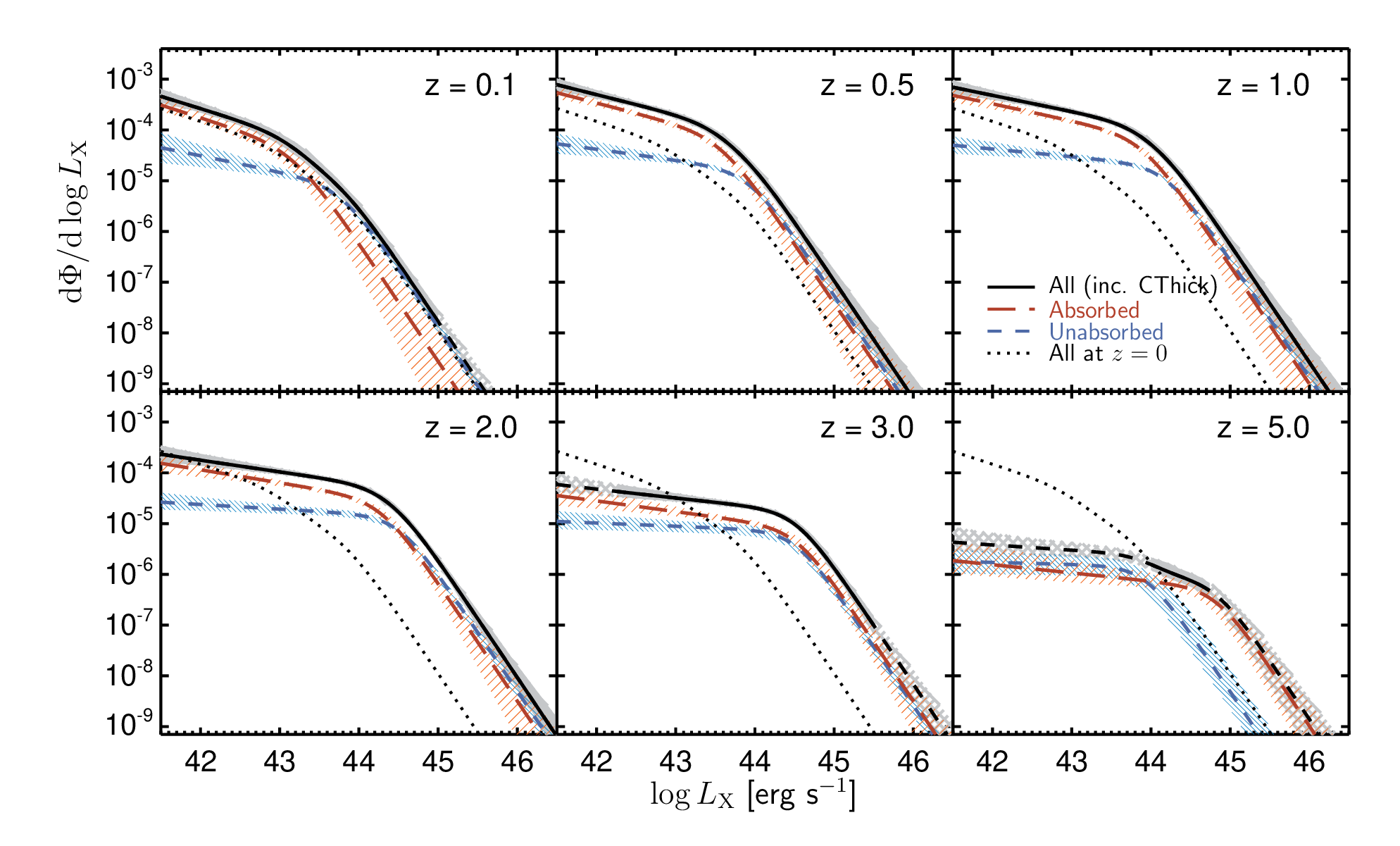}
\caption{
Our best model for the total XLF of AGNs (black solid line) and the constituent XLFs of the unabsorbed ($20<\log \nh<22$: blue short-dashed lines) and absorbed ($22<\log \nh<24$: red long-dashed lines) AGNs evaluated at a number of redshifts, based on our simultaneous fitting of both the hard- and soft-band X-ray samples to our XLAF model.
The shaded regions indicate the 99 per cent confidence interval based on the posterior distribution of the model parameters.
\reftwo{The hatched grey regions (and dashed black lines) indicate luminosity ranges where we lack either hard- or soft-band sources and our constraints are driven by extrapolation of our functional form.}
At $z\lesssim2$, unabsorbed AGNs tend to dominate at high luminosties, whereas absorbed AGNs---due to the lower $L_*$ of their XLF---dominate at lower luminosities. Both populations undergo a strong luminosity evolution (generally shifting to higher luminosities as redshift increases) and an overall density evolution, changing the ratio of absorbed to unabsorbed AGNs at different luminosities and resulting in a complex evolution in the shape of the total XLF. 
Extrapolating our model to the highest redshifts ($z\sim5$) indicates that absorbed AGNs may start to dominate at the highest luminosities; however, our observational constraints are weak in this regime and thus this behaviour may be an artefact of our particular model parametrization. 
}
\label{fig:xlf_abs_unabs}
\end{figure*}

\begin{figure*}
\includegraphics[width=\textwidth]{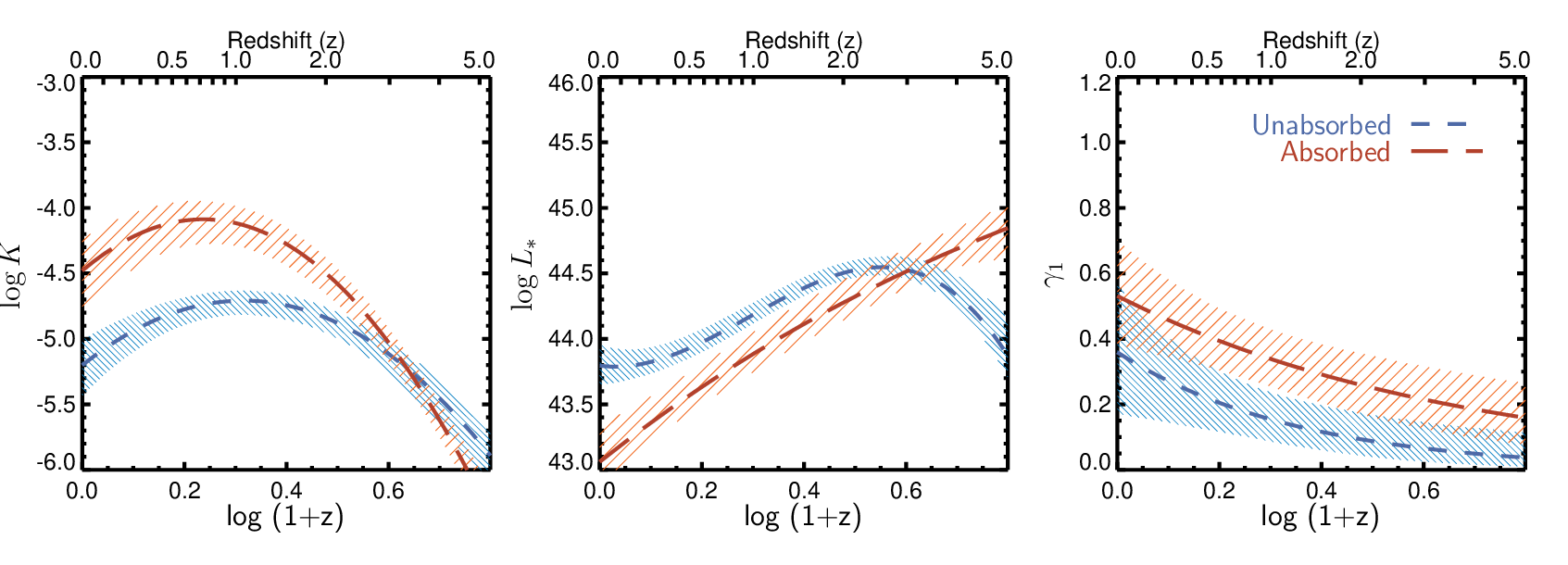}
\caption{
Shape parameters (normalization $K$, characteristic break luminosity $L_*$, and faint-end slope $\gamma_1$) for the XLFs of unabsorbed (blue short-dashed line) and absorbed (red long-dashed line) as a function of redshift, calculated by fitting both the hard- and soft-band X-ray samples to our full XLAF model. The shaded regions indicate the 99 per cent confidence interval based on the posterior distribution of the model parameters and thus may under-represent the uncertainty in the observations at a particular redshift.
}
\label{fig:pars_vs_z_abs}
\end{figure*}

\begin{figure*}
\includegraphics[width=\textwidth, trim=0 10 0 0]{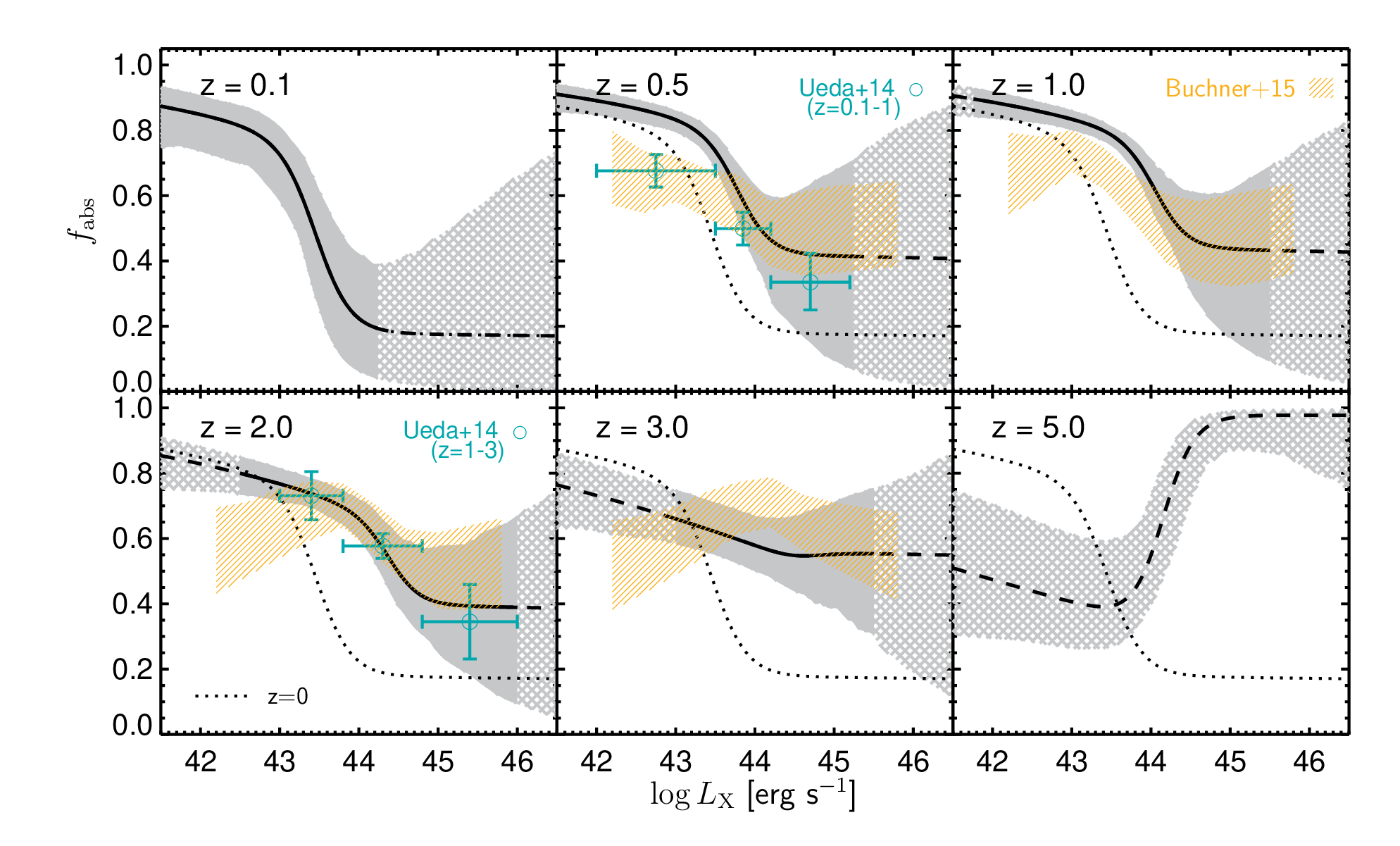}
\caption{
Fraction of absorbed AGNs, $f_\mathrm{abs}$ (defined here as the fraction of AGNs with $22<\log \nh<24$ relative to all Compton-thin, $\log \nh<24$, AGNs) based on our best-fit model of the XLAF evaluated at a number of redshifts (solid black lines). The grey regions indicate the 99 per cent confidence interval based on the posterior distribution of the model parameters. 
\reftwo{The hatched grey regions (and dashed black lines) indicate luminosity ranges where we lack hard-band sources and thus our constraints on \fabs\ are poor and driven by extrapolation of our functional form.}
The model evaluated at $z=0$ is shown by the dotted line in all panels. 
\secdraft{We also show direct estimates of the absorbed fraction as a function of luminosity from \citet[green circles, for the indicated redshift ranges]{Ueda14} and 
\citet[orange shaded regions indicate their 90 per cent confidence intervals for bins that approximately span each redshift]{Buchner15}.}
Our flexible model reproduces a luminosity dependence of $f_\mathrm{abs}$, which plateaus at low and high luminosities, with a rapid transition in between.
The position of this transition shifts to higher luminosities at higher redshifts due to the luminosity evolution of both the unabsorbed and absorbed AGN XLFs. 
\secdraft{This pattern is roughly consistent with the \citet{Ueda14} and \citet{Buchner15} measurements, although our estimates are slightly higher at low luminosities.}
At $z=3$ we find that the absorbed fraction is roughly constant as a function of luminosity. At higher redshifts ($z=5$), absorbed AGNs may dominate at high luminosities, 
\reftwo{although this behaviour is based on extrapolation of our model XLFs to these high redshifts where we lack hard-band sources and are thus unable to directly constrain \fabs.}
}
\label{fig:fabs}
\end{figure*}

Figure \ref{fig:xlf_abs_unabs} compares our model XLFs of the unabsorbed (blue) and absorbed (red) AGNs at a range of redshifts, as well as showing the total XLF of the full population (black). 
The XLF of absorbed AGNs generally has a lower break luminosity ($L_*$), a higher normalization, and a steeper faint-end slope than the XLF of unabsorbed AGNs. 
Hence, the absorbed AGN population is dominant at low luminosities, whereas unabsorbed AGN are increasingly important at higher luminosities.
Both the absorbed and unabsorbed AGN XLFs undergo a strong luminosity evolution, shifting to higher luminosities from $z\sim0$ out to $z\sim3$. 
Both XLFs also evolve in overall density, with the absorbed AGNs evolving more strongly and thus making up a larger fraction of the total AGN population at $z=1$, even at high luminosities, but dropping away more rapidly at higher redshifts. 
The evolution of the XLF shape parameters with redshift is presented in Figure \ref{fig:pars_vs_z_abs}. 
The differing evolution of the absorbed and unabsorbed AGN XLFs and thus their relative contributions to the total space density of AGNs results in a complex evolution in the shape of the total XLF of AGNs (solid black line in Figure \ref{fig:xlf_abs_unabs}).

In Figure \ref{fig:fabs} we plot the fraction of absorbed AGNs (relative to the total Compton-thin population), $f_\mathrm{abs}$, as a function of \LX\ at various redshifts (i.e. the ratio of the absorbed AGN XLF to the sum of the absorbed and unabsorbed XLFs).
The absorbed fraction roughly plateaus at high and low luminosities, with a fairly rapid transition in between due to the difference in the break luminosities of the XLFs. 
The strong luminosity evolution of both XLFs causes this transition to shift to higher luminosities at higher redshifts.  
The stronger overall density evolution of the absorbed AGNs also leads to an increase in $f_\mathrm{abs}$ at high luminosities ($\lx \gtrsim10^{45}$ \ergs), where absorbed AGNs increasingly dominate, although our constraints on the bright end of the absorbed AGN XLF are fairly poor.
\secdraft{
Our model reproduces a luminosity and redshift dependence of the absorbed fraction and is generally consistent with direct estimates based on X-ray spectral classifications \citep[e.g.][]{Ueda14,Buchner15}.
We do not find any evidence for the \emph{decrease} in $f_\mathrm{abs}$ with decreasing luminosity, which is seen by \citet{Buchner15} and some other studies \citep[e.g.][]{Burlon11,Brightman11b}, although this may reflect the limited diagnostic power of our method to distinguish absorbed and unabsorbed AGNs at the lowest luminosities.}

Our model indicates that the positive luminosity evolution of the unabsorbed AGN population slows down by $z\sim3$ and appears to start a \emph{negative} luminosity evolution to higher redshifts (see central panel of Figure \ref{fig:pars_vs_z_abs}).  
At $z\gtrsim3$, the negative luminosity evolution of the unabsorbed AGN XLF combined with the density evolution of the absorbed AGN XLF leads to a change in the pattern of \fabs. At $z=3$, \fabs\ is approximately constant as a function of \LX\footnote{Such a lack of luminosity dependence in the absorbed fraction at $z\gtrsim3$ has also been found by \citet{Vito14b}, albeit classifying column densities of $\nh>10^{23}$ cm$^{-2}$ as absorbed.}, whereas at higher redshifts the absorbed AGNs appear to dominate at the highest luminosities.
These results hint at a complex and differing evolution of absorbed and unabsorbed AGNs in the early Universe.
However, our observational constraints on the total XLF at $z\sim 4 - 7$ are generally fairly poor.
Furthermore, our diagnostic power to determine the absorbed fraction of AGN---comparing the hard- and soft-band X-ray samples---is substantially reduced at these redshifts.
The hard-band sample is small at these redshifts and both the hard and soft bands probe high rest-frame energies and are thus only weakly affected by absorption. 
\reftwo{In Figure \ref{fig:fabs} we indicate the range of luminosities where we detect both hard and soft band sources by the solid black line and grey confidence region. 
The dashed line and hatched grey region indicate where our constraints on \fabs\ are primarily determined by extrapolation of our functional form and should be treated with caution.
Constraining the evolution of the bright-end of the XLF---in particular testing whether the absorbed XLF evolves less rapidly than the unabsorbed XLF at $z\gtrsim3$, leading to the high \fabs\ at $\lx\gtrsim10^{45}$ \ergs\ indicated by our extrapolation---requires large area, hard-band selected samples of X-ray sources, which will be provided by the \emph{eROSITA} mission \citep{Merloni12}.
}

Nonetheless, our model is able to reproduce the overall evolution of the XLF and by accounting for absorption can reconcile our hard- and soft-band X-ray samples. 
Furthermore, we naturally recover the luminosity dependence of the absorbed fraction at $z\sim0-2$ and its evolution.

\begin{table}
\caption{\secdraft{Best \emph{a posteriori} estimates of parameters describing the galaxy luminosity function.}}
\begin{center}
\begin{minipage}[c]{0.63\columnwidth}
\begin{tabular}{c c}
\hline
{Parameter} & {Value} \\
\hline
$\log A$ (Mpc$^{-3}$ dex$^{-1}$) & $-3.59 \pm 0.08$\\
$\alpha$					& $\;0.81\pm 0.03$\\
$\log L_0$ (\ergs)			& $41.12\pm0.07$\\
$\beta$					& $\;2.66\pm0.24$\\
$z_c$					& $\;0.82\pm 0.10$\\
\hline
\end{tabular}

{N.B. All parameters are partially constrained by our informative priors, specified in Table \ref{tab:galpriors}}
\end{minipage}
\end{center}
\label{tab:galresults}
\end{table}

\section{Discusssion}
\label{sec:discuss}

\subsection{Comparison with previous studies of the X-ray luminosity function}
\label{sec:comp_xlf}

Our work allows us to interpret the evolution of the total XLF of AGNs as due to the combination of the XLFs of unabsorbed and absorbed (as well as Compton-thick) AGNs. 
We find that both XLFs can be described by our FDPL model, albeit with different sets of parameters. 
Our approach differs from most prior studies of the evolution of the XLF, which instead attempt to find a model that directly describes the \emph{total} XLF, either using an LDDE parametrization or some form of luminosity and density evolution
\citep[e.g.][]{Ebrero09,Yencho09,Aird10}.

\begin{figure*}
\begin{center}
\includegraphics[width=\textwidth,trim=0 20 0 0]{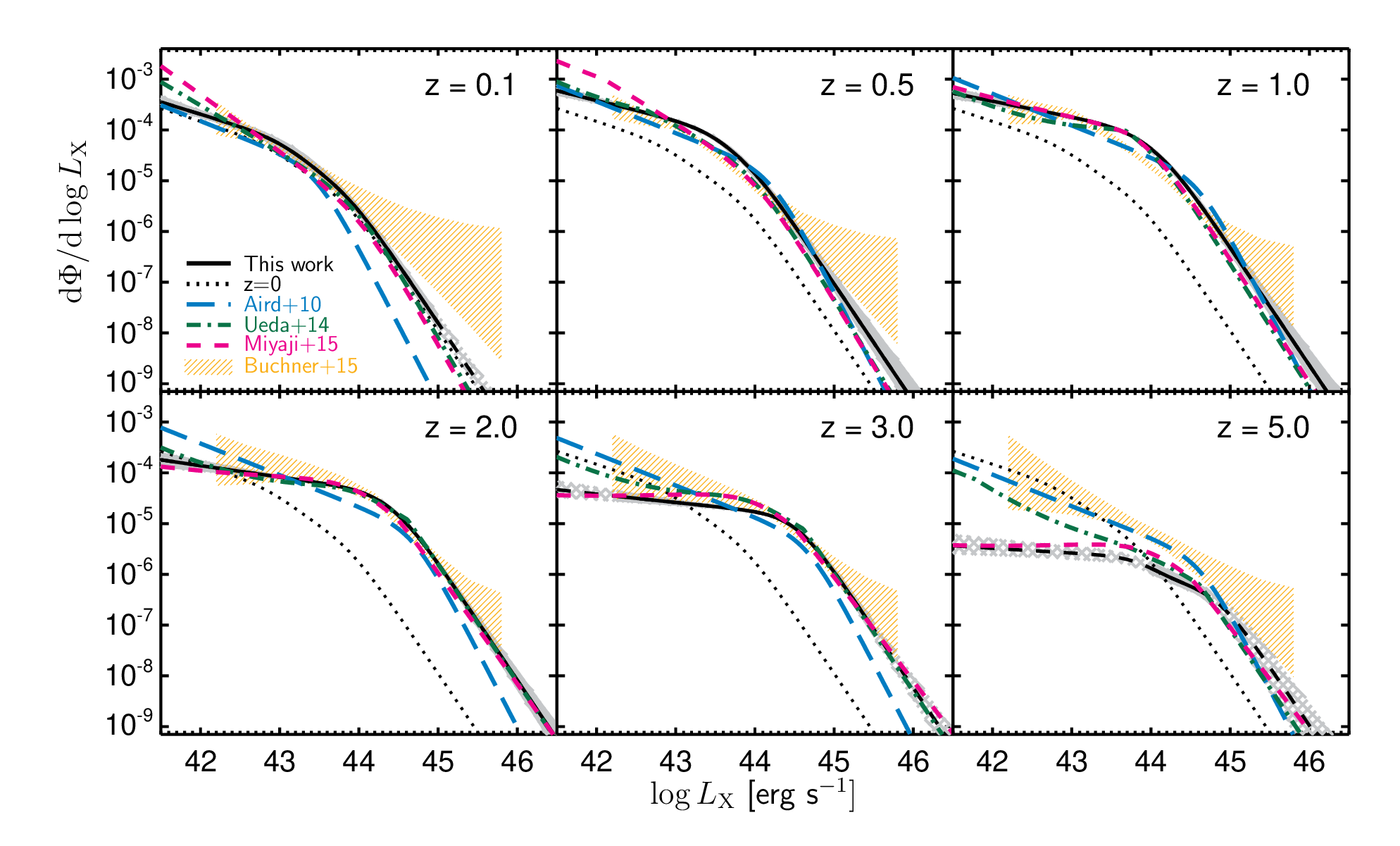}
\end{center}
\caption{
\secdraft{
Our model for the total XLF of all Compton-thin ($\nh<10^{24}$ cm$^{-2}$) AGNs (black solid line, grey region indicates 99 per cent confidence interval in model parameters, \reftwo{dashed line and grey hatching indicate where we lack data and are extrapolating our model}) compared to prior model fits: 
the LADE model from \citet[blue long-dashed line]{Aird10}; 
the LDDE2 model from \citet[green dot-dashed line]{Ueda14};
and the LDDE model from \citet[pink short-dashed line]{Miyaji15}.
We also show estimates from \citet{Buchner15} who used a non-parametric method with simple smoothness assumptions to estimate the space density in fixed bins of \LX, \NH\ and $z$ (orange hatched region indicate their 90 per cent confidence intervals for bins that approximately span the indicated redshift and include all Compton-thin AGNs). 
}
}
\label{fig:xlf_comp}
\end{figure*}

In Figure \ref{fig:xlf_comp} we compare the total XLF based on our final model (black solid line) to models from three recent studies: 
the LADE model from A10; the LDDE2 model put forward by \citet[hereafter U14]{Ueda14};
\refone{and the LDDE parametrization used in recent work by \citet[hereafter M15]{Miyaji15}.}
\secdraft{We also compare estimates from \citet[herafter B15]{Buchner15} who used a non-parametric method to estimate the space density of AGNs in fixed bins of \LX, \NH \ and $z$ with only simple smoothness assumptions used to link values between bins.}

At $z\lesssim3$ there is good agreement between our work and the previous studies, although there are a number of important differences.
The bright end of our XLF is generally in good agreement with previous studies, although our updated work predicts a substantially higher space density then the A10 model at both the lowest ($z=0.1$) and highest ($z\gtrsim2$) redshifts.
\secdraft{
The B15 estimates are higher than our model at the bright end, especially at lower redshifts.
However, this work lacked the very large-area ($\gtrsim50$ deg$^{2}$) samples included in our study and other work. 
Thus, the B15 estimates at high luminosities and low redshifts may be driven by the smoothness assumptions inherent to their method.
}

At $z\approx0.5-1$ the faint end of our total XLF generally lies slightly above the A10 model. 
The differences between A10 and our updated study are likely due to a combination of absorption effects (neglected in A10), the limitations of the LADE model, and the smaller sample in the A10 work.
The LDDE models from U14 and M15 are in better agreement with our work, although there are slight differences in the shapes that may reflect differences in the parametrizations.
For example, the M15 XLF is also somewhat higher than our model at the faintest luminosities at $z\le0.5$, which may be due to the LDDE parametrization requiring a steep faint-end slope (that is altered by the luminosity-dependent evolution). 
Our study probes to lower luminosities than M15, accounting for normal galaxies at low luminosities, and can thus rule out such a steep slope.
We also predict a slightly higher space density than U14 or M15 at 
$\lx\sim10^{43-44}$ \ergs and $z=0.5$, and a slightly lower space density at moderate luminosities ($\lx\lesssim10^{44}$ \ergs) at $z=3$.
\secdraft{
The B15 estimates are generally in good agreement with our results at fainter luminosities out to $z=3$, although our model tends to lie towards the lower end of their confidence intervals.
There are small differences close to $L_*$ at $z=0.5$ and $z=1.0$ whereby B15 predicts slightly lower space densities, which may also be driven by the assumed functional form and differences in the methodology.}

At $\lx\sim10^{44-45}$ \ergs\ and $z=5.0$ our model is below the extrapolation of the A10 work or the B15 estimates but is consistent with the recent work of U14 and M15 (which did include some sources at these redshifts and luminosities).
\refone{We find a strong decline in the space density in of AGNs to high redshifts in this moderate luminosity range, which is consistent with previous studies \citep[e.g.][]{Civano11,Vito13}.}
However, the high-redshift behaviour of our model may, in part, be driven by the extrapolation of the functional form that best describes the lower redshift XLF, where the bulk of our sample lies. 
\secdraft{
We will present a focused study of the evolution of the XLF at high redshifts in future work (Georgakakis et al. in preparation).}

\begin{figure*}
\begin{center}
\includegraphics[width=0.75\textwidth, trim=0 20 0 0]{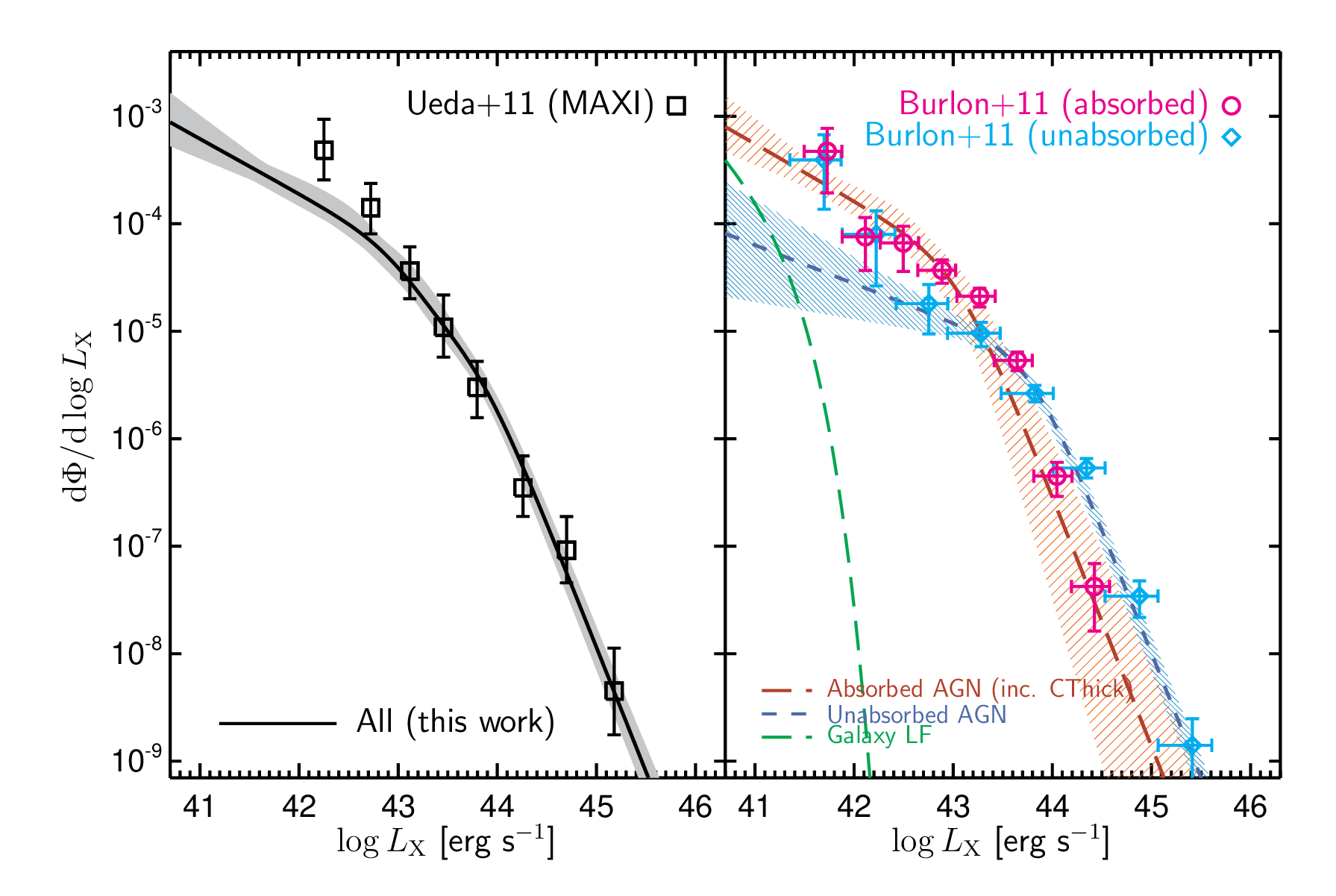}
\end{center}
\caption{
\textit{Left:} 
Our best fit model for the total XLF of AGNs (black line) extrapolated to $z=0$, compared to measurements of the local 2--10 keV XLF of AGNs from \citet{Ueda11} using MAXI (black squares).
\textit{Right:}
Our best fit model for the XLF of unabsorbed (blue short-dashed line) and absorbed (red long-dashed line) AGNs extrapolated to $z=0$, compared to measurements from \citet{Burlon11} from \textit{Swift}/BAT (converted to 2--10 keV luminosities assuming $\Gamma=1.9$) where the populations were divided based on X-ray spectral fits (cyan diamonds and pink circles for unabsorbed and absorbed AGNs, respectively). 
The grey, blue and red regions indicate the 99 per cent confidence interval in our model extrapolation based on the posterior distribution of model parameters. 
Our model extrapolation, based on predominantly higher redshift samples and indirectly inferring the absorption distribution, is generally in good agreement with these local measurements. 
Our results indicate that the steep slope of the total XLF at $z\sim0$ observed by \citet[see also \citealt{Ballantyne14}]{Ueda11} may be due the gradual change in the mix of absorbed and unabsorbed AGNs.
However, our model does not predict the increase in the unabsorbed AGN XLF at $\lx\lesssim10^{43}$ \ergs\ found by \citet{Burlon11}, which leads to a decreased absorbed fraction at low luminosities.
}
\label{fig:xlf_z0}
\end{figure*}

As an additional check, in Figure \ref{fig:xlf_z0} we compare the extrapolation of out model XLF to $z=0$ with measurements based on local AGN populations.
Our total XLF (left panel) does not undergo a sharp break, but flattens gradually below $\lx\sim10^{44}$ \ergs\ due to the mixing of the unabsorbed and absorbed AGN XLFs. 
This mixing results in a relatively steep slope in the $\lx\sim10^{43-44}$ \ergs\ luminosity range, consistent with the \citet{Ueda11} measurements 
shown by the black squares.
The right panel compares direct measurements of the XLFs of unabsorbed ($\nh<10^{22}$ cm$^{-2}$) and absorbed ($\nh>10^{22}$ cm$^{-2}$) from \citet{Burlon11} using the \textit{Swift}/BAT sample,
where the populations were divided based on the results of X-ray spectral fitting for individual sources (cf. our statistical method).
The extrapolation based on our baseline model for unabsorbed and absorbed AGNs is in good agreement with these measurements, which is reassuring given that our model is primarily constrained by much higher redshift sources and we do not directly measure \NH\ for individual sources.
We note that the lowest two luminosity bins for the unabsorbed AGNs from \citet{Burlon11} lie above our model (although only by $\sim1-2 \sigma$). 
Thus, \citet{Burlon11} found that the absorbed fraction decreases at both high and \emph{low} luminosities \citep[see also][]{Brightman11b,Buchner15}. 
Our model does not generally predict such a behaviour in the absorbed fraction\footnote{We note that fitting the unabsorbed and absorbed XLFs as indpendent broken double power laws could reproduce such a pattern, as in \citet{Burlon11}, thus the lack of such behaviour is not due to a limitation of our XLAF model.}.
We find that $f_\mathrm{abs}$ plateaus at $\sim85$ per cent at low luminosities (see Figure \ref{fig:fabs}).
In Figure \ref{fig:xlf_z0} (right panel) we also show our final best fit GLF. 
Normal galaxies have much lower luminosities than the unabsorbed AGN found by \citet{Burlon11}. 
Thus, the lack of a decrease in $f_\mathrm{abs}$ at low luminosities does not appear to be due to us (statistically) misclassifying some fraction of our X-ray sources as galaxies rather than unabsorbed AGNs.

In conclusion,  our work is mostly consistent with recent estimates of the total XLF of AGNs at $z\approx0-4$.
However, our new model provides a simpler way of interpreting the evolution of the XLF and the absorbed fraction as due to the combination of the unabsorbed and absorbed AGN XLFs and their slightly different evolution. 
We discuss the possible physical origin of this behaviour in Section \ref{sec:drives} below.

\subsection{Does the faint end of the XLF flatten at high redshifts?}
\label{sec:faintendslope}

How the shape of the XLF evolves with redshift, and in particular whether the faint-end slope becomes progressively flatter at higher redshifts, remains a major question in studies of the evolution of the AGN population \citep[e.g.][]{Hopkins06,  Aird10}.
In our study, we do not directly constrain the faint-end slope of the total XLF of AGNs -- instead we describe the total XLF as a combination of the XLFs of the unabsorbed and absorbed AGNs and infer different faint-end slopes for each population (hereafter referred to as \gunabs\ and \gabs\ respectively). 
Our baseline model allows both \gunabs\ and \gabs\ to change with redshift and our results indicate that both slopes become flatter as redshift increases (see Figure \ref{fig:pars_vs_z_abs}). 
At all redshifts, \gunabs\ is flatter than \gabs.
However, based purely on our data and analyses, we are unable to rule out a simpler model where neither \gunabs\ or \gabs\ change with redshift (hereafter the ``no-change" model). \refone{We still require that \gunabs\ is flatter than \gabs\ in the no-change model.}

Despite these issues, we find that the faint-end slope of the \emph{total} XLF does appear to change with redshift.
This can be seen in Figure \ref{fig:xlf_abs_unabs}.
At $z=0.1$ there is a gradual transition in the $\lx\sim10^{43-44}$ \ergs\ luminosity range between the unabsorbed AGN XLF (dominant at high luminosities) and the absorbed AGN XLF (dominant at low luminosities) which results in in a relatively steep slope at these intermediate luminosities. 
It is only at fainter luminosities that the absorbed AGN XLF starts to dominate and a flatter overall slope is seen\footnote{Indeed, an advantage of our own study is that we probe down to very faint luminosities ($\lx\sim10^{39}$ \ergs) at low redshifts using deep \textit{Chandra} fields, carefully considering the contribution from normal galaxies, and improve the overall constraints on the faint end of the AGN XLF.}.
At higher redshifts, the absorbed AGN XLF is increasingly dominant at moderate luminosities and thus a clearer break and a flatter faint-end slope is seen in the total XLF at $\lx\sim10^{42-44}$ \ergs\ by $z\sim1$. 
The underlying flattening of \gabs\ in our baseline model contributes to this effect but the mixing of the populations appears to be more important. 

\begin{figure*}
\includegraphics[width=\textwidth]{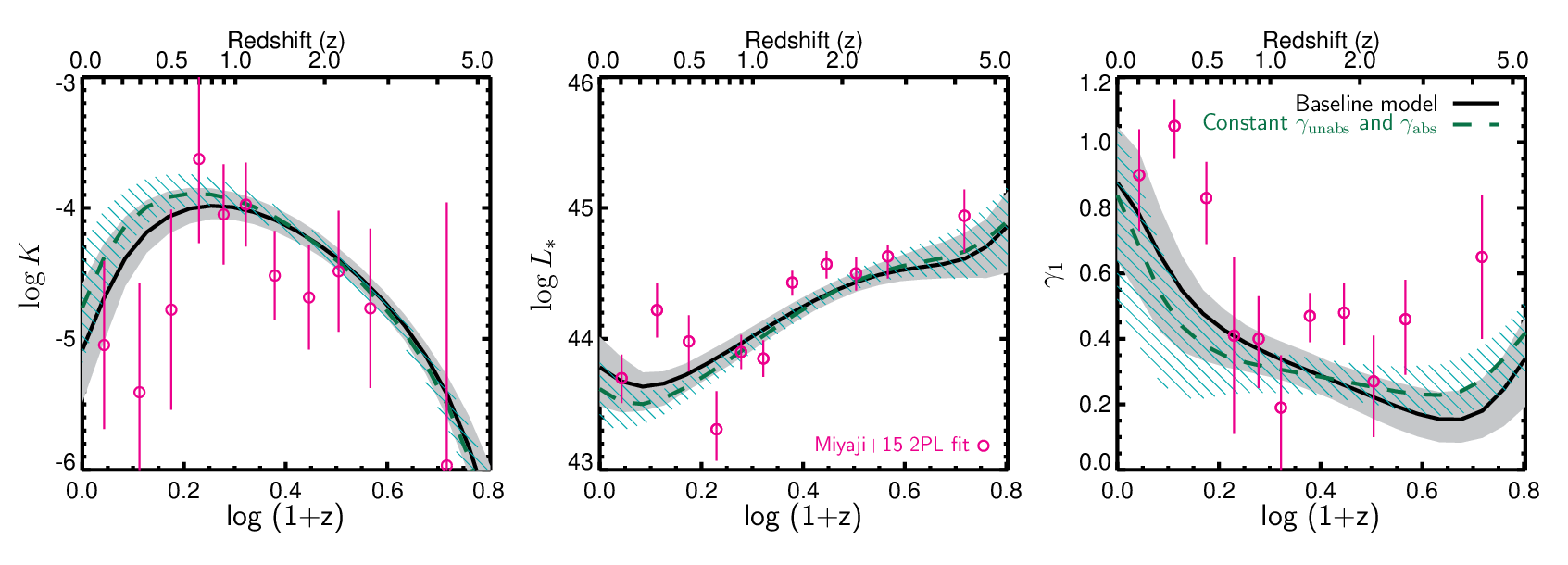}
\caption{
\refone{
The apparent double power-law shape parameters of the \emph{total} XLF of AGN as a function of redshift. These estimates are based on a simple least-squares fit at various redshifts and in the luminosity range $42.5<\log \lx<46.5$ to the predictions of our model (which describes the XLF as the combination of independent XLFs of the unabsorbed and absorbed AGNs; see Figure \ref{fig:pars_vs_z_abs} for the redshift dependence of the parameters of the individual XLFs that are directly constrained by our data). 
The solid black line indicates our baseline model where both \gunabs\ and \gabs\ of the underlying unabsorbed and absorbed AGN XLFs flatten with redshift, whereas the dashed green line uses our ``no-change" model where \gunabs\ and \gabs\ remain constant.
The shaded regions represent 99 per cent confidence regions in the underlying models. 
In both cases, the apparent faint-end slope ($\gamma_1$) undergoes a strong flattening with increasing redshift.
The similarity in the parameters describing the total XLF for both models illustrates how distinguishing between these models is difficult without additional information.
Pink data points show estimates of the double power-law shape parameters of the total XLF in thin redshift shells from \citet{Miyaji15} and reveal a similar evolutionary pattern.
}
}
\label{fig:gapparent}
\end{figure*}

\refone{
To further examine this effect, in Figure \ref{fig:gapparent} we show estimates of the \emph{apparent} shape of the total XLF as a function of redshift.
We estimate double power-law shape parameters by taking our overall model at various redshifts, creating fake data points in 0.25 dex wide redshift bins for $42.5<\log \lx <46$, and performing a simple least squares fit to the standard double power law form.
The solid black lines show the shape parameters of the total XLF for our baseline model, whereas the dashed green line shows the parameters for the no-change model.
In both cases, the evolution of the total XLF is very similar -- described by a combination of density evolution, positive luminosity evolution, and an \emph{apparent} flattening of the faint-end slope as redshift increases.
The apparent flattening is driven by the mixing of the constituent unabsorbed and absorbed AGN XLFs and their relative evolution.
Both the baseline and no-change models predict a similar evolution of the total XLF, including this apparent flattening. 
We also show estimates of the shape parameters of the total XLF from M15, which they calculated by directly fitting the XLF with a double power-law in thin redshift shells.
Their observed evolutionary pattern, including a strong flattening in the faint-end slope of the total XLF, is consistent with our findings.
}

Additional evidence in support of our baseline model rather than the no-change scenario may be provided by the data shown in Figure \ref{fig:xlf_z0} (right).
These data are in good agreement with our baseline model, extrapolated to $z=0$ (although the lowest luminosity bins for the absorbed XLF are systematically above our model). 
However, in the no-change model, \gabs\ and \gunabs\ are are driven towards the values required by our higher redshift data.
Thus, the extrapolation of the no-change model to $z=0$ does not provide a good agreement with the \citet{Burlon11} measurements, lending support to the baseline model. 
We note that in both cases our total XLF is in good agreement with the \citet{Ueda11} $z=0$ measurements (see Figure \ref{fig:xlf_z0} left). 
Incorporating these data sets into our Bayesian analysis remains beyond the scope of this paper.

\refone{
In conclusion, our own analysis requires that \gunabs\ is flatter than \gabs\ at any given redshift but does not provide definitive evidence that the individual slopes change with redshift. 
However, direct measurements of the XLFs of unabsorbed and absorbed AGNs at $z=0$ indicate that redshift evolution of \gunabs\ and \gabs\ \emph{is} required when compared to our higher redshift study.
Regardless of whether \gunabs\ and \gabs\ are changing, the apparent faint-end slope of the \emph{total} XLF does flatten with increasing redshift. 
This behaviour is primarily driven by the changing mix of unabsorbed and absorbed AGNs.}

\subsection{Evolution of the absorbed fraction of AGNs}
\label{sec:fabsevol}

\begin{figure}
\includegraphics[width=0.98\columnwidth]{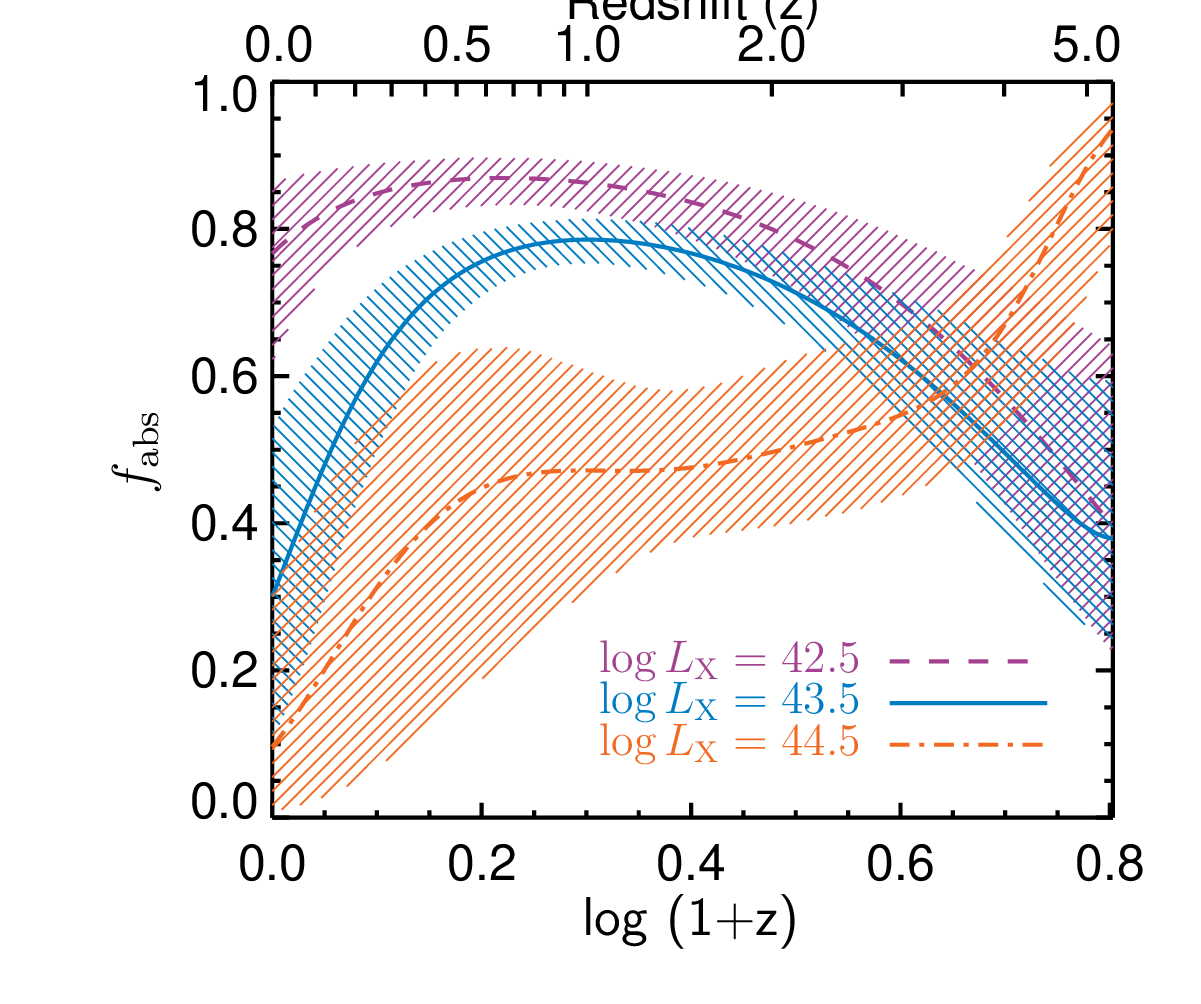}
\includegraphics[width=0.98\columnwidth,trim=0 15 0 0]{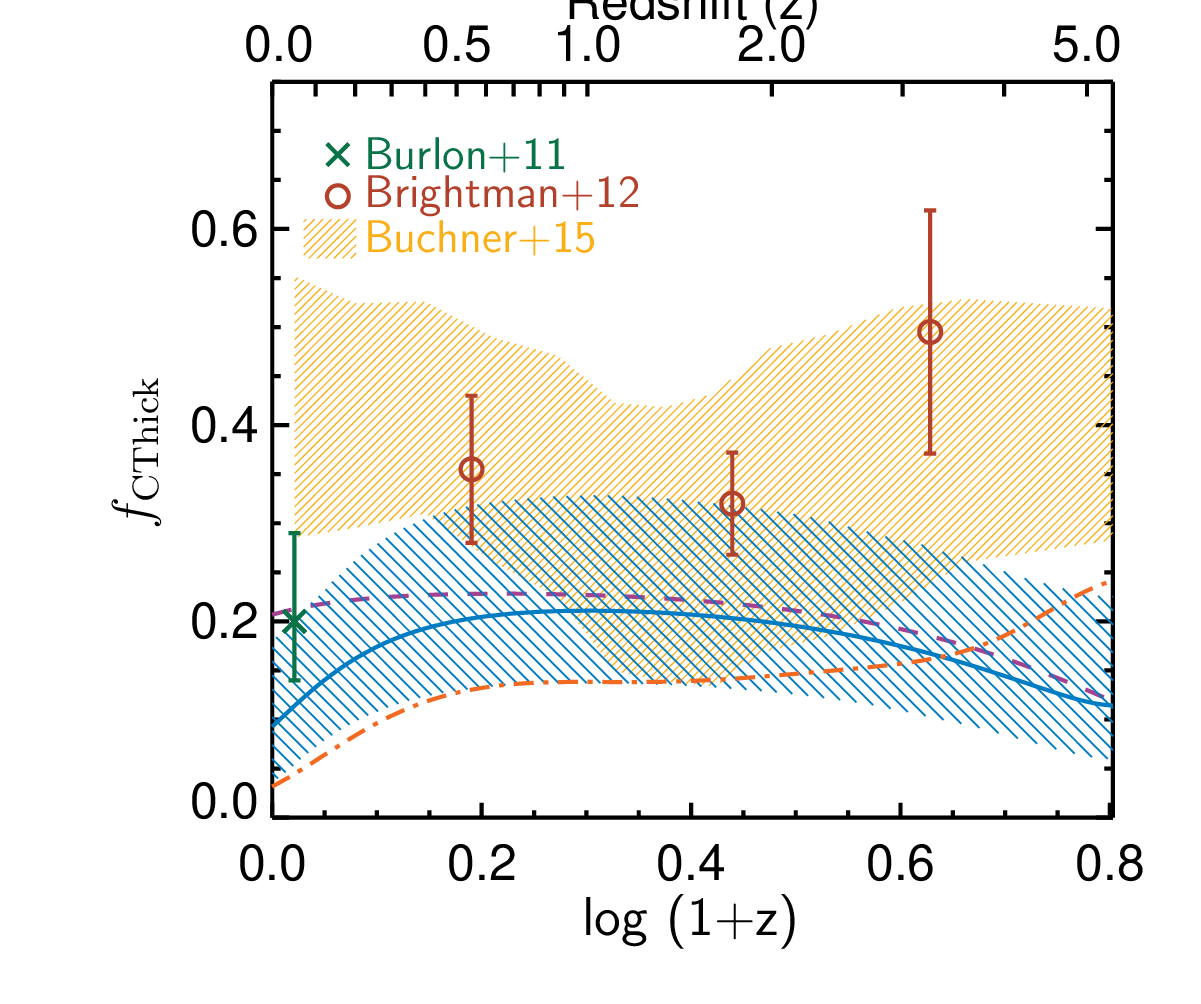}
\caption{
\textit{Top:} Absorbed fraction of Compton-thin AGNs ($f_\mathrm{abs}$) versus redshift based on our model, evaluated at three different luminosities. Shaded regions indicate 99 per cent confidence interval in the model parameters.
The absorbed fraction changes most rapidly at $\log \lx=43.5$ due to the evolution of $L_*$ for the unabsorbed and absorbed AGN XLFs. 
At high redshifts ($z\gtrsim3$) we have tentative evidence of a drop in $f_\mathrm{abs}$ at low and moderate luminosities and a rise at higher luminosities.
\textit{Bottom:} Fraction of Compton-thick AGNs (relative to all AGNs at a given luminosity) versus redshift from our model. For clarity, we omit the confidence regions for the low and high luminosities. 
\secdraft{
We compare to the estimates of the Compton-thick fraction from \citet[green crosses]{Burlon11}, \citet[red circles]{Brightman12b}, and \citet[orange shaded region indicates 90 per cent confidence interval]{Buchner15} that used X-ray spectral fitting to directly identify Compton-thick AGNs.}
While our estimate of $f_\mathrm{CThick}$ at $\log \lx=43.5$ is systematically lower than prior works, we generally agree within the uncertainties. 
}
\label{fig:fabs_vs_z}
\end{figure}

An increase in the fraction of absorbed AGNs with redshift (at a fixed luminosity) is now well established \citep[e.g.][]{LaFranca05,Hasinger08,Treister09}.
Our model reproduces such a behaviour (see Figure \ref{fig:fabs_vs_z}, top). 
However, we are able to attribute the underlying cause of this evolution to the luminosity evolution the XLFs of both the unabsorbed and absorbed AGNs.
This luminosity evolution shifts the XLFs of both populations to higher luminosities at higher redshifts, and thus the transition between the lower luminosity regime where absorbed AGNs dominate and the higher luminosity regime where unabsorbed AGNs dominate shifts to higher luminosities at higher redshifts (see Figure \ref{fig:fabs}).
Thus, close to the break luminosities of the XLFs ($\lx\sim10^{43.5-44.5}$ \ergs) a strong evolution in the absorbed fraction is observed, whereas at higher and lower luminosities the evolution is weaker (see Figure \ref{fig:fabs_vs_z}, top). 

Our results also suggest (albeit at low significance) that the absorbed fraction at high luminosities may increase slightly between $z\sim0$ and $z\sim1$ as the overall normalization of the absorbed AGN XLF increases more rapidly than the normalization of the unabsorbed XLF. 
While such a redshift dependence is much weaker than the strong evolution around $L_*$, it may explain why some studies have found substantial populations of obscured, luminous AGNs (Type-2 QSOs) at these high redshifts  \citep[e.g.][]{Brusa10,Stern12,Banerji15}. 
The normalization of the absorbed AGN XLF peaks at $z\approx1$ and declines towards higher redshifts. By $z\sim3$ the continuing, strong luminosity evolution of the absorbed AGNs appears to take over, again leading to a high absorbed fraction at the highest luminosities (conversely the absorbed fraction of lower luminosity AGNs may decline at $z\gtrsim2$).
\refone{It should be noted, however, that our best-fitting model is primarily constrained by lower redshift data, where the bulk of our sample lies. The high-redshift behaviour may simply reflect an extrapolation of this model to higher redshifts where our observational constraints are poor.}
Nevertheless, our findings do hint at a potentially complex evolution in the absorbed fraction of high luminosity AGNs.

\subsection{The $\mathrm{N_H}$ distribution, the fraction of Compton-thick AGNs and their contribution to the cosmic X-ray background}
\label{sec:cxb}

\begin{figure*}
\includegraphics[width=\columnwidth,trim=0 20 0 0]{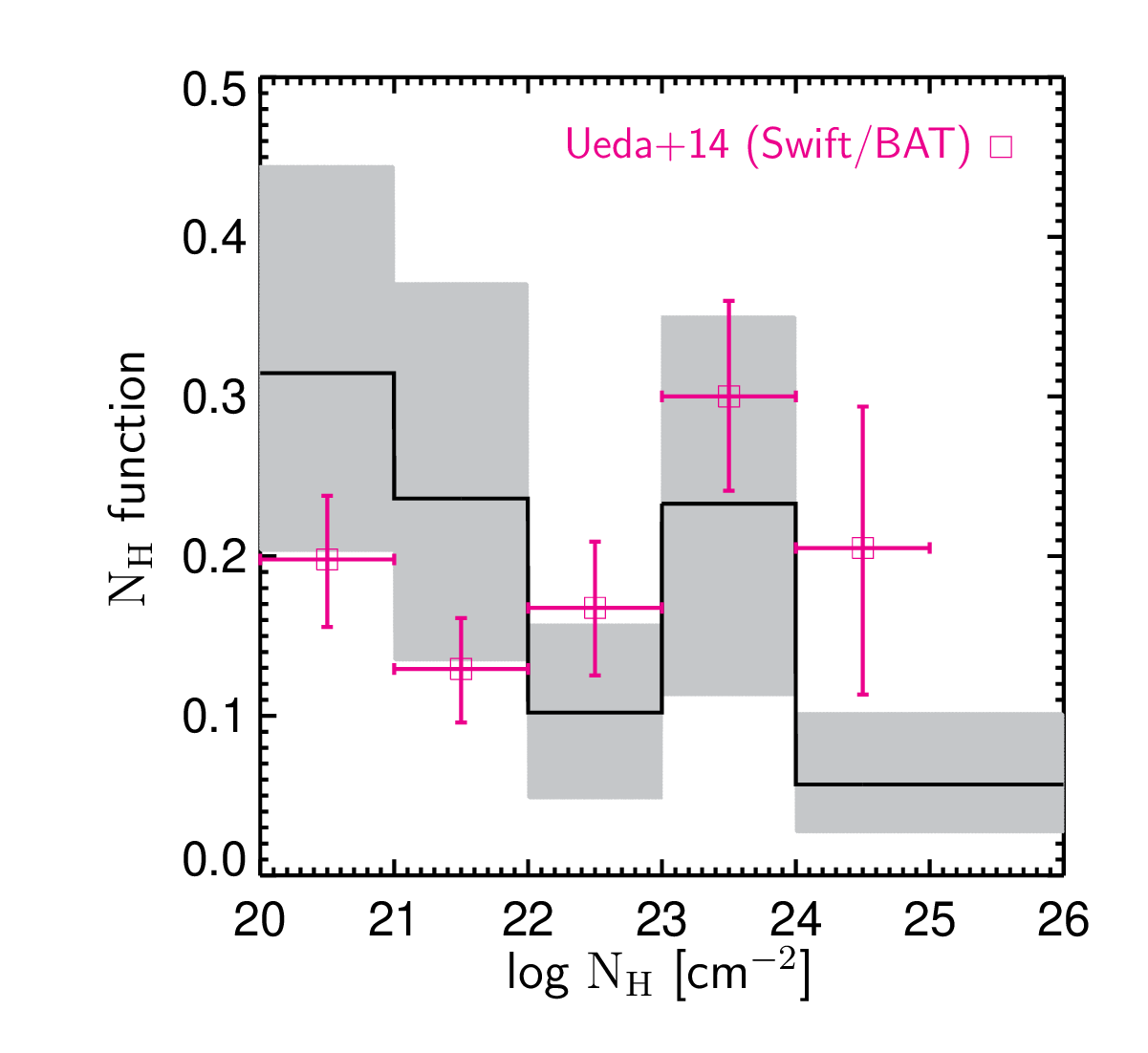}
\includegraphics[width=\columnwidth,trim=0 20 0 0]{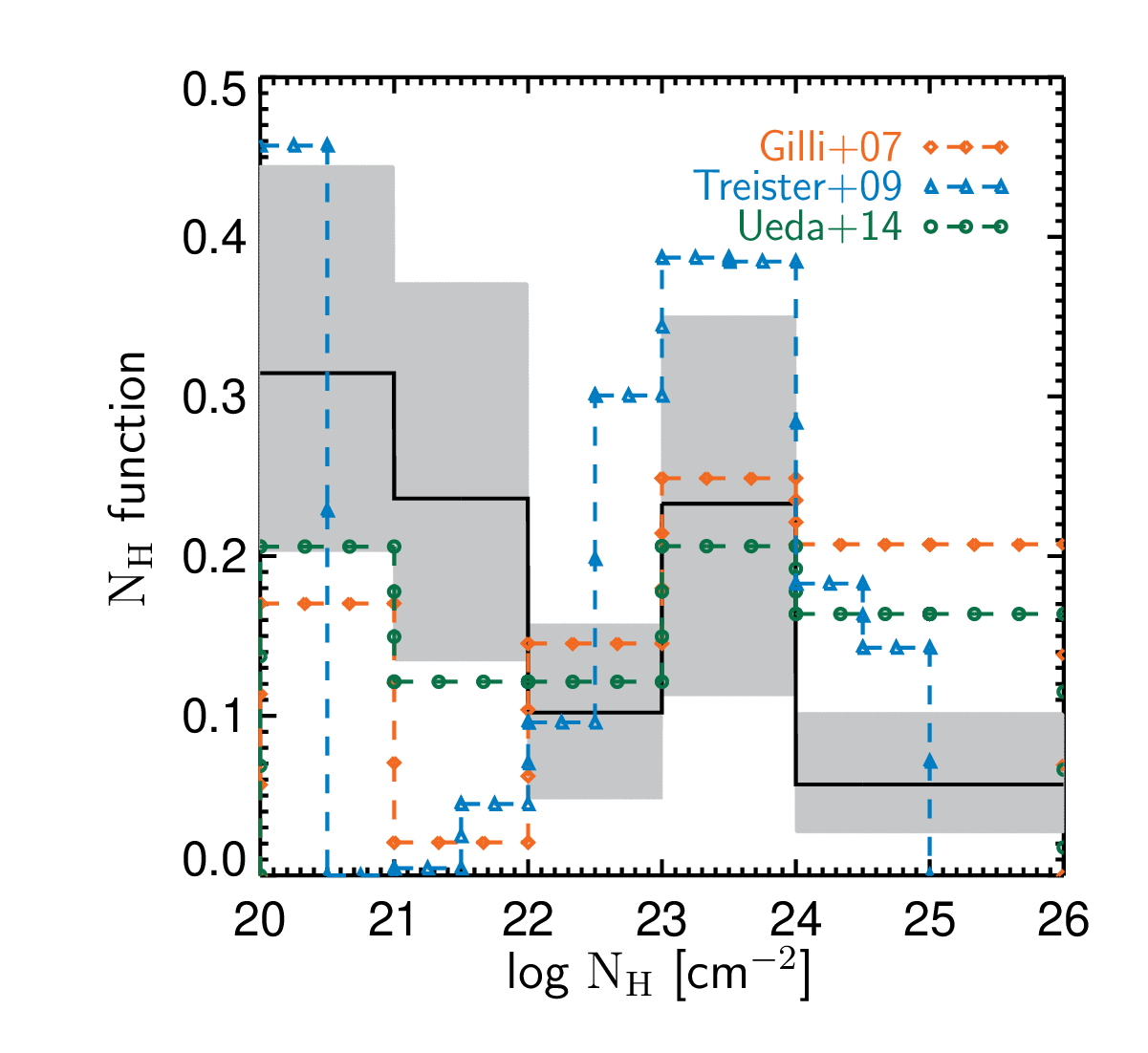}
\caption{
Our inferred distribution of absorption column densities (\NH) evaluated at $z=0.05$  and $\log \lx=43.5$ (black line, shaded region indicates 99 per cent confidence interval in the model) and compared to direct measurements based on X-ray spectral fits of the \textit{Swift}/BAT AGNs \citep[\textit{left:}][]{Ueda14} 
\refone{and that assumed in previous CXB synthesis models \citep[\textit{right:} ][]{Gilli07,Treister09,Ueda14}. }
}
\label{fig:nhdist}
\end{figure*}

While we do not directly estimate absorption column densities for individual sources, our study does predict an overall distribution of \NH\ based on our statistical approach.
In Figure \ref{fig:nhdist} we show our inferred distribution of \NH\ (the \NH\ function) evaluated at $z=0.05$ and $\log \lx=43.5$. 
\refone{In the left panel,} we compare to estimates of the \NH\ function from U14, based on direct measurements of \NH\ in the \textit{Swift}/BAT local AGN sample \citep{Tueller08,Ichikawa12} and corrected for absorption-dependent biases. 
Our inferred \NH\ function has a similar overall shape to the U14 measurements, dropping between $\log \nh\sim20-21$ and $\log \nh \sim 22-23$ and peaking again at $\log \nh \approx 23-24$. 
Our absorbed fraction ($f_\mathrm{abs}$) is slightly lower at this specific luminosity and redshift, thus our \NH\ function predicts slightly more unabsorbed ($\log \nh =20-22$) sources and fewer absorbed ($\log \nh=22-24$) sources. 
Furthermore, our prediction of the \NH\ function in the Compton-thick regime ($\log \nh>24$) is lower than the U14 measurements and corresponds to an overall Compton-thick fraction ($f_\mathrm{CThick}$) at $z=0.05$ and $\log \lx=43.5$ of just $11.4^{+7.7}_{-5.6}$ per cent.
Even so, the differences between our model and the \textit{Swift}/BAT measurements from U14 are not significant.

\refone{
The right panel of Figure \ref{fig:nhdist} compares our results to the \NH\ functions assumed in various AGN population synthesis models, used to estimate the integrated cosmic X-ray background (CXB).
While the overall shape of our \NH-function is similar, there are significant differences from prior studies.
The \citet[hereafter G07]{Gilli07} and \citet[hereafter T09]{Treister09} models predict fewer AGNs with $\nh\approx 10^{21-22}$ cm$^{-2}$), although neither our study nor the CXB models are particularly sensitive to the exact distribution in this range.
More importantly, 
our \NH-function is significantly different in the Compton-thick regime. 
The normalisation is significantly lower than the G07 or U14 models across the $\nh=10^{20-26}$ \cmsq\ range, ruling out these distributions (and the correspondingly higher $f_\mathrm{CThick}$) at the $>99$ per cent confidence level.
The T09 \NH-function, on the other hand, is higher than our model at $\nh=10^{24-25}$ \cmsq\ but assumes no sources with $\nh>10^{25}$ \cmsq;
thus, their overall $f_\mathrm{CThick}$ is consistent with our estimates where we assume a constant distribution up to $\nh=10^{26}$ \cmsq.
The exact shape of the \NH\ function in the Compton-thick regime, the relative number of $\nh\gtrsim 10^{25}$ \cmsq\ AGNs, and the precise X-ray spectral properties of such sources remain major uncertainties in estimates of $f_\mathrm{CThick}$ and could account for the differences between our estimates and the prior studies shown in Figure \ref{fig:nhdist} (right).
For example, if the most heavily Compton-thick sources ($\log \nh\gtrsim 25$) are also deeply buried and do not exhibit the same level of soft, scattered emission as less absorbed sources \citep[as suggested by some studies e.g.][]{Brightman14} then our measurements would be biased to below the true value.}

\secdraft{Figure \ref{fig:fabs_vs_z} (bottom) also compares our model for $f_\mathrm{CThick}$ to estimates from \citet{Burlon11}, \citet{Brightman12b} and B15 that used X-ray spectral fitting to directly identify Compton-thick AGNs in samples from \textit{Swift}/BAT, \textit{Chandra}, or \textit{XMM-Newton} respectively.}
While our model is systematically lower than all of these measurements, it is generally consistent given the uncertainties in our model and the data.
The B15 estimates are significantly higher at low ($z\lesssim0.2$) and high ($z\gtrsim3$) redshifts, although these regimes are not well probed by their samples and thus the estimates may be partly driven by their priors.
\refone{In the key $0.2<z<3$ range, where the B15 estimates are most reliable, our estimates are statistically consistent, although the B15 confidence intervals do allow for the potential of a higher $f_\mathrm{CThick}$.}

It is worth noting that our inferences regarding the Compton-thick population are \emph{indirect}. 
Our estimates of the underlying distribution of \NH\ rely on global comparisons of the hard- and soft-band samples and the changing sensitivities of these energy bands to different levels of absorption as a function of redshift.
With this approach, our ability to distinguish between the effects of Compton-thick sources and heavily absorbed, yet Compton-thin ($\nh\approx10^{23-24}$ \cmsq) sources is limited, which could affect our estimates of $f_\mathrm{CThick}$. 
We also assume that the XLF of Compton-thick AGNs traces the absorbed AGN XLF exactly (apart from a single scale factor) at all redshifts
and assume a completely flat distribution over the $\nh=10^{24-26}$ \cmsq\ range.
Given these issues, our estimates of the Compton-thick fraction should be treated with caution.

\begin{figure}
\begin{center}
\includegraphics[width=\columnwidth,trim=15 20 0 0]{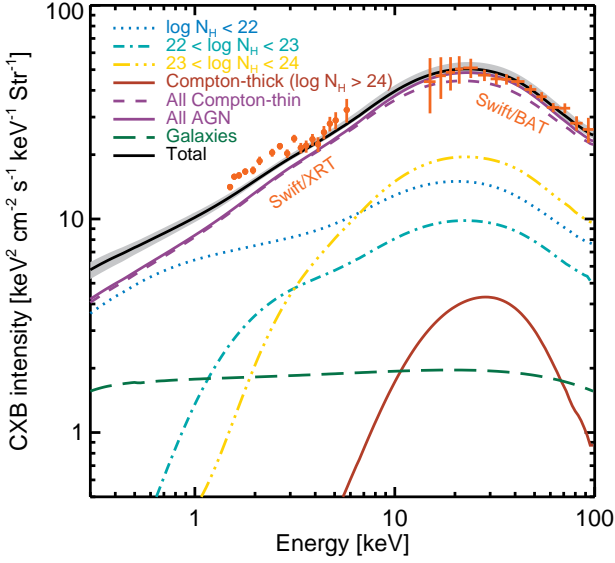}
\end{center}
\caption{
\secdraft{
Predicted Cosmic X-ray Background (CXB) spectrum based on our final model of the XLAF and GLF (black solid line, grey region indicates 99 per cent confidence interval) 
compared to recent measurements from \emph{Swift}/XRT \citep[orange circles at $<7$ keV,][]{Moretti09}
and \textit{Swift}/BAT \citep[orange crosses at $>10$ keV,][]{Ajello08b}.
The blue dotted, light-green dot-dashed, yellow triple-dot-dashed, and dark-red solid lines  show the contributions from AGNs with different absorption column densities.
The dashed purple line indicates the contribution of all Compton-thin ($\nh<10^{24}$ cm$^{-2}$) AGNs, while the solid purple line is the total AGN contribution including Compton-thick sources. 
The contribution of Compton-thick AGNs in our model to the peak of the CXB at $\sim30$ keV is minimal but does bring the model closer to the observational data.  
The dark green long-dashed line indicates the contribution of normal galaxies to the CXB (based on our estimates of the GLF) and accounts for $\gtrsim10$ per cent of the total CXB flux in our model at $\lesssim3$ keV. 
}  
}

\label{fig:estcxb}
\end{figure}

As an additional consistency check of our results, in Figure \ref{fig:estcxb} we use our model to estimate the overall contribution of AGNs to the CXB, retaining our assumed X-ray spectral model described in Section \ref{sec:singlesource}. 
Our model prediction for the CXB is in good agreement with the measurements at energies $>10$ keV that we do not directly probe with our source samples (we discuss the low energy CXB and the contribution of galaxies in Section \ref{sec:galdiscuss} below).
Despite our low Compton-thick fraction compared to U14 or G07, we are still able to reproduce the observed peak in the CXB at $\sim20-30$ keV. 
At these energies, we find that Compton-thick AGNs make a small, but important 
contribution to the integrated CXB emission. 

Part of the reason for our successful reproduction of the CXB may be the comparatively strong reflection that we allow for in unabsorbed ($\log \nh<22$) and moderately absorbed ($\log \nh\approx22-23$) sources. 
The extent and strength of reflection in such sources, especially those with $\gtrsim L_*$ luminosities and at $z\gtrsim1$ that make a large contribution to the CXB flux, is still a matter of debate \citep[e.g.][]{Vasudevan13b,DelMoro14}. 
Indeed, the CXB can be reproduced with a wide variety of models that assume different X-ray spectral models for the constituent AGNs, as well as different XLFs, \NH\ distributions, and---perhaps most crucially---Compton-thick fractions \citep[e.g.][]{Draper09,Ballantyne11,Akylas12}.
While fully exploring CXB synthesis models for different spectral models and Compton-thick fractions is beyond the scope of this work, Figure \ref{fig:estcxb} does indicate that our relatively low Compton-thick fraction should not be ruled out.

\subsection{What drives the absorption-dependent evolution of the AGN population?}
\label{sec:drives}

The shape and evolution of the XLF of AGNs provides important constraints on the evolution of the underlying physical properties of growing SMBHs, namely their masses and Eddington ratios (a tracer of the rate of accretion relative to the SMBH mass).
Our study---revealing the independent XLFs of unabsorbed and absorbed AGNs \refone{out to high redshifts}---provides further insights into how the evolution of the AGN population relates to their absorption properties.

We find that, at least to $z\sim2$, the evolution of the unabsorbed XLF and the evolution of the absorbed XLF are fairly similar. 
The bulk of the change in both XLFs is driven by moderately strong luminosity evolution,
which could be attributed to a change in the underlying distribution of SMBH masses, whereby the average mass of an accreting SMBH is higher at earlier cosmic times. 
Alternatively, the luminosity evolution may reflect an increase in the average accretion rates with increasing redshift.
Both XLFs also undergo an overall density evolution, whereby their normalizations increase between $z=0$ and $z\sim1$ then drop rapidly again at higher redshifts.
This evolutionary pattern may reflect changes in the rate at which AGN activity is triggered within the galaxy population. 
Finally, the faint-end slopes of both XLFs may also be changing with redshift.
Such changes could indicate that the underlying distribution of Eddington ratios is changing, with more rapidly accreting sources become relatively more predominant at higher redshifts, 
although recent studies find a fairly constant, power-law distribution of Eddington ratios to $z\sim1-2$ \citep{Aird12,Bongiorno12}. The observed flattening of the total XLF may instead be more closely related to the changing mix of unabsorbed and absorbed AGNs (see Section \ref{sec:faintendslope} above).

While the overall evolution of the XLFs of unabsorbed and absorbed AGNs follow similar patterns, the individual XLFs are significantly different at a given redshift.
The most important differences (at least at $z\lesssim2$) appear to be
1) the lower $L_*$ of the absorbed AGNs, and
2) the steeper faint-end slope of the absorbed AGN XLF.
These effects combine with the overall evolution of the XLFs to produce the luminosity and redshift dependence of the absorbed fraction seen in Figure \ref{fig:fabs}. 

The differences in $L_*$ may reflect the life cycle of AGN activity.
Luminous, unabsorbed AGNs could correspond to short periods in high accretion rate phases
that have the power to eject material from the galactic nucleus. 
Exhaustion of the fuel supply then leads to a comparatively rapid fading of the AGN, that is traced by the faint end of the unabsorbed AGN XLF. 
Absorbed AGN, conversely, may correspond to longer periods when the SMBH grows at lower  accretion rates but has little impact on the surrounding obscuring material. 
The two phases could correspond to
separate AGN events with
different triggering mechanisms
(e.g. major mergers versus secular processes within a host galaxy). 
Changes in the extent and efficiency of different triggering mechanisms could then lead to the slight differences in the evolution of the unabsorbed and absorbed XLFs at $z\lesssim2$.
Alternatively, the unabsorbed and absorbed phases could reflect long-term variations related to a single AGN triggering event.

At $z\gtrsim2$, our results hint at a more complex evolutionary behaviour.
The luminosity evolution of the unabsorbed XLF appears to slow at $z\sim2-3$ and then decline to higher redshifts, whereas the absorbed XLF continues to evolve positively in $L_*$ but the normalization ($K$) is rapidly declining (see Figure \ref{fig:pars_vs_z_abs}).
\secdraft{
This pattern suggests that absorbed growth phases continue to shift to higher accretion rates or SMBH masses at high redshifts, yet the triggering of such events becomes increasingly rarer. 
Conversely, the most luminous, unabsorbed growth phases may be suppressed 
at the earliest times.}
However, our power to distinguish between the unabsorbed and absorbed populations is limited at the these redshifts.

\begin{figure}
\includegraphics[width=\columnwidth,trim=0 20 0 -10]{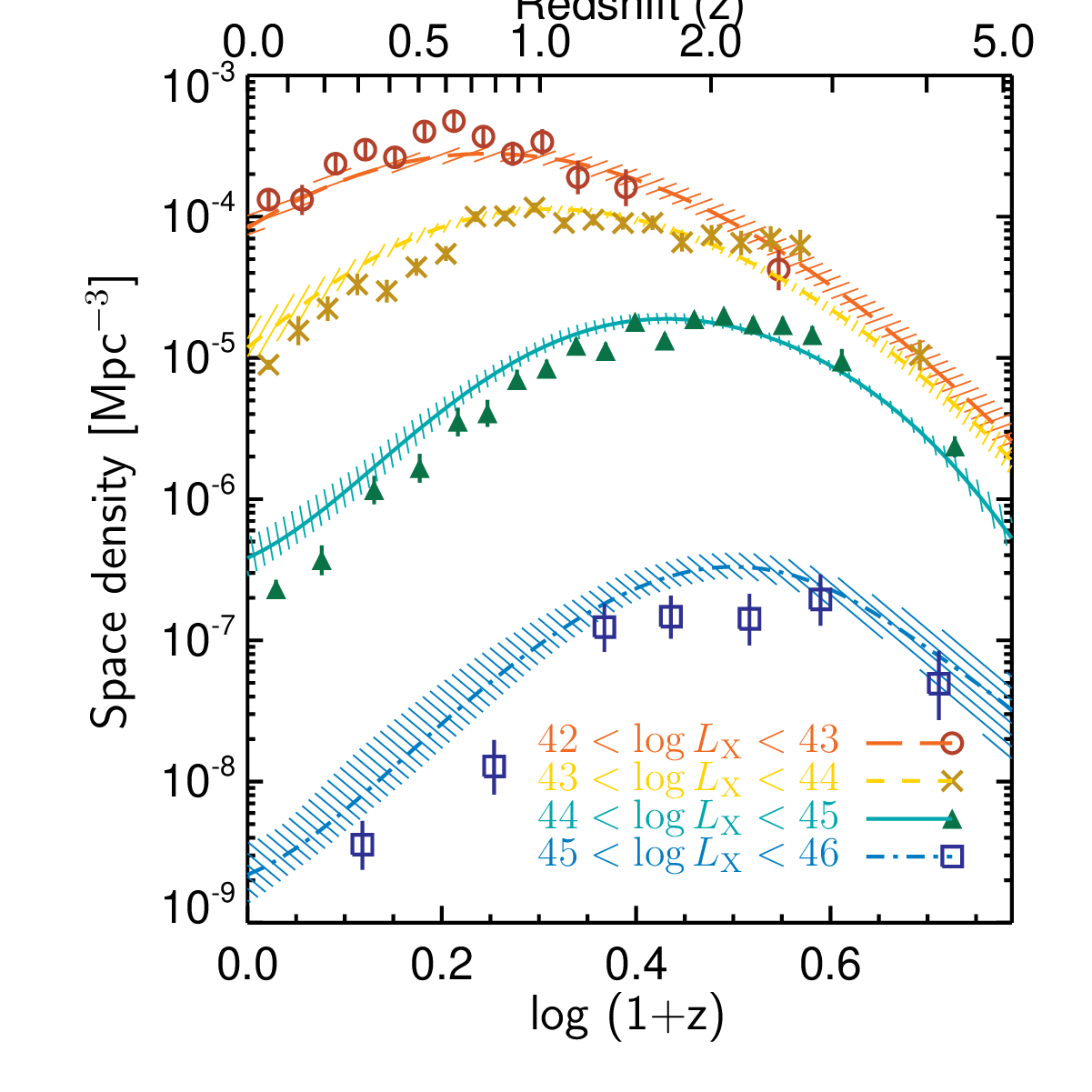}
\caption{
Total space density of AGNs for different ranges of X-ray luminosity based on our model (coloured lines).
Shaded regions indicate the 99 per cent confidence interval in our model parameters. 
We see clear ``downsizing'' in terms of AGN luminosity, whereby higher luminosity AGNs peak in terms of there space density at higher redshifts than lower luminosity sources.
\refone{The data points are taken from the recent work of \citet{Miyaji15} and 
reveal the same pattern; the small discrepancies with our model are most likely due to differences in the modelling of the XLF and the \NH\ distribution.}
}

\label{fig:spacedensity}
\end{figure}

The combined evolution of the unabsorbed and absorbed XLFs across our entire redshift range results in strong, luminosity-dependent evolution in the overall space densities of AGNs (see Figure \ref{fig:spacedensity}). 
Higher luminosity AGNs peak, in terms of their space densities, at higher redshifts than lower luminosity AGNs.
\refone{A similar pattern has been seen in a number of previous works, most recently by M15 (data points in Figure \ref{fig:spacedensity}). The minor differences between the M15 estimates and our results are likely due to the different modelling of the underlying XLF and the correction for absorption.}\footnote{\refone{M15 use the \NH-function from U14 to correct for absorption effects (as a function of \LX\ and $z$), which differs from our results.}}
This pattern in space densities, described as ``downsizing" of AGN luminosities \citep[e.g.][]{Ueda03,Barger05}, is often attributed purely to changes in the average SMBH mass with redshift.
However, our constraints on the underlying evolution of the XLFs indicate that the observed pattern in the evolution of total space densities (Figure \ref{fig:spacedensity}) may be due to a combination of changes in SMBH mass, the distributions of Eddington ratios, and the life cycles of obscured and unobscured AGN phases.

\subsection{The contribution of normal galaxies to the observed luminosity function and the cosmic X-ray background}
\label{sec:galdiscuss}

\secdraft{
An important advance in our work is the inclusion of normal, X-ray detected galaxies in our analysis of the XLF.
Instead of applying a strict luminosity cut (e.g. $\lx>10^{42}$ \ergs), we develop an approach to statistically classify X-ray sources as galaxies or AGNs.
We find that AGNs still dominate the space density of X-ray selected sources at $\lx\approx10^{42}$ \ergs\ at low redshifts.
At $z\approx0.1$ the space density of galaxies is a factor $\sim 1000$ lower than the space density of AGNs at $\lx=10^{42}$ \ergs. 
Galaxies only begin to make a substantial contribution ($>10$ per cent in space density) at $\lx\lesssim 2.5 \times 10^{41}$ \ergs and start to dominate ($>50$ per cent in space density) at $\lx\lesssim2\times10^{40}$ \ergs. 
}

\secdraft{
We find that the GLF undergoes strong luminosity evolution up to $z\approx0.8$, whereby the characteristic luminosity increases as approximately $(1+z)^{2.7}$ \citep[consistent with previous studies of the evolution of the GLF at X-ray wavelengths, e.g.][]{Ptak07,Georgakakis07,Tzanavaris08}.
This evolution increases the contribution of galaxies at low luminosities. 
At $z\approx1$, the space density of galaxies corresponds to $\sim10$ per cent of the AGN space density at $\lx=10^{42}$ \ergs\ and galaxies dominate at $\lx\lesssim1.6\times10^{41}$ \ergs. 
Careful consideration of contamination by normal galaxies is thus vital in studies of faint AGN at these redshifts. 
}

\secdraft{
At higher redshifts, our data are not deep enough to accurately probe the GLF and its evolution. 
However, out model indicates that by $z\approx3$ galaxies may dominate over AGNs at $\lx=10^{42}$ \ergs\ \citep[see also][]{Laird06,Aird08}.
Thus, galaxies may make a substantial contribution to the X-ray population at high redshifts and should be carefully considered in future studies of the high-redshift XLF of AGNs. 
}

\secdraft{
Galaxies also make a significant contribution to the CXB at low energies. 
Previous studies have shown that normal galaxies can account for $\sim4-20$ per cent of the total CXB emission at $\sim 1-6$ keV energies, based on stacking of the X-ray data at the optical positions of known galaxies \citep[e.g.][]{Worsley06,Hickox07,Xue12}.
Here, we can account for the contribution of normal galaxies to the CXB 
 based on our modeling of the \emph{resolved} X-ray populations. 
Figure \ref{fig:estcxb} shows that galaxies contribute $\gtrsim10$ per cent of the total CXB emission at energies $\lesssim3$ keV.
However, our estimate of the total CXB 
 remains $\sim 10-20$ per cent lower than the recent \textit{Swift}/XRT measurements at $\lesssim3$ keV, 
although measurements of the absolute flux of the CXB vary substantially between missions in this regime, probably due to calibration uncertainties \citep[e.g.][]{Barcons00,Moretti09}.
Indeed, our model of the total CXB is in excellent agreement with earlier measurements of the CXB flux with \textit{ASCA} \citep{Gendreau95} over the full $\sim1-7$ keV range.
Thus, the significance of any discrepancy is uncertain. 
}

\subsection{The accretion history of the Universe}
\label{sec:acchistory}

\begin{figure}
\includegraphics[width=\columnwidth,trim=0 20 0 -10]{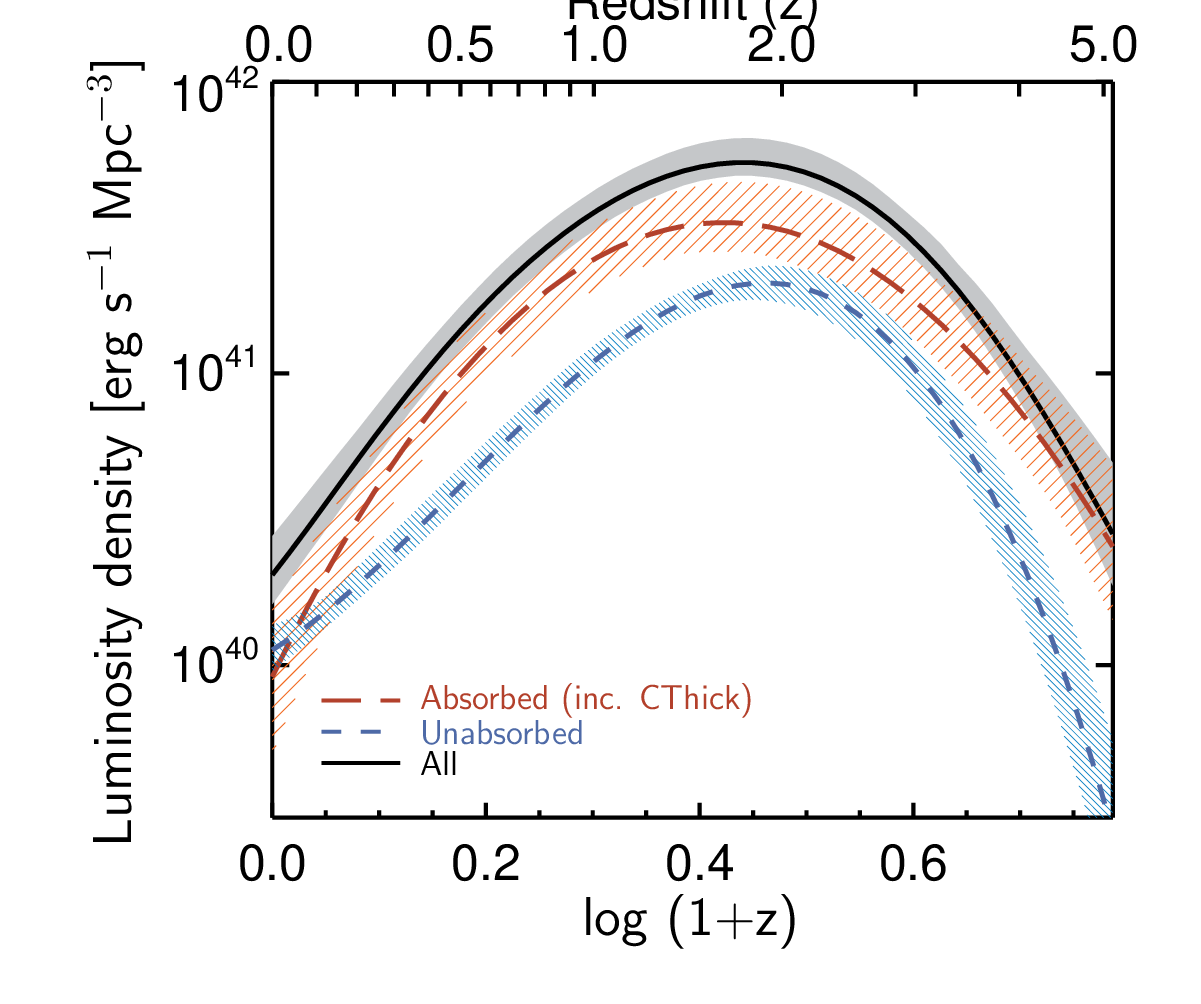}
\caption{
Bolometric luminosity density, a tracer of the total accretion density, for unabsorbed, absorbed (including Compton-thick) and all AGNs as a function redshift, based on our final model of the XLAF.
Shaded regions indicate the 99 per cent confidence interval in our model parameters.
The total rate of SMBH growth peaks at $z\sim2$ and is dominated by absorbed and Compton-thick AGNs.
}

\label{fig:lumdensity}
\end{figure}

Our new measurements of the XLF of AGNs out to $z\sim5$ enable us to place improved constraints on the overall accretion history of the Universe. 
In Figure \ref{fig:lumdensity} we plot the total luminosity density of AGNs, 
\begin{equation}
\int d \log L_\mathrm{bol} \phi_\mathrm{bol}(L_\mathrm{bol}, z \giv \nhfpars)
\end{equation}
where $L_\mathrm{bol}$ is the bolometric luminosity and $\phi_\mathrm{bol}(L_\mathrm{bol}, z \giv \nhfpars)$ is the bolometric luminosity function of AGNs, found by integrating our XLAF model over $\log \nh$ and converting the resulting XLF to bolometric values.
We adopt the luminosity-dependent bolometric corrections from \citet{Hopkins07b}.
We find a clear peak in the luminosity density at $z=1.76\pm0.05$, 
\refone{consistent with prior studies \citep[e.g. U14,][]{Delvecchio14}.}
We also show the luminosity density for the absorbed (including Compton-thick) and unabsorbed populations. 
At $z\gtrsim0.1$, the majority of SMBH growth is taking place in absorbed or Compton-thick AGNs. The proportion increases rapidly between $z=0$ and the overall peak at $z\sim1-2$.
The contribution from unabsorbed AGNs increases more gradually and peaks at a slightly higher redshift ($z=1.89\pm0.05$). 

By integrating the AGN luminosity density across our redshift range, we can estimate the total SMBH mass density that has been built up by accretion \citep{Soltan82}.
Adopting a radiative efficiency of $\eta=0.1$,
we estimate a relic SMBH mass density of $4.20^{+0.29}_{-0.14} \times 10^5\;\mathcal{M}_\odot$ Mpc$^{-3}$ at $z=0$, where the error represents the uncertainty in our XLF model parameters.
Of this mass density, $\sim30$ per cent has been built up in unabsorbed growth phases, $\sim50$ per cent in absorbed AGNs, and a further $\sim20$ per cent in Compton-thick sources. 
Our estimate is higher than in A10 ($2.2\pm0.2 \times 10^5\;\mathcal{M}_\odot$ Mpc$^{-3}$), likely due to our improved accounting for absorbed and Compton-thick sources, and is in much better agreement with the estimate by \citet{Marconi04} based on the velocity dispersions of local galaxies and the $\mathcal{M}_\mathrm{BH}-\sigma$ relation ($4.6^{+1.9}_{-1.4} \times 10^5 ;\mathcal{M}_\odot$ Mpc$^{-3}$).
However, uncertainties in the bolometric correction and radiative efficiency dominate over our XLF model uncertainties. 
A higher radiative efficiency (e.g. if a large fraction of SMBHs have high spins) would reduce our estimate of the relic SMBH mass density. 
Alternatively, a large population of deeply buried, highly Compton-thick sources could bring our estimate of the relic SMBH mass density more in line with recent revisions to the $\mathcal{M}_\mathrm{BH}-\sigma$ relation \citep[e.g.][]{Comastri15}.

\begin{figure*}
\begin{center}
\includegraphics[width=0.64\textwidth,trim=0 30 0 -10]{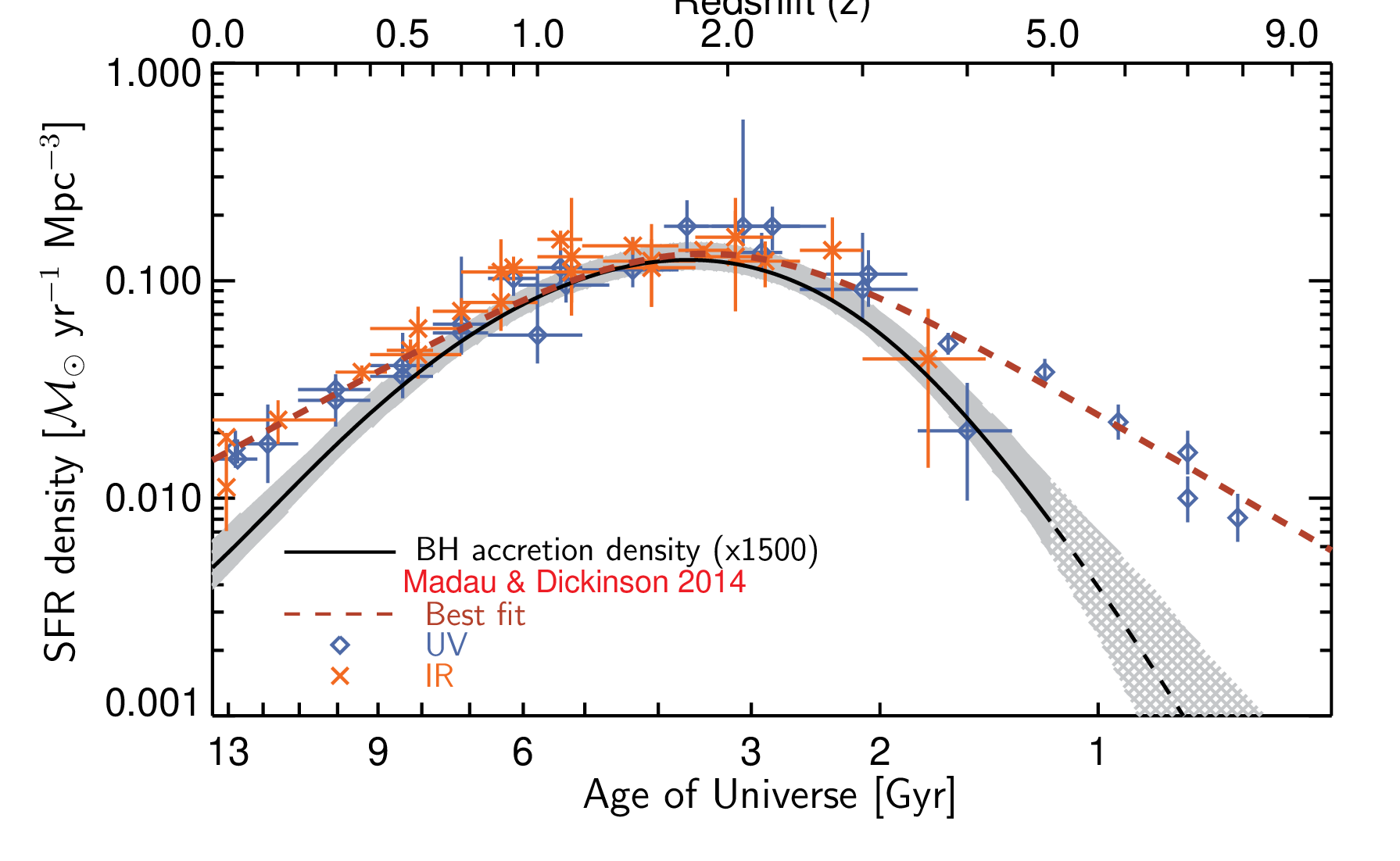}
\end{center}
\caption{
Total star-formation rate (SFR) density compared to our estimate of the total SMBH accretion density.
We show the best overall fit (dashed dark-red line) for the evolution of the SFR density from the recent review by \citet{Madau14} along with their compilation of measurements based on rest-frame ultraviolet (blue diamonds) and infrared (orange crosses) observations. 
Our estimate of the SMBH accretion density (solid black line) is scaled up by an arbitrary factor of 1500. 
Shaded regions indicate the 99 per cent confidence interval in our model parameters and may under-represent the overall uncertainties in the bolometric correction, radiative efficiency and observational data at a given redshift.
The dashed black line and grey hatching indicate where we are substantially extrapolating our model outside the coverage of our data.
Both galaxy and SMBH growth peak at $z\sim2$. However, the SMBH accretion density appears to evolve more rapidly, with a much stronger decline to higher redshifts. 
}
\label{fig:sfrd}
\end{figure*}

In Figure \ref{fig:sfrd} we compare estimates of the total star formation rate (SFR) density \citep[from a recent review by][]{Madau14} to our estimates of the SMBH accretion density, converted to $\mathcal{M}_\odot$ yr$^{-1}$ Mpc$^{-3}$ using the same bolometric corrections and radiative efficiency assumptions described above and scaled up by an additional factor of 1500.
We extrapolate our model to $z>5$ to enable comparisons with estimates of the SFR density in the first $\sim 1$ Gyr of cosmic time.
The SFR density and SMBH accretion rate both peak at $z\sim2$. 
However, the total rate of SMBH growth rises slight more rapidly from $z=0$ to $z\sim2$, indicating that the relative efficiency of SMBH growth relative to galaxy growth may change over this period.
The SMBH accretion density drops much more rapidly to higher redshifts than the SFR density. 
Thus, galaxy growth may precede the build up of their central SMBHs in the early Universe, or the relative efficiency with which SMBHs form and grow may be substantially lower.
However, our estimates are extrapolations at the highest redshifts.
\refone{New and ongoing surveys with \chandra\ (e.g. COSMOS-Legacy, P.I. Civano; CDFS-7Ms, P.I. Brandt) may improve constraints in this regime.}
Nonetheless, obtaining direct and accurate measurements of the XLF of AGNs at the highest redshifts ($z\gtrsim6$) requires $\sim2$ orders of magnitude improvement in survey speed and is one of the principal aims of ESA's next large mission, the \textit{Athena} X-ray observatory \citep{Nandra13,Aird13b}.

\section{Summary}
\label{sec:summary}

In this paper we present new measurements of the XLF of AGNs out to $z\sim5$.
In this section we first summarize our work and then present our conclusions.

We compile large samples of uniformally selected X-ray sources in the hard (2--7 keV) and soft (0.5--2 keV) energy bands from a number of deep and wide \chandra\ surveys (CDFS, CDFN, EGS, COSMOS and Bootes), supplemented by wider area surveys from \textit{ASCA} and \textit{ROSAT}.
We also compile multiwavelength photometry across our fields that we use to robustly identify counterparts to our X-ray sources and calculate photometric redshifts.
Next, we present a Bayesian methodology that allows us to incorporate a range of X-ray spectral shapes and statistically account for the effects of absorption, 
as well as accounting for photometric redshift uncertainties, flux uncertainties, and the Eddington bias.
We also introduce an approach to statistically account for the contamination of our samples by normal, X-ray detected galaxies. 
This allows us to place improved constraints on the XLF of AGNs down to much lower luminosities than prior studies.

We present initial measurements of the XLF based on our hard-band and soft-band samples separately, neglecting absorption effects. 
We introduce a new, flexible model to describe the XLF of AGNs as an evolving double power-law, where the break luminosity ($L_*$), normalization ($K$) and faint-end slope ($\gamma_1$) all evolve with redshift. 
There is significant evidence (based on Bayesian model comparison) favouring this model over luminosity-dependent density evolution (LDDE) parametrizations of the XLF based on either the hard-band or soft-band samples. 
\refone{However, differences in the space densities at low luminosities based on the hard-band compared to the soft-band samples indicate that absorption has a substantial effect and requires further consideration.}

To investigate the extent and distribution of absorption, we model the XLFs of unabsorbed ($\nh<10^{22}$ cm$^{-2}$) and absorbed ($\nh>10^{22}$ cm$^{-2}$) AGNs as independent, flexible double power-law functions. 
We use the combination of the hard- and soft-band samples to constrain our flexible model, seeking a model that can reconcile both samples assuming no \emph{a priori} knowledge of the extent of absorption for any individual source.
We find that the absorbed AGN population has an XLF that generally has a lower break luminosity, a higher normalization, and a steeper faint-end slope than the unabsorbed AGNs \refone{at all redshifts out to $z\sim2$}. 
Both the unabsorbed and absorbed AGN XLFs evolve in luminosity and density.
Differences in the extent of the luminosity and density evolution lead to a comparatively complex evolution in the shape of the total XLF of AGNs.

Our work provides new insight into the nature and evolution of the AGN population. 
Our main conclusions are as follows:
\begin{enumerate}[itemsep=5pt,topsep=3pt,
leftmargin=*,labelindent=5pt,labelsep=5pt,itemindent=1pt,label=\arabic*.]

\item
Differences between the break luminosities ($L_*$), normalizations ($K$), and faint-end slopes ($\gamma_1$) of the unabsorbed and absorbed AGN XLFs explain the strong luminosity dependence in the fraction of absorbed AGNs at $z\lesssim 2$.
Absorbed AGNs dominate at low luminosities, with the absorbed fraction falling rapidly as luminosity increases above a characteristic value.

\item
Both the unabsorbed and absorbed XLFs undergo strong luminosity evolution up to $z\sim2$, 
shifting the transition in the absorbed fraction to higher luminosities at higher redshifts.
A strong evolution in the absorbed fraction is thus observed at $\lx \sim 10~^{43.5-44.5}$ \ergs, close to the break luminosities of the XLFs. 
At higher redshifts ($z\gtrsim2$), the luminosity evolution of the unabsorbed AGN XLF slows and may become negative, although this behaviour may be driven by the model parametrization at lower redshifts.

\item
The normalization of the absorbed AGN XLF evolves more rapidly than the normalization of the unabsorbed AGN XLF, peaking at $z\sim1$ and declining to higher redshifts.
This leads to a mild increase in the absorbed fraction at high luminosities ($\lx\sim10^{45}$ \ergs) between $z\sim0$ and $z\sim1$.

\item
The total XLF of AGNs (combining unabsorbed and absorbed populations) undergoes a complicated evolution with redshift, including substantial flattening of the faint-end slope. \emph{This behaviour is primarily driven by the changing mix of unabsorbed and absorbed AGNs.}

\item
Our inferred distribution of \NH\ generally exhibits two peaks at $\nh\sim10^{20-21}$ cm$^{-2}$ and $\nh\sim10^{23-24}$, although the exact shape depends on luminosity and redshift. 
We infer a relatively low fraction of Compton-thick AGNs ($\sim10-20$ per cent at $\lx=10^{43.5}$ \ergs). Nevertheless, our model reproduces the peak of the cosmic X-ray background at $\sim20-30$ keV.

\item
The majority of SMBH growth takes place in absorbed or Compton-thick AGNs at all redshifts $z\gtrsim0.1$, although the proportion increases rapidly between $z=0$ and $z\sim2$.
The contribution from unabsorbed AGNs increases more gradually and peaks at a higher redshift ($z\sim2$). Unabsorbed phases of activity constitute a substantial fraction ($\sim30$ per cent) of the overall accretion density.

\item
At low redshift ($z\approx0.1$) the space density of X-ray selected galaxies is negligible compared to AGNs at $\lx=10^{42}$ \ergs. 
At $z\approx1$, however, galaxies correspond to $\sim10$ per cent of the AGN space density at $\lx=10^{42}$ \ergs\ and dominate at $\lx\lesssim1.6\times10^{41}$ \ergs. 
Normal galaxies contribute $\gtrsim10$ per cent of the total integrated CXB emission at energies $\lesssim3$ keV.

\item
The total AGN accretion density generally tracks the rise in the overall SFR density from $z\sim0$ to $z\sim2$, indicating that SMBH growth and galaxy growth proceed together over this period. The accretion density appears to fall off substantially faster than the SFR density to higher redshifts, indicating that galaxy growth may precede the build up of their central SMBHs in the early Universe or the relative efficiency of SMBH growth may be lower.

\end{enumerate}

\secdraft{
Our work provides new insight into the nature of the evolution of the AGN population and the underlying cause of the luminosity and redshift dependence of the absorbed fraction. 
However, to fully reveal the underlying physical processes that drive both unabsorbed and absorbed growth phases requires further study to connect the distributions of AGN accretion rates and their absorption properties to the key physical properties of the evolving galaxy population. 
}

\section*{acknowledgements}
We thank the referee for helpful comments that have improved
this paper.
We also thank Johannes Buchner for helpful discussions and providing machine-readable results.
We acknowledge helpful discussions with Ranjan Vasudevan.
The scientific results reported in this article are based to a significant degree on observations made by the \textit{Chandra} X-ray Observatory.
This work has made use of the Rainbow Cosmological Surveys Database, which is operated by the Universidad Complutense de Madrid (UCM).
This study makes use of data from AEGIS, a multiwavelength sky survey conducted with the \textit{Chandra}, \textit{GALEX}, \textit{Hubble}, Keck, CFHT, MMT, Subaru, Palomar, \textit{Spitzer}, VLA, and other telescopes and supported in part by the NSF, NASA, and the STFC.
Funding for the DEEP2 Galaxy Redshift Survey has been
provided by NSF grants AST-95-09298, AST-0071048, AST-0507428, and AST-0507483 as well as NASA LTSA grant NNG04GC89G.
Funding for the DEEP3 Galaxy Redshift Survey has been
provided by NSF grants AST-0808133, AST-0807630, and AST-0806732.
This work made use of images and/or data products provided by the NOAO Deep Wide-Field Survey \citep{Jannuzi99}, which is supported by the National Optical Astronomy Observatory (NOAO). NOAO is operated by AURA, Inc., under a cooperative agreement with the National Science Foundation.
Funding for SDSS-III has been provided by the Alfred P. Sloan Foundation, the Participating Institutions, the National Science Foundation, and the U.S. Department of Energy Office of Science. The SDSS-III web site is http://www.sdss3.org/.
SDSS-III is managed by the Astrophysical Research Consortium for the Participating Institutions of the SDSS-III Collaboration.
Based in part on data collected at Subaru Telescope, which is operated by the National Astronomical Observatory of Japan.
This work is based in part on observations made with the Spitzer Space Telescope, which is operated by the Jet Propulsion Laboratory, California Institute of Technology under a contract with NASA.
Some of the data used in this paper were obtained from the Mikulski Archive for Space Telescopes (MAST). STScI is operated by the Association of Universities for Research in Astronomy, Inc., under NASA contract NAS5-26555. Support for MAST for non-HST data is provided by the NASA Office of Space Science via grant NNX13AC07G and by other grants and contracts.
Based in part on observations made with ESO Telescopes at the La Silla or Paranal Observatories.
JA acknowledges support from a COFUND Junior Research Fellowship from the Institute of Advanced Study, Durham University, and ERC Advanced Grant FEEDBACK at the University of Cambridge. 
ALC acknowledges support from NSF CAREER award AST-1055081.
AG acknowledges the {\sc thales} project 383549 that is jointly funded by the European Union  and the  Greek Government  in  the framework  of the  programme ``Education and lifelong learning''. 
PGP-G acknowledges support from the Spanish Government through MINECO Grant AYA2012-31277.

\appendix
\section{Multiwavelength photometric datasets}\label{sec:photoappendix}

In this appendix we provide details of the photometric datasets that are used to identify counterparts and calculate photometric redshifts for X-ray sources in our \textit{Chandra} fields. 
The table below gives the name of all bandpasses and their effective wavelengths. The prioritization in our likilihood ratio counterpart matching (where the band with priority 1 is used first and then the subsequent priority bands are used to identify additional, robust counterparts) is given in the third column.
The final column gives the number of counterparts identified in each band and the corresponding fraction of the total X-ray sample.  

\small

\tablefirsthead{\hline {Band}  & {$\lambda_\mathrm{eff}$ (\AA)} &  {$LR$ priority} & {$N_\mathrm{cntrprt}$} &\\ \hline}
\tablehead{\hline {Band}  & {$\lambda_\mathrm{eff}$ (\AA)} &  {$LR$ priority} & {$N_\mathrm{cntrprt}$} &\\ \hline}
\begin{supertabular}{lcccc}
\hline\multicolumn{5}{c}{CDFS} \\
\hline
IRAC [3.6\micron] &      35416.6 &  1 &  593 (95.6\%) & \\
IRAC [4.5\micron] &      44826.2 & ... & ... & \\
IRAC [5.8\micron] &      56457.2 & ... & ... & \\
IRAC [8.0\micron] &      78264.8 & ... & ... & \\
VIMOS $U$ &       3711.4 & ... & ... & \\
VIMOS $R$ &       6414.0 &  2 &    7 (1.1\%) & \\
WFC3 F125W &      12425.8 & ... & ... & \\
WFC3 F160W &      15324.7 &  3 &    0 (0.0\%) & \\
ACS $B$ &       4297.6 & ... & ... & \\
ACS $V$ &       5840.7 & ... & ... & \\
ACS $I$ &       7668.3 &  4 &    0 (0.0\%) & \\
ACS $Z$ &       9021.7 & ... & ... & \\
ACS F814W &       7993.3 & ... & ... & \\
ESO WFI $U$ &       3414.4 & ... & ... & \\
ESO WFI $B$ &       4558.5 & ... & ... & \\
ESO WFI $V$ &       5355.9 & ... & ... & \\
ESO WFI $R$ &       6451.0 & ... & ... & \\
ESO WFI $I$ &       8568.1 &  5 &    5 (0.8\%) & \\
CTIO $z$ &       8966.4 & ... & ... & \\
HAWK-I $Y$ &      10194.5 & ... & ... & \\
HAWK-I $J$ &      12556.2 & ... & ... & \\
HAWK-I $K$ &      21423.2 & ... & ... & \\
HAWK-I 1.061\micron &      10619.2 & ... & ... & \\
SOFI $J$ &      12444.9 & ... & ... & \\
ISAAC $J$ &      12459.9 & ... & ... & \\
ISAAC $H$ &      16453.2 & ... & ... & \\
ISAAC $K$ &      21594.9 &  6 &    0 (  0.0\%) & \\
SOFI $K_\mathrm{S}$ &      21591.4 & ... & ... & \\
ISAAC $K_\mathrm{S}$ &      21594.9 & ... & ... & \\
GALEX NUV &       2271.1 & ... & ... & \\
GALEX FUV &       1528.1 & ... & ... & \\
COMBO17 $B$ &       4554.4 & ... & ... & \\
COMBO17 $V$ &       5358.1 & ... & ... & \\
COMBO17 $R$ &       6432.4 & ... & ... & \\
COMBO17 $I$ &       8523.6 & ... & ... & \\
COMBO17 F420M &       4177.7 & ... & ... & \\
COMBO17 F464M &       4616.8 & ... & ... & \\
COMBO17 F485M &       4858.5 & ... & ... & \\
COMBO17 F518M &       5188.1 & ... & ... & \\
COMBO17 F571M &       5716.1 & ... & ... & \\
COMBO17 F604M &       6044.3 & ... & ... & \\
COMBO17 F646M &       6450.9 & ... & ... & \\
COMBO17 F696M &       6958.9 & ... & ... & \\
COMBO17 F753M &       7530.5 & ... & ... & \\
COMBO17 F815M &       8157.2 & ... & ... & \\
COMBO17 F855M &       8556.4 & ... & ... & \\
COMBO17 F915M &       9140.0 & ... & ... & \\

\hline\multicolumn{5}{c}{CDFN} \\
\hline
IRAC [3.6\micron] &      35416.6 &  1 &  446 (90.1\%) & \\
IRAC [4.5\micron] &      44826.2 & ... & ... & \\
IRAC [5.8\micron] &      56457.2 & ... & ... & \\
IRAC [8.0\micron] &      78264.8 & ... & ... & \\
KPNO $U$ &       3566.7 & ... & ... & \\
SUBARU $B$ &       4366.4 & ... & ... & \\
SUBARU $V$ &       5429.6 & ... & ... & \\
SUBARU $R$ &       6486.4 & ... & ... & \\
SUBARU $I$ &       7952.0 &  2 &   13 (2.6\%) & \\
SUBARU $z'$ &       9146.7 & ... & ... & \\
ACS $B$ &       4297.6 & ... & ... & \\
ACS $V$ &       5840.7 & ... & ... & \\
ACS $I$ &       7668.3 &  3 &    0 (0.0\%) & \\
ACS $z$ &       9021.7 & ... & ... & \\
QUIRC $HK$ &      18264.0 & ... & ... & \\
CFHT $J$ &      12518.7 & ... & ... & \\
CAHA $J$ &      12029.7 & ... & ... & \\
CAHA $K$ &      22070.8 & ... & ... & \\
CFHT $K$ &      21530.8 &  4 &    5 (1.0\%) & \\
MOIRCS $K$ &      21354.4 & ... & ... & \\
GALEX FUV &       1528.1 & ... & ... & \\
GALEX NUV &       2271.1 & ... & ... & \\
WFC3 F105W &      10582.4 & ... & ... & \\
WFC3 F125W &      12425.8 & ... & ... & \\
WFC3 F160W &      15324.7 & ... & ... & \\

\hline\multicolumn{5}{c}{EGS} \\
\hline
IRAC [$3.6\mu$m] &      35416.6 &  1 &  961 (92.5\%) & \\
IRAC [$4.5\mu$m] &      44826.2 & ... & ... & \\
IRAC [$5.8\mu$m] &      56457.2 & ... & ... & \\
IRAC [8\micron] &      78264.8 & ... & ... & \\
Subaru $R_c$ &       6486.4 &  2 &   51 (4.9\%) & \\
ACS $V$ &       5796.7 & ... & ... & \\
ACS $I$ &       8234.0 &  3 &    0 (0.0\%) & \\
Subaru $K_S$ &      21354.4 &  4 &    1 (0.1\%) & \\
CFHTLS $u^*$ &       3805.6 & ... & ... & \\
CFHTLS $g'$ &       4833.7 & ... & ... & \\
CFHTLS $r'$ &       6234.1 & ... & ... & \\
CFHTLS $i'$ &       7659.1 &  5 &    1 (0.1\%) & \\
CFHTLS $z'$ &       8820.9 & ... & ... & \\
MMT $u'$ &       3604.1 & ... & ... & \\
MMT $g'$ &       4763.5 & ... & ... & \\
MMT $i'$ &       7770.5 &  6 &    1 (0.1\%) & \\
MMT $z'$ &       9030.9 & ... & ... & \\
DEEP $B$ &       4402.0 & ... & ... & \\
DEEP $R$ &       6595.1 &  7 &   10 (1.0\%) & \\
DEEP $I$ &       8118.7 & ... & ... & \\
Palomar $J$ &      12435.0 & ... & ... & \\
Palomar $K_S$ &      21353.0 &  8 &    4 (0.4\%) & \\
GALEX FUV &       1528.1 & ... & ... & \\
GALEX NUV &       2271.1 & ... & ... & \\
NICMOS F110W &      10622.4 & ... & ... & \\
NICMOS F160W &      15819.6 & ... & ... & \\
CAHA $J$ &      12029.7 & ... & ... & \\
WFC3 F125W &      12425.8 & ... & ... & \\
WFC3 F160W &      15324.7 & ... & ... & \\
NEWFIRM $J_1$ &      10441.4 & ... & ... & \\
NEWFIRM $J_2$ &      11930.0 & ... & ... & \\
NEWFIRM $J_3$ &      12764.0 & ... & ... & \\
NEWFIRM $H_1$ &      15585.4 & ... & ... & \\
NEWFIRM $H_2$ &      17048.8 & ... & ... & \\
NEWFIRM $K$ &      21643.9 & ... & ... & \\

\hline\multicolumn{5}{c}{COSMOS} \\
\hline
CFHT $u^*$ &       3805.0 & ... & ... & \\
Subaru $B_J$ &       4427.2 & ... & ... & \\
Subaru $V_J$ &       5454.9 & ... & ... & \\
Subaru $g^+$ &       4728.6 & ... & ... & \\
Subaru $r^+$ &       6249.1 & ... & ... & \\
Subaru $i^+$ &       7646.0 &  1 & 1338 ( 85.2\%) & \\
Subaru $z^+$ &       9011.0 & ... & ... & \\
IRAC [$3.6\mu$m] &      35416.6 &  2 &  200 ( 12.7\%) & \\
IRAC [$4.5\mu$m] &      44826.2 & ... & ... & \\
IRAC [$5.8\mu$m] &      56457.2 & ... & ... & \\
IRAC [$8\mu$m] &      78264.8 & ... & ... & \\
CFHT $i^*$ &       7582.6 & ... & ... & \\
Subaru IB427 &       4262.0 & ... & ... & \\
Subaru IB464 &       4633.7 & ... & ... & \\
Subaru IB505 &       5060.9 & ... & ... & \\
Subaru IB574 &       5762.8 & ... & ... & \\
Subaru IB709 &       7071.5 & ... & ... & \\
Subaru IB827 &       8242.5 & ... & ... & \\
CFHT $K_S$ &      21530.8 & ... & ... & \\
UKIRT $J$ &      12464.8 & ... & ... & \\
Subaru IB484 &       4847.6 & ... & ... & \\
Subaru IB527 &       5259.4 & ... & ... & \\
Subaru IB624 &       6230.9 & ... & ... & \\
Subaru IB679 &       6778.6 & ... & ... & \\
Subaru IB738 &       7359.5 & ... & ... & \\
Subaru IB767 &       7682.4 & ... & ... & \\
GALEX NUV &       2271.1 & ... & ... & \\
GALEX FUV &       1528.1 & ... & ... & \\

\hline\multicolumn{5}{c}{Bootes} \\
\hline
NDWFS $B_W$ &       4184.9 & ... & ... & \\
NDWFS $R$ &       6501.8 & ... & ... & \\
NDWFS $I$ &       8030.0 &  1 &  763 (91.7\%) & \\
SDSS u &       3546.0 & ... & ... & \\
SDSS g &       4669.6 & ... & ... & \\
SDSS r &       6156.2 & ... & ... & \\
SDSS i &       7471.6 &  2 &    6 (  0.7\%) & \\
SDSS z &       8917.4 & ... & ... & \\
NDWFS $K$ &      22078.4 &  3 &    9 (1.1\%) & \\
FLAMINGOS $J$ &      12442.5 & ... & ... & \\
FLAMINGOS $K_S$ &      21483.9 & ... & ... & \\
IRAC [$3.6\mu$m] &      35416.6 &  4 &   40 (4.8\%) & \\
IRAC [$4.5\mu$m] &      44826.2 & ... & ... & \\
IRAC [$5.8\mu$m] &      56457.2 & ... & ... & \\
IRAC [8\micron] &      78264.8 & ... & ... & \\
GALEX FUV &       1528.1 & ... & ... & \\
GALEX NUV &       2271.1 &  5 &    3 (0.4\%) & \\
\hline
\end{supertabular}

 \label{lastpage}

{\footnotesize

\begin{thebibliography}{149}
\expandafter\ifx\csname natexlab\endcsname\relax\def\natexlab#1{#1}\fi

\bibitem[{{Ahn} {et~al.}(2012){Ahn}, {Alexandroff}, {Allende Prieto},
  {Anderson}, {Anderton}, {Andrews}, {Aubourg}, {Bailey}, {Balbinot}, {Barnes},
  {et~al.}}]{Ahn12}
{Ahn}, C.~P., {et~al.} 2012, \apjs, 203, 21

\bibitem[{{Aird} {et~al.}(2012){Aird}, {Coil}, {Moustakas}, {Blanton},
  {Burles}, {Cool}, {Eisenstein}, {Smith}, {Wong}, \& {Zhu}}]{Aird12}
{Aird}, J., {et~al.} 2012, \apj, 746, 90

\bibitem[{{Aird} {et~al.}(2013{\natexlab{a}}){Aird}, {Coil}, {Moustakas},
  {Diamond-Stanic}, {Blanton}, {Cool}, {Eisenstein}, {Wong}, \& {Zhu}}]{Aird13}
---. 2013{\natexlab{a}}, \apj, 775, 41

\bibitem[{{Aird} {et~al.}(2013{\natexlab{b}}){Aird}, {Comastri}, {Brusa},
  {Cappelluti}, {Moretti}, {Vanzella}, {Volonteri}, {Alexander}, {Afonso},
  {Fiore}, {Georgantopoulos}, {Iwasawa}, {Merloni}, {Nandra}, {Salvaterra},
  {Salvato}, {Severgnini}, {Schawinski}, {Shankar}, {Vignali}, \&
  {Vito}}]{Aird13b}
---. 2013{\natexlab{b}}, An Athena+ Supporting Paper (arXiv:1306.2325)

\bibitem[{{Aird} {et~al.}(2008){Aird}, {Nandra}, {Georgakakis}, {Laird},
  {Steidel}, \& {Sharon}}]{Aird08}
{Aird}, J., {Nandra}, K., {Georgakakis}, A., {Laird}, E.~S., {Steidel}, C.~C.,
  \& {Sharon}, C. 2008, \mnras, 387, 883

\bibitem[{{Aird} {et~al.}(2010){Aird}, {Nandra}, {Laird}, {Georgakakis},
  {Ashby}, {Barmby}, {Coil}, {Huang}, {Koekemoer}, {Steidel}, \&
  {Willmer}}]{Aird10}
{Aird}, J., {et~al.} 2010, \mnras, 401, 2531

\bibitem[{{Ajello} {et~al.}(2008){Ajello}, {Greiner}, {Sato}, {Willis},
  {Kanbach}, {Strong}, {Diehl}, {Hasinger}, {Gehrels}, {Markwardt}, \&
  {Tueller}}]{Ajello08b}
{Ajello}, M., {et~al.} 2008, \apj, 689, 666

\bibitem[{{Akiyama} {et~al.}(2000){Akiyama}, {Ohta}, {Yamada}, {Kashikawa},
  {Yagi}, {Kawasaki}, {Sakano}, {Tsuru}, {Ueda}, {Takahashi}, {Lehmann},
  {Hasinger}, \& {Voges}}]{Akiyama00}
{Akiyama}, M., {et~al.} 2000, \apj, 532, 700

\bibitem[{{Akiyama} {et~al.}(2003){Akiyama}, {Ueda}, {Ohta}, {Takahashi}, \&
  {Yamada}}]{Akiyama03}
{Akiyama}, M., {Ueda}, Y., {Ohta}, K., {Takahashi}, T., \& {Yamada}, T. 2003,
  \apjs, 148, 275

\bibitem[{{Akylas} {et~al.}(2012){Akylas}, {Georgakakis}, {Georgantopoulos},
  {Brightman}, \& {Nandra}}]{Akylas12}
{Akylas}, A., {Georgakakis}, A., {Georgantopoulos}, I., {Brightman}, M., \&
  {Nandra}, K. 2012, \aap, 546, A98

\bibitem[{{Akylas} {et~al.}(2006){Akylas}, {Georgantopoulos}, {Georgakakis},
  {Kitsionas}, \& {Hatziminaoglou}}]{Akylas06}
{Akylas}, A., {Georgantopoulos}, I., {Georgakakis}, A., {Kitsionas}, S., \&
  {Hatziminaoglou}, E. 2006, \aap, 459, 693

\bibitem[{{Alexander} {et~al.}(2003){Alexander}, {Bauer}, {Brandt},
  {Schneider}, {Hornschemeier}, {Vignali}, {Barger}, {Broos}, {Cowie},
  {Garmire}, {Townsley}, {Bautz}, {Chartas}, \& {Sargent}}]{Alexander03}
{Alexander}, D.~M., {et~al.} 2003, \aj, 126, 539

\bibitem[{{Appenzeller} {et~al.}(1998){Appenzeller}, {Thiering}, {Zickgraf},
  {Krautter}, {Voges}, {Chavarria}, {Kneer}, {Mujica}, {Pakull}, {Rosso},
  {Ruzicka}, {Serrano}, \& {Ziegler}}]{Appenzeller98}
{Appenzeller}, I., {et~al.} 1998, \apjs, 117, 319

\bibitem[{{Ashby} {et~al.}(2009){Ashby}, {Stern}, {Brodwin}, {Griffith},
  {Eisenhardt}, {Koz{\l}owski}, {Kochanek}, {Bock}, {Borys}, {Brand}, {Brown},
  {Cool}, {Cooray}, {Croft}, {Dey}, {Eisenstein}, {Gonzalez}, {Gorjian},
  {Grogin}, {Ivison}, {Jacob}, {Jannuzi}, {Mainzer}, {Moustakas},
  {R\"{o}ttgering}, {Seymour}, {Smith}, {Stanford}, {Stauffer}, {Sullivan},
  {van Breugel}, {Willner}, \& {Wright}}]{Ashby09}
{Ashby}, M.~L.~N., {et~al.} 2009, \apj, 701, 428

\bibitem[{{Assef} {et~al.}(2011){Assef}, {Kochanek}, {Ashby}, {Brodwin},
  {Brown}, {Cool}, {Forman}, {Gonzalez}, {Hickox}, {Jannuzi}, {Jones}, {Le
  Floc'h}, {Moustakas}, {Murray}, \& {Stern}}]{Assef11}
{Assef}, R.~J., {et~al.} 2011, \apj, 728, 56

\bibitem[{{Ballantyne}(2014)}]{Ballantyne14}
{Ballantyne}, D.~R. 2014, \mnras, 437, 2845

\bibitem[{{Ballantyne} {et~al.}(2011){Ballantyne}, {Draper}, {Madsen}, {Rigby},
  \& {Treister}}]{Ballantyne11}
{Ballantyne}, D.~R., {Draper}, A.~R., {Madsen}, K.~K., {Rigby}, J.~R., \&
  {Treister}, E. 2011, \apj, 736, 56

\bibitem[{{Banerji} {et~al.}(2015){Banerji}, {Alaghband-Zadeh}, {Hewett}, \&
  {McMahon}}]{Banerji15}
{Banerji}, M., {Alaghband-Zadeh}, S., {Hewett}, P.~C., \& {McMahon}, R.~G.
  2015, \mnras, 447, 3368

\bibitem[{{Barcons} {et~al.}(2000){Barcons}, {Mateos}, \&
  {Ceballos}}]{Barcons00}
{Barcons}, X., {Mateos}, S., \& {Ceballos}, M.~T. 2000, \mnras, 316, L13

\bibitem[{{Barger} {et~al.}(2003){Barger}, {Cowie}, {Capak}, {Alexander},
  {Bauer}, {Fernandez}, {Brandt}, {Garmire}, \& {Hornschemeier}}]{Barger03c}
{Barger}, A.~J., {et~al.} 2003, \aj, 126, 632

\bibitem[{{Barger} {et~al.}(2005){Barger}, {Cowie}, {Mushotzky}, {Yang},
  {Wang}, {Steffen}, \& {Capak}}]{Barger05}
{Barger}, A.~J., {Cowie}, L.~L., {Mushotzky}, R.~F., {Yang}, Y., {Wang}, W.-H.,
  {Steffen}, A.~T., \& {Capak}, P. 2005, \aj, 129, 578

\bibitem[{{Barger} {et~al.}(2008){Barger}, {Cowie}, \& {Wang}}]{Barger08}
{Barger}, A.~J., {Cowie}, L.~L., \& {Wang}, W.-H. 2008, \apj, 689, 687

\bibitem[{{Barro} {et~al.}(2011{\natexlab{a}}){Barro},
  {P\'{e}rez-Gonz\'{a}lez}, {Gallego}, {Ashby}, {Kajisawa}, {Miyazaki},
  {Villar}, {Yamada}, \& {Zamorano}}]{Barro11}
{Barro}, G., {et~al.} 2011{\natexlab{a}}, \apjs, 193, 13

\bibitem[{{Barro} {et~al.}(2011{\natexlab{b}}){Barro},
  {P\'{e}rez-Gonz\'{a}lez}, {Gallego}, {Ashby}, {Kajisawa}, {Miyazaki},
  {Villar}, {Yamada}, \& {Zamorano}}]{Barro11b}
---. 2011{\natexlab{b}}, \apjs, 193, 30

\bibitem[{{Bongiorno} {et~al.}(2012){Bongiorno}, {Merloni}, {Brusa},
  {Magnelli}, {Salvato}, {Mignoli}, {Zamorani}, {Fiore}, {Rosario}, {Mainieri},
  {Hao}, {Comastri}, {Vignali}, {Balestra}, {Bardelli}, {Berta}, {Civano},
  {Kampczyk}, {Le Floc'h}, {Lusso}, {Lutz}, {Pozzetti}, {Pozzi}, {Riguccini},
  {Shankar}, \& {Silverman}}]{Bongiorno12}
{Bongiorno}, A., {et~al.} 2012, \mnras, 427, 3103

\bibitem[{{Brammer} {et~al.}(2008){Brammer}, {van Dokkum}, \&
  {Coppi}}]{Brammer08}
{Brammer}, G.~B., {van Dokkum}, P.~G., \& {Coppi}, P. 2008, \apj, 686, 1503

\bibitem[{{Brand} {et~al.}(2006){Brand}, {Brown}, {Dey}, {Jannuzi}, {Kochanek},
  {Kenter}, {Fabricant}, {Fazio}, {Forman}, {Green}, {Jones}, {McNamara},
  {Murray}, {Najita}, {Rieke}, {Shields}, \& {Vikhlinin}}]{Brand06}
{Brand}, K., {et~al.} 2006, \apj, 641, 140

\bibitem[{{Brandt} \& {Alexander}(2015)}]{Brandt15}
{Brandt}, W.~N., \& {Alexander}, D.~M. 2015, \aapr, 23, 1

\bibitem[{{Brightman} \& {Nandra}(2011{\natexlab{a}})}]{Brightman11}
{Brightman}, M., \& {Nandra}, K. 2011{\natexlab{a}}, \mnras, 413, 1206

\bibitem[{{Brightman} \& {Nandra}(2011{\natexlab{b}})}]{Brightman11b}
---. 2011{\natexlab{b}}, \mnras, 414, 3084

\bibitem[{{Brightman} {et~al.}(2014){Brightman}, {Nandra}, {Salvato}, {Hsu},
  {Aird}, \& {Rangel}}]{Brightman14}
{Brightman}, M., {Nandra}, K., {Salvato}, M., {Hsu}, L.-T., {Aird}, J., \&
  {Rangel}, C. 2014, \mnras, 443, 1999

\bibitem[{{Brightman} \& {Ueda}(2012)}]{Brightman12b}
{Brightman}, M., \& {Ueda}, Y. 2012, \mnras, 423, 702

\bibitem[{{Brusa} {et~al.}(2010){Brusa}, {Civano}, {Comastri}, {Miyaji},
  {Salvato}, {Zamorani}, {Cappelluti}, {Fiore}, {Hasinger}, {Mainieri},
  {Merloni}, {Bongiorno}, {Capak}, {Elvis}, {Gilli}, {Hao}, {Jahnke},
  {Koekemoer}, {Ilbert}, {Le Floc'h}, {Lusso}, {Mignoli}, {Schinnerer},
  {Silverman}, {Treister}, {Trump}, {Vignali}, {Zamojski}, {Aldcroft},
  {Aussel}, {Bardelli}, {Bolzonella}, {Cappi}, {Caputi}, {Contini},
  {Finoguenov}, {Fruscione}, {Garilli}, {Impey}, {Iovino}, {Iwasawa},
  {Kampczyk}, {Kartaltepe}, {Kneib}, {Knobel}, {Kovac}, {Lamareille},
  {Leborgne}, {Le Brun}, {Le Fevre}, {Lilly}, {Maier}, {McCracken}, {Pello},
  {Peng}, {Perez-Montero}, {de Ravel}, {Sanders}, {Scodeggio}, {Scoville},
  {Tanaka}, {Taniguchi}, {Tasca}, {de la Torre}, {Tresse}, {Vergani}, \&
  {Zucca}}]{Brusa10}
{Brusa}, M., {et~al.} 2010, \apj, 716, 348

\bibitem[{{Brusa} {et~al.}(2007){Brusa}, {Zamorani}, {Comastri}, {Hasinger},
  {Cappelluti}, {Civano}, {Finoguenov}, {Mainieri}, {Salvato}, {Vignali},
  {Elvis}, {Fiore}, {Gilli}, {Impey}, {Lilly}, {Mignoli}, {Silverman}, {Trump},
  {Urry}, {Bender}, {Capak}, {Huchra}, {Kneib}, {Koekemoer}, {Leauthaud},
  {Lehmann}, {Massey}, {Matute}, {McCarthy}, {McCracken}, {Rhodes}, {Scoville},
  {Taniguchi}, \& {Thompson}}]{Brusa07}
---. 2007, \apjs, 172, 353

\bibitem[{{Buchner} {et~al.}(2015){Buchner}, {Georgakakis}, {Nandra},
  {Brightman}, {Menzel}, {Liu}, {Hsu}, {Salvato}, {Rangel}, {Aird}, {Merloni},
  \& {Ross}}]{Buchner15}
{Buchner}, J., {et~al.} 2015, \apj, in press

\bibitem[{{Buchner} {et~al.}(2014){Buchner}, {Georgakakis}, {Nandra}, {Hsu},
  {Rangel}, {Brightman}, {Merloni}, {Salvato}, {Donley}, \&
  {Kocevski}}]{Buchner14}
---. 2014, \aap, 564, A125

\bibitem[{{Burlon} {et~al.}(2011){Burlon}, {Ajello}, {Greiner}, {Comastri},
  {Merloni}, \& {Gehrels}}]{Burlon11}
{Burlon}, D., {Ajello}, M., {Greiner}, J., {Comastri}, A., {Merloni}, A., \&
  {Gehrels}, N. 2011, \apj, 728, 58

\bibitem[{{Calzetti} {et~al.}(2000){Calzetti}, {Armus}, {Bohlin}, {Kinney},
  {Koornneef}, \& {Storchi-Bergmann}}]{Calzetti00}
{Calzetti}, D., {Armus}, L., {Bohlin}, R.~C., {Kinney}, A.~L., {Koornneef}, J.,
  \& {Storchi-Bergmann}, T. 2000, \apj, 533, 682

\bibitem[{{Capak} {et~al.}(2007){Capak}, {Aussel}, {Ajiki}, {McCracken},
  {Mobasher}, {Scoville}, {Shopbell}, {Taniguchi}, {Thompson}, {Tribiano},
  {Sasaki}, {Blain}, {Brusa}, {Carilli}, {Comastri}, {Carollo}, {Cassata},
  {Colbert}, {Ellis}, {Elvis}, {Giavalisco}, {Green}, {Guzzo}, {Hasinger},
  {Ilbert}, {Impey}, {Jahnke}, {Kartaltepe}, {Kneib}, {Koda}, {Koekemoer},
  {Komiyama}, {Leauthaud}, {Le Fevre}, {Lilly}, {Liu}, {Massey}, {Miyazaki},
  {Murayama}, {Nagao}, {Peacock}, {Pickles}, {Porciani}, {Renzini}, {Rhodes},
  {Rich}, {Salvato}, {Sanders}, {Scarlata}, {Schiminovich}, {Schinnerer},
  {Scodeggio}, {Sheth}, {Shioya}, {Tasca}, {Taylor}, {Yan}, \&
  {Zamorani}}]{Capak07}
{Capak}, P., {et~al.} 2007, \apjs, 172, 99

\bibitem[{{Cardamone} {et~al.}(2010{\natexlab{a}}){Cardamone}, {Urry},
  {Schawinski}, {Treister}, {Brammer}, \& {Gawiser}}]{Cardamone10b}
{Cardamone}, C.~N., {Urry}, C.~M., {Schawinski}, K., {Treister}, E., {Brammer},
  G., \& {Gawiser}, E. 2010{\natexlab{a}}, \apjl, 721, L38

\bibitem[{{Cardamone} {et~al.}(2010{\natexlab{b}}){Cardamone}, {van Dokkum},
  {Urry}, {Taniguchi}, {Gawiser}, {Brammer}, {Taylor}, {Damen}, {Treister},
  {Cobb}, {Bond}, {Schawinski}, {Lira}, {Murayama}, {Saito}, \&
  {Sumikawa}}]{Cardamone10}
{Cardamone}, C.~N., {et~al.} 2010{\natexlab{b}}, \apjs, 189, 270

\bibitem[{{Ciliegi} {et~al.}(2003){Ciliegi}, {Zamorani}, {Hasinger}, {Lehmann},
  {Szokoly}, \& {Wilson}}]{Ciliegi03}
{Ciliegi}, P., {Zamorani}, G., {Hasinger}, G., {Lehmann}, I., {Szokoly}, G., \&
  {Wilson}, G. 2003, \aap, 398, 901

\bibitem[{{Civano} {et~al.}(2011){Civano}, {Brusa}, {Comastri}, {Elvis},
  {Salvato}, {Zamorani}, {Capak}, {Fiore}, {Gilli}, {Hao}, {Ikeda}, {Kakazu},
  {Kartaltepe}, {Masters}, {Miyaji}, {Mignoli}, {Puccetti}, {Shankar},
  {Silverman}, {Vignali}, {Zezas}, \& {Koekemoer}}]{Civano11}
{Civano}, F., {et~al.} 2011, \apj, 741, 91

\bibitem[{{Civano} {et~al.}(2012){Civano}, {Elvis}, {Brusa}, {Comastri},
  {Salvato}, {Zamorani}, {Aldcroft}, {Bongiorno}, {Capak}, {Cappelluti},
  {Cisternas}, {Fiore}, {Fruscione}, {Hao}, {Kartaltepe}, {Koekemoer}, {Gilli},
  {Impey}, {Lanzuisi}, {Lusso}, {Mainieri}, {Miyaji}, {Lilly}, {Masters},
  {Puccetti}, {Schawinski}, {Scoville}, {Silverman}, {Trump}, {Urry},
  {Vignali}, \& {Wright}}]{Civano12}
---. 2012, \apjs, 201, 30

\bibitem[{{Coil} {et~al.}(2011){Coil}, {Blanton}, {Burles}, {Cool},
  {Eisenstein}, {Moustakas}, {Wong}, {Zhu}, {Aird}, {Bernstein}, {Bolton}, \&
  {Hogg}}]{Coil11}
{Coil}, A.~L., {et~al.} 2011, \apj, 741, 8

\bibitem[{{Comastri} {et~al.}(2015){Comastri}, {Gilli}, {Marconi}, {Risaliti},
  \& {Salvati}}]{Comastri15}
{Comastri}, A., {Gilli}, R., {Marconi}, A., {Risaliti}, G., \& {Salvati}, M.
  2015, \aap, 574, L10

\bibitem[{{Comastri} {et~al.}(2011){Comastri}, {Ranalli}, {Iwasawa}, {Vignali},
  {Gilli}, {Georgantopoulos}, {Barcons}, {Brandt}, {Brunner}, {Brusa},
  {Cappelluti}, {Carrera}, {Civano}, {Fiore}, {Hasinger}, {Mainieri},
  {Merloni}, {Nicastro}, {Paolillo}, {Puccetti}, {Rosati}, {Silverman},
  {Tozzi}, {Zamorani}, {Balestra}, {Bauer}, {Luo}, \& {Xue}}]{Comastri11}
{Comastri}, A., {et~al.} 2011, \aap, 526, L9

\bibitem[{{Cooper} {et~al.}(2011){Cooper}, {Aird}, {Coil}, {Davis}, {Faber},
  {Juneau}, {Lotz}, {Nandra}, {Newman}, {Willmer}, \& {Yan}}]{Cooper11}
{Cooper}, M.~C., {et~al.} 2011, \apjs, 193, 14

\bibitem[{{Cooper} {et~al.}(2012){Cooper}, {Yan}, {Dickinson}, {Juneau},
  {Lotz}, {Newman}, {Papovich}, {Salim}, {Walth}, {Weiner}, \&
  {Willmer}}]{Cooper12b}
---. 2012, \mnras, 425, 2116

\bibitem[{{Cowie} {et~al.}(2004){Cowie}, {Barger}, {Hu}, {Capak}, \&
  {Songaila}}]{Cowie04}
{Cowie}, L.~L., {Barger}, A.~J., {Hu}, E.~M., {Capak}, P., \& {Songaila}, A.
  2004, \aj, 127, 3137

\bibitem[{{Del Moro} {et~al.}(2014){Del Moro}, {Mullaney}, {Alexander},
  {Comastri}, {Bauer}, {Treister}, {Stern}, {Civano}, {Ranalli}, {Vignali},
  {Aird}, {Ballantyne}, {Balokovi\'{c}}, {Boggs}, {Brandt}, {Christensen},
  {Craig}, {Gandhi}, {Gilli}, {Hailey}, {Harrison}, {Hickox}, {LaMassa},
  {Lansbury}, {Luo}, {Puccetti}, {Urry}, \& {Zhang}}]{DelMoro14}
{Del Moro}, A., {et~al.} 2014, \apj, 786, 16

\bibitem[{{Delvecchio} {et~al.}(2014){Delvecchio}, {Gruppioni}, {Pozzi},
  {Berta}, {Zamorani}, {Cimatti}, {Lutz}, {Scott}, {Vignali}, {Cresci},
  {Feltre}, {Cooray}, {Vaccari}, {Fritz}, {Le Floc'h}, {Magnelli}, {Popesso},
  {Oliver}, {Bock}, {Carollo}, {Contini}, {Le F\'{e}vre}, {Lilly}, {Mainieri},
  {Renzini}, \& {Scodeggio}}]{Delvecchio14}
{Delvecchio}, I., {et~al.} 2014, \mnras, 439, 2736

\bibitem[{{Dickey} \& {Lockman}(1990)}]{Dickey90}
{Dickey}, J.~M., \& {Lockman}, F.~J. 1990, \araa, 28, 215

\bibitem[{{Donley} {et~al.}(2008){Donley}, {Rieke}, {P\'{e}rez-Gonz\'{a}lez},
  \& {Barro}}]{Donley08}
{Donley}, J.~L., {Rieke}, G.~H., {P\'{e}rez-Gonz\'{a}lez}, P.~G., \& {Barro},
  G. 2008, \apj, 687, 111

\bibitem[{{Draper} \& {Ballantyne}(2009)}]{Draper09}
{Draper}, A.~R., \& {Ballantyne}, D.~R. 2009, \apj, 707, 778

\bibitem[{{Dwelly} \& {Page}(2006)}]{Dwelly06}
{Dwelly}, T., \& {Page}, M.~J. 2006, \mnras, 372, 1755

\bibitem[{{Ebrero} {et~al.}(2009){Ebrero}, {Carrera}, {Page}, {Silverman},
  {Barcons}, {Ceballos}, {Corral}, {Della Ceca}, \& {Watson}}]{Ebrero09}
{Ebrero}, J., {et~al.} 2009, \aap, 493, 55

\bibitem[{{Elston} {et~al.}(2006){Elston}, {Gonzalez}, {McKenzie}, {Brodwin},
  {Brown}, {Cardona}, {Dey}, {Dickinson}, {Eisenhardt}, {Jannuzi}, {Lin},
  {Mohr}, {Raines}, {Stanford}, \& {Stern}}]{Elston06}
{Elston}, R.~J., {et~al.} 2006, \apj, 639, 816

\bibitem[{{Elvis} {et~al.}(2009){Elvis}, {Civano}, {Vignali}, {Puccetti},
  {Fiore}, {Cappelluti}, {Aldcroft}, {Fruscione}, {Zamorani}, {Comastri},
  {Brusa}, {Gilli}, {Miyaji}, {Damiani}, {Koekemoer}, {Finoguenov}, {Brunner},
  {Urry}, {Silverman}, {Mainieri}, {Hasinger}, {Griffiths}, {Carollo}, {Hao},
  {Guzzo}, {Blain}, {Calzetti}, {Carilli}, {Capak}, {Ettori}, {Fabbiano},
  {Impey}, {Lilly}, {Mobasher}, {Rich}, {Salvato}, {Sanders}, {Schinnerer},
  {Scoville}, {Shopbell}, {Taylor}, {Taniguchi}, \& {Volonteri}}]{Elvis09}
{Elvis}, M., {et~al.} 2009, \apjs, 184, 158

\bibitem[{{Feroz} {et~al.}(2009){Feroz}, {Hobson}, \& {Bridges}}]{Feroz09}
{Feroz}, F., {Hobson}, M.~P., \& {Bridges}, M. 2009, \mnras, 398, 1601

\bibitem[{{Fiore} {et~al.}(2012){Fiore}, {Puccetti}, {Grazian}, {Menci},
  {Shankar}, {Santini}, {Piconcelli}, {Koekemoer}, {Fontana}, {Boutsia},
  {Castellano}, {Lamastra}, {Malacaria}, {Feruglio}, {Mathur}, {Miller}, \&
  {Pannella}}]{Fiore12}
{Fiore}, F., {et~al.} 2012, \aap, 537, A16

\bibitem[{{Fischer} {et~al.}(1998){Fischer}, {Hasinger}, {Schwope}, {Brunner},
  {Boller}, {Tr\"{u}mper}, {Voges}, \& {Neizvestnyj}}]{Fischer98}
{Fischer}, J.-U., {Hasinger}, G., {Schwope}, A.~D., {Brunner}, H., {Boller},
  T., {Tr\"{u}mper}, J., {Voges}, W., \& {Neizvestnyj}, S. 1998, Astronomische
  Nachrichten, 319, 347

\bibitem[{{Gendreau} {et~al.}(1995){Gendreau}, {Mushotzky}, {Fabian}, {Holt},
  {Kii}, {Serlemitsos}, {Ogasaka}, {Tanaka}, {Bautz}, {Fukazawa}, {Ishisaki},
  {Kohmura}, {Makishima}, {Tashiro}, {Tsusaka}, {Kunieda}, {Ricker}, \&
  {Vanderspek}}]{Gendreau95}
{Gendreau}, K.~C., {et~al.} 1995, \pasj, 47, L5

\bibitem[{{Georgakakis} {et~al.}(2008){Georgakakis}, {Nandra}, {Laird}, {Aird},
  \& {Trichas}}]{Georgakakis08}
{Georgakakis}, A., {Nandra}, K., {Laird}, E.~S., {Aird}, J., \& {Trichas}, M.
  2008, \mnras, 388, 1205

\bibitem[{{Georgakakis} {et~al.}(2007){Georgakakis}, {Rowan-Robinson},
  {Babbedge}, \& {Georgantopoulos}}]{Georgakakis07}
{Georgakakis}, A., {Rowan-Robinson}, M., {Babbedge}, T.~S.~R., \&
  {Georgantopoulos}, I. 2007, \mnras, 377, 203

\bibitem[{{Georgakakis} {et~al.}(2006){Georgakakis}, {Chavushyan}, {Plionis},
  {Georgantopoulos}, {Koulouridis}, {Leonidaki}, \& {Mercado}}]{Georgakakis06b}
{Georgakakis}, A.~E., {Chavushyan}, V., {Plionis}, M., {Georgantopoulos}, I.,
  {Koulouridis}, E., {Leonidaki}, I., \& {Mercado}, A. 2006, \mnras, 367, 1017

\bibitem[{{Georgantopoulos} {et~al.}(2013){Georgantopoulos}, {Comastri},
  {Vignali}, {Ranalli}, {Rovilos}, {Iwasawa}, {Gilli}, {Cappelluti}, {Carrera},
  {Fritz}, {Brusa}, {Elbaz}, {Mullaney}, {Castello-Mor}, {Barcons}, {Tozzi},
  {Balestra}, \& {Falocco}}]{Georgantopoulos13}
{Georgantopoulos}, I., {et~al.} 2013, \aap, 555, A43

\bibitem[{{Gilli} {et~al.}(2007){Gilli}, {Comastri}, \& {Hasinger}}]{Gilli07}
{Gilli}, R., {Comastri}, A., \& {Hasinger}, G. 2007, \aap, 463, 79

\bibitem[{{Goulding} {et~al.}(2012){Goulding}, {Forman}, {Hickox}, {Jones},
  {Kraft}, {Murray}, {Vikhlinin}, {Coil}, {Cooper}, {Davis}, \&
  {Newman}}]{Goulding12}
{Goulding}, A.~D., {et~al.} 2012, \apjs, 202, 6

\bibitem[{{Grazian} {et~al.}(2006){Grazian}, {Fontana}, {de Santis}, {Nonino},
  {Salimbeni}, {Giallongo}, {Cristiani}, {Gallozzi}, \& {Vanzella}}]{Grazian06}
{Grazian}, A., {et~al.} 2006, \aap, 449, 951

\bibitem[{{Hasinger}(2008)}]{Hasinger08}
{Hasinger}, G. 2008, \aap, 490, 905

\bibitem[{{Hasinger} {et~al.}(2005){Hasinger}, {Miyaji}, \&
  {Schmidt}}]{Hasinger05}
{Hasinger}, G., {Miyaji}, T., \& {Schmidt}, M. 2005, \aap, 441, 417

\bibitem[{{Hickox} \& {Markevitch}(2007)}]{Hickox07}
{Hickox}, R.~C., \& {Markevitch}, M. 2007, \apjl, 661, L117

\bibitem[{{Hiroi} {et~al.}(2012){Hiroi}, {Ueda}, {Akiyama}, \&
  {Watson}}]{Hiroi12}
{Hiroi}, K., {Ueda}, Y., {Akiyama}, M., \& {Watson}, M.~G. 2012, \apj, 758, 49

\bibitem[{{Hopkins} {et~al.}(2006){Hopkins}, {Hernquist}, {Cox}, {Robertson},
  {Di Matteo}, \& {Springel}}]{Hopkins06}
{Hopkins}, P.~F., {Hernquist}, L., {Cox}, T.~J., {Robertson}, B., {Di Matteo},
  T., \& {Springel}, V. 2006, \apj, 639, 700

\bibitem[{{Hopkins} {et~al.}(2007){Hopkins}, {Richards}, \&
  {Hernquist}}]{Hopkins07b}
{Hopkins}, P.~F., {Richards}, G.~T., \& {Hernquist}, L. 2007, \apj, 654, 731

\bibitem[{{Hsu} {et~al.}(2014){Hsu}, {Salvato}, {Nandra}, {Brusa}, {Bender},
  {Buchner}, {Donley}, {Kocevski}, {Guo}, {Hathi}, {Rangel}, {Willner},
  {Brightman}, {Georgakakis}, {Budav\'{a}ri}, {Szalay}, {Ashby}, {Barro},
  {Dahlen}, {Faber}, {Ferguson}, {Galametz}, {Grazian}, {Grogin}, {Huang},
  {Koekemoer}, {Lucas}, {McGrath}, {Mobasher}, {Peth}, {Rosario}, \&
  {Trump}}]{Hsu14}
{Hsu}, L.-T., {et~al.} 2014, \apj, 796, 60

\bibitem[{{Ichikawa} {et~al.}(2012){Ichikawa}, {Ueda}, {Terashima}, {Oyabu},
  {Gandhi}, {Matsuta}, \& {Nakagawa}}]{Ichikawa12}
{Ichikawa}, K., {Ueda}, Y., {Terashima}, Y., {Oyabu}, S., {Gandhi}, P.,
  {Matsuta}, K., \& {Nakagawa}, T. 2012, \apj, 754, 45

\bibitem[{{Ilbert} {et~al.}(2009){Ilbert}, {Capak}, {Salvato}, {Aussel},
  {McCracken}, {Sanders}, {Scoville}, {Kartaltepe}, {Arnouts}, {Le Floc'h},
  {Mobasher}, {Taniguchi}, {Lamareille}, {Leauthaud}, {Sasaki}, {Thompson},
  {Zamojski}, {Zamorani}, {Bardelli}, {Bolzonella}, {Bongiorno}, {Brusa},
  {Caputi}, {Carollo}, {Contini}, {Cook}, {Coppa}, {Cucciati}, {de la Torre},
  {de Ravel}, {Franzetti}, {Garilli}, {Hasinger}, {Iovino}, {Kampczyk},
  {Kneib}, {Knobel}, {Kovac}, {Le Borgne}, {Le Brun}, {F\`{e}vre}, {Lilly},
  {Looper}, {Maier}, {Mainieri}, {Mellier}, {Mignoli}, {Murayama}, {Pell\`{o}},
  {Peng}, {P\'{e}rez-Montero}, {Renzini}, {Ricciardelli}, {Schiminovich},
  {Scodeggio}, {Shioya}, {Silverman}, {Surace}, {Tanaka}, {Tasca}, {Tresse},
  {Vergani}, \& {Zucca}}]{Ilbert09}
{Ilbert}, O., {et~al.} 2009, \apj, 690, 1236

\bibitem[{{Jannuzi} \& {Dey}(1999)}]{Jannuzi99}
{Jannuzi}, B.~T., \& {Dey}, A. 1999, in {Astronomical Society of the Pacific
  Conference Series}, Vol. 191, {Photometric Redshifts and the Detection of
  High Redshift Galaxies}, ed. R.~{Weymann}, L.~{Storrie-Lombardi},
  M.~{Sawicki}, \& R.~{Brunner}, 111

\bibitem[{Jeffreys(1961)}]{Jeffreys61}
Jeffreys, H. 1961, {Theory of probability} (Clarendon Press)

\bibitem[{Kass \& Raftery(1995)}]{Kass95}
Kass, R.~E., \& Raftery, A.~E. 1995, Journal of the American Statistical
  Association, 90, 773

\bibitem[{{Kochanek} {et~al.}(2012){Kochanek}, {Eisenstein}, {Cool},
  {Caldwell}, {Assef}, {Jannuzi}, {Jones}, {Murray}, {Forman}, {Dey}, {Brown},
  {Eisenhardt}, {Gonzalez}, {Green}, \& {Stern}}]{Kochanek12}
{Kochanek}, C.~S., {et~al.} 2012, \apjs, 200, 8

\bibitem[{{La Franca} {et~al.}(2005){La Franca}, {Fiore}, {Comastri}, {Perola},
  {Sacchi}, {Brusa}, {Cocchia}, {Feruglio}, {Matt}, {Vignali}, {Carangelo},
  {Ciliegi}, {Lamastra}, {Maiolino}, {Mignoli}, {Molendi}, \&
  {Puccetti}}]{LaFranca05}
{La Franca}, F., {et~al.} 2005, \apj, 635, 864

\bibitem[{{Laird} {et~al.}(2009){Laird}, {Nandra}, {Georgakakis}, {Aird},
  {Barmby}, {Conselice}, {Coil}, {Davis}, {Faber}, {Fazio}, {Guhathakurta},
  {Koo}, {Sarajedini}, \& {Willmer}}]{Laird09}
{Laird}, E.~S., {et~al.} 2009, \apjs, 180, 102

\bibitem[{{Laird} {et~al.}(2006){Laird}, {Nandra}, {Hobbs}, \&
  {Steidel}}]{Laird06}
{Laird}, E.~S., {Nandra}, K., {Hobbs}, A., \& {Steidel}, C.~C. 2006, \mnras,
  373, 217

\bibitem[{{Lasker} {et~al.}(2008){Lasker}, {Lattanzi}, {McLean}, {Bucciarelli},
  {Drimmel}, {Garcia}, {Greene}, {Guglielmetti}, {Hanley}, {Hawkins},
  {Laidler}, {Loomis}, {Meakes}, {Mignani}, {Morbidelli}, {Morrison},
  {Pannunzio}, {Rosenberg}, {Sarasso}, {Smart}, {Spagna}, {Sturch},
  {Volpicelli}, {White}, {Wolfe}, \& {Zacchei}}]{Lasker08}
{Lasker}, B.~M., {et~al.} 2008, \aj, 136, 735

\bibitem[{{Lehmer} {et~al.}(2005){Lehmer}, {Brandt}, {Alexander}, {Bauer},
  {Schneider}, {Tozzi}, {Bergeron}, {Garmire}, {Giacconi}, {Gilli}, {Hasinger},
  {Hornschemeier}, {Koekemoer}, {Mainieri}, {Miyaji}, {Nonino}, {Rosati},
  {Silverman}, {Szokoly}, \& {Vignali}}]{Lehmer05}
{Lehmer}, B.~D., {et~al.} 2005, \apjs, 161, 21

\bibitem[{{Lehmer} {et~al.}(2012){Lehmer}, {Xue}, {Brandt}, {Alexander},
  {Bauer}, {Brusa}, {Comastri}, {Gilli}, {Hornschemeier}, {Luo}, {Paolillo},
  {Ptak}, {Shemmer}, {Schneider}, {Tozzi}, \& {Vignali}}]{Lehmer12}
---. 2012, \apj, 752, 46

\bibitem[{{Lilly} {et~al.}(2009){Lilly}, {Le Brun}, {Maier}, {Mainieri},
  {Mignoli}, {Scodeggio}, {Zamorani}, {Carollo}, {Contini}, {Kneib}, {Le
  F\`{e}vre}, {Renzini}, {Bardelli}, {Bolzonella}, {Bongiorno}, {Caputi},
  {Coppa}, {Cucciati}, {de la Torre}, {de Ravel}, {Franzetti}, {Garilli},
  {Iovino}, {Kampczyk}, {Kovac}, {Knobel}, {Lamareille}, {Le Borgne}, {Pello},
  {Peng}, {P\'{e}rez-Montero}, {Ricciardelli}, {Silverman}, {Tanaka}, {Tasca},
  {Tresse}, {Vergani}, {Zucca}, {Ilbert}, {Salvato}, {Oesch}, {Abbas},
  {Bottini}, {Capak}, {Cappi}, {Cassata}, {Cimatti}, {Elvis}, {Fumana},
  {Guzzo}, {Hasinger}, {Koekemoer}, {Leauthaud}, {Maccagni}, {Marinoni},
  {McCracken}, {Memeo}, {Meneux}, {Porciani}, {Pozzetti}, {Sanders},
  {Scaramella}, {Scarlata}, {Scoville}, {Shopbell}, \& {Taniguchi}}]{Lilly09}
{Lilly}, S.~J., {et~al.} 2009, \apjs, 184, 218

\bibitem[{{Loredo}(2004)}]{Loredo04}
{Loredo}, T.~J. 2004, in {American Institute of Physics Conference Series},
  Vol. 735, {American Institute of Physics Conference Series}, ed.
  R.~{Fischer}, R.~{Preuss}, \& U.~V. {Toussaint}, 195--206

\bibitem[{{Luo} {et~al.}(2010){Luo}, {Brandt}, {Xue}, {Brusa}, {Alexander},
  {Bauer}, {Comastri}, {Koekemoer}, {Lehmer}, {Mainieri}, {Rafferty},
  {Schneider}, {Silverman}, \& {Vignali}}]{Luo10}
{Luo}, B., {et~al.} 2010, \apjs, 187, 560

\bibitem[{{Madau} \& {Dickinson}(2014)}]{Madau14}
{Madau}, P., \& {Dickinson}, M. 2014, \araa, 52, 415

\bibitem[{{Magdziarz} \& {Zdziarski}(1995)}]{Magdziarz95}
{Magdziarz}, P., \& {Zdziarski}, A.~A. 1995, \mnras, 273, 837

\bibitem[{{Marconi} {et~al.}(2004){Marconi}, {Risaliti}, {Gilli}, {Hunt},
  {Maiolino}, \& {Salvati}}]{Marconi04}
{Marconi}, A., {Risaliti}, G., {Gilli}, R., {Hunt}, L.~K., {Maiolino}, R., \&
  {Salvati}, M. 2004, \mnras, 351, 169

\bibitem[{{McCracken} {et~al.}(2010){McCracken}, {Capak}, {Salvato}, {Aussel},
  {Thompson}, {Daddi}, {Sanders}, {Kneib}, {Willott}, {Mancini}, {Renzini},
  {Cook}, {Le F\`{e}vre}, {Ilbert}, {Kartaltepe}, {Koekemoer}, {Mellier},
  {Murayama}, {Scoville}, {Shioya}, \& {Tanaguchi}}]{McCracken10}
{McCracken}, H.~J., {et~al.} 2010, \apj, 708, 202

\bibitem[{{Mendez} {et~al.}(2013){Mendez}, {Coil}, {Aird}, {Diamond-Stanic},
  {Moustakas}, {Blanton}, {Cool}, {Eisenstein}, {Wong}, \& {Zhu}}]{Mendez13}
{Mendez}, A.~J., {et~al.} 2013, \apj, 770, 40

\bibitem[{{Merloni} {et~al.}(2012){Merloni}, {Predehl}, {Becker},
  {B\"{o}hringer}, {Boller}, {Brunner}, {Brusa}, {Dennerl}, {Freyberg},
  {Friedrich}, {Georgakakis}, {Haberl}, {Hasinger}, {Meidinger}, {Mohr},
  {Nandra}, {Rau}, {Reiprich}, {Robrade}, {Salvato}, {Santangelo}, {Sasaki},
  {Schwope}, {Wilms}, \& {German eROSITA Consortium}}]{Merloni12}
{Merloni}, A., {et~al.} 2012, (arXiv:1209.3114)

\bibitem[{{Mineo} {et~al.}(2014){Mineo}, {Gilfanov}, {Lehmer}, {Morrison}, \&
  {Sunyaev}}]{Mineo14}
{Mineo}, S., {Gilfanov}, M., {Lehmer}, B.~D., {Morrison}, G.~E., \& {Sunyaev},
  R. 2014, \mnras, 437, 1698

\bibitem[{{Miyaji} {et~al.}(2015){Miyaji}, {Hasinger}, {Salvato}, {Brusa},
  {Cappelluti}, {Civano}, {Puccetti}, {Elvis}, {Brunner}, {Fotopoulou}, {Ueda},
  {Griffiths}, {Koekemoer}, {Akiyama}, {Comastri}, {Gilli}, {Lanzuisi},
  {Merloni}, \& {Vignali}}]{Miyaji15}
{Miyaji}, T., {et~al.} 2015, ApJ in press (arXiv:1503.00056)

\bibitem[{{Miyaji} {et~al.}(2000){Miyaji}, {Hasinger}, \& {Schmidt}}]{Miyaji00}
{Miyaji}, T., {Hasinger}, G., \& {Schmidt}, M. 2000, \aap, 353, 25

\bibitem[{{Miyaji} {et~al.}(2001){Miyaji}, {Hasinger}, \& {Schmidt}}]{Miyaji01}
---. 2001, \aap, 369, 49

\bibitem[{Moretti {et~al.}(2009)Moretti, Pagani, Cusumano, Campana, Perri,
  Abbey, Ajello, Beardmore, Burrows, Chincarini, {et~al.}}]{Moretti09}
Moretti, A., {et~al.} 2009, Astronomy \& Astrophysics, 493, 501

\bibitem[{{Murray} {et~al.}(2005){Murray}, {Kenter}, {Forman}, {Jones},
  {Green}, {Kochanek}, {Vikhlinin}, {Fabricant}, {Fazio}, {Brand}, {Brown},
  {Dey}, {Jannuzi}, {Najita}, {McNamara}, {Shields}, \& {Rieke}}]{Murray05}
{Murray}, S.~S., {et~al.} 2005, \apjs, 161, 1

\bibitem[{{Nandra} {et~al.}(2013){Nandra}, {Barret}, {Barcons}, {Fabian}, {den
  Herder}, {Piro}, {Watson}, {Adami}, {Aird}, {Afonso}, {et~al.}}]{Nandra13}
{Nandra}, K., {et~al.} 2013, (arXiv:1306.2307)

\bibitem[{{Nandra} {et~al.}(2015){Nandra}, {Laird}, {Aird}, {Salvato},
  {Georgakakis}, {Barro}, {Perez Gonzalez}, {Barmby}, {Chary}, {Coil},
  {Cooper}, {Davis}, {Dickinson}, {Faber}, {Fazio}, {Guhathakurta}, {Gwyn},
  {Hsu}, {Huang}, {Ivison}, {Koo}, {Newman}, {Rangel}, {Yamada}, \&
  {Willmer}}]{Nandra15}
---. 2015, ApJS in press, (arXiv:1503.09078)

\bibitem[{{Nandra} {et~al.}(2007){Nandra}, {O'Neill}, {George}, \&
  {Reeves}}]{Nandra07}
{Nandra}, K., {O'Neill}, P.~M., {George}, I.~M., \& {Reeves}, J.~N. 2007,
  \mnras, 382, 194

\bibitem[{{P\'{e}rez-Gonz\'{a}lez} {et~al.}(2005){P\'{e}rez-Gonz\'{a}lez},
  {Rieke}, {Egami}, {Alonso-Herrero}, {Dole}, {Papovich}, {Blaylock}, {Jones},
  {Rieke}, {Rigby}, {Barmby}, {Fazio}, {Huang}, \& {Martin}}]{Perez-Gonzlez05}
{P\'{e}rez-Gonz\'{a}lez}, P.~G., {et~al.} 2005, \apj, 630, 82

\bibitem[{{P\'{e}rez-Gonz\'{a}lez} {et~al.}(2008){P\'{e}rez-Gonz\'{a}lez},
  {Rieke}, {Villar}, {Barro}, {Blaylock}, {Egami}, {Gallego}, {Gil de Paz},
  {Pascual}, {Zamorano}, \& {Donley}}]{Perez-Gonzlez08}
---. 2008, \apj, 675, 234

\bibitem[{{Polletta} {et~al.}(2007){Polletta}, {Tajer}, {Maraschi},
  {Trinchieri}, {Lonsdale}, {Chiappetti}, {Andreon}, {Pierre}, {Le F\`{e}vre},
  {Zamorani}, {Maccagni}, {Garcet}, {Surdej}, {Franceschini}, {Alloin},
  {Shupe}, {Surace}, {Fang}, {Rowan-Robinson}, {Smith}, \&
  {Tresse}}]{Polletta07}
{Polletta}, M., {et~al.} 2007, \apj, 663, 81

\bibitem[{{Ptak} {et~al.}(2007){Ptak}, {Mobasher}, {Hornschemeier}, {Bauer}, \&
  {Norman}}]{Ptak07}
{Ptak}, A., {Mobasher}, B., {Hornschemeier}, A., {Bauer}, F., \& {Norman}, C.
  2007, \apj, 667, 826

\bibitem[{{Puccetti} {et~al.}(2009){Puccetti}, {Vignali}, {Cappelluti},
  {Fiore}, {Zamorani}, {Aldcroft}, {Elvis}, {Gilli}, {Miyaji}, {Brunner},
  {Brusa}, {Civano}, {Comastri}, {Damiani}, {Fruscione}, {Finoguenov},
  {Koekemoer}, \& {Mainieri}}]{Puccetti09}
{Puccetti}, S., {et~al.} 2009, \apjs, 185, 586

\bibitem[{{Ranalli} {et~al.}(2003){Ranalli}, {Comastri}, \&
  {Setti}}]{Ranalli03}
{Ranalli}, P., {Comastri}, A., \& {Setti}, G. 2003, \aap, 399, 39

\bibitem[{{Reddy} {et~al.}(2006){Reddy}, {Steidel}, {Erb}, {Shapley}, \&
  {Pettini}}]{Reddy06}
{Reddy}, N.~A., {Steidel}, C.~C., {Erb}, D.~K., {Shapley}, A.~E., \& {Pettini},
  M. 2006, \apj, 653, 1004

\bibitem[{{Ross} {et~al.}(2013){Ross}, {McGreer}, {White}, {Richards}, {Myers},
  {Palanque-Delabrouille}, {Strauss}, {Anderson}, {Shen}, {Brandt},
  {Y\`{e}che}, {Swanson}, {Aubourg}, {Bailey}, {Bizyaev}, {Bovy}, {Brewington},
  {Brinkmann}, {DeGraf}, {Di Matteo}, {Ebelke}, {Fan}, {Ge}, {Malanushenko},
  {Malanushenko}, {Mandelbaum}, {Maraston}, {Muna}, {Oravetz}, {Pan},
  {P\^{a}ris}, {Petitjean}, {Schawinski}, {Schlegel}, {Schneider}, {Silverman},
  {Simmons}, {Snedden}, {Streblyanska}, {Suzuki}, {Weinberg}, \&
  {York}}]{Ross12}
{Ross}, N.~P., {et~al.} 2013, \apj, 773, 14

\bibitem[{{Ross} \& {Fabian}(2005)}]{Ross05}
{Ross}, R.~R., \& {Fabian}, A.~C. 2005, \mnras, 358, 211

\bibitem[{{Salvato} {et~al.}(2009){Salvato}, {Hasinger}, {Ilbert}, {Zamorani},
  {Brusa}, {Scoville}, {Rau}, {Capak}, {Arnouts}, {Aussel}, {Bolzonella},
  {Buongiorno}, {Cappelluti}, {Caputi}, {Civano}, {Cook}, {Elvis}, {Gilli},
  {Jahnke}, {Kartaltepe}, {Impey}, {Lamareille}, {Le Floc'h}, {Lilly},
  {Mainieri}, {McCarthy}, {McCracken}, {Mignoli}, {Mobasher}, {Murayama},
  {Sasaki}, {Sanders}, {Schiminovich}, {Shioya}, {Shopbell}, {Silverman},
  {Smol{\v c}i\'{c}}, {Surace}, {Taniguchi}, {Thompson}, {Trump}, {Urry}, \&
  {Zamojski}}]{Salvato09}
{Salvato}, M., {et~al.} 2009, \apj, 690, 1250

\bibitem[{{Salvato} {et~al.}(2011){Salvato}, {Ilbert}, {Hasinger}, {Rau},
  {Civano}, {Zamorani}, {Brusa}, {Elvis}, {Vignali}, {Aussel}, {Comastri},
  {Fiore}, {Le Floc'h}, {Mainieri}, {Bardelli}, {Bolzonella}, {Bongiorno},
  {Capak}, {Caputi}, {Cappelluti}, {Carollo}, {Contini}, {Garilli}, {Iovino},
  {Fotopoulou}, {Fruscione}, {Gilli}, {Halliday}, {Kneib}, {Kakazu},
  {Kartaltepe}, {Koekemoer}, {Kovac}, {Ideue}, {Ikeda}, {Impey}, {Le Fevre},
  {Lamareille}, {Lanzuisi}, {Le Borgne}, {Le Brun}, {Lilly}, {Maier},
  {Manohar}, {Masters}, {McCracken}, {Messias}, {Mignoli}, {Mobasher}, {Nagao},
  {Pello}, {Puccetti}, {Perez-Montero}, {Renzini}, {Sargent}, {Sanders},
  {Scodeggio}, {Scoville}, {Shopbell}, {Silvermann}, {Taniguchi}, {Tasca},
  {Tresse}, {Trump}, \& {Zucca}}]{Salvato11}
---. 2011, \apj, 742, 61

\bibitem[{{Sanders} {et~al.}(2007){Sanders}, {Salvato}, {Aussel}, {Ilbert},
  {Scoville}, {Surace}, {Frayer}, {Sheth}, {Helou}, {Brooke}, {Bhattacharya},
  {Yan}, {Kartaltepe}, {Barnes}, {Blain}, {Calzetti}, {Capak}, {Carilli},
  {Carollo}, {Comastri}, {Daddi}, {Ellis}, {Elvis}, {Fall}, {Franceschini},
  {Giavalisco}, {Hasinger}, {Impey}, {Koekemoer}, {Le F\`{e}vre}, {Lilly},
  {Liu}, {McCracken}, {Mobasher}, {Renzini}, {Rich}, {Schinnerer}, {Shopbell},
  {Taniguchi}, {Thompson}, {Urry}, \& {Williams}}]{Sanders07}
{Sanders}, D.~B., {et~al.} 2007, \apjs, 172, 86

\bibitem[{{Schulze} {et~al.}(2015){Schulze}, {Bongiorno}, {Gavignaud},
  {Schramm}, {Silverman}, {Merloni}, {Zamorani}, {Hirschmann}, {Mainieri},
  {Wisotzki}, {Shankar}, {Fiore}, {Koekemoer}, \& {Temporin}}]{Schulze15}
{Schulze}, A., {et~al.} 2015, \mnras, 447, 2085

\bibitem[{{Schwope} {et~al.}(2000){Schwope}, {Hasinger}, {Lehmann}, {Schwarz},
  {Brunner}, {Neizvestny}, {Ugryumov}, {Balega}, {Tr\"{u}mper}, \&
  {Voges}}]{Schwope00}
{Schwope}, A., {et~al.} 2000, Astronomische Nachrichten, 321, 1

\bibitem[{{Shankar} {et~al.}(2013){Shankar}, {Weinberg}, \&
  {Miralda-Escud\'{e}}}]{Shankar13}
{Shankar}, F., {Weinberg}, D.~H., \& {Miralda-Escud\'{e}}, J. 2013, \mnras,
  428, 421

\bibitem[{{Silverman} {et~al.}(2008){Silverman}, {Green}, {Barkhouse}, {Kim},
  {Kim}, {Wilkes}, {Cameron}, {Hasinger}, {Jannuzi}, {Smith}, {Smith}, \&
  {Tananbaum}}]{Silverman08}
{Silverman}, J.~D., {et~al.} 2008, \apj, 679, 118

\bibitem[{{Skilling}(2004)}]{Skilling04}
{Skilling}, J. 2004, in {AIP Conference Proceedings of the 24th International
  Workshop on Bayesian Inference and Maximum Entropy Methods in Science and
  Engineering}, Vol. 735, 395--405

\bibitem[{{Soltan}(1982)}]{Soltan82}
{Soltan}, A. 1982, \mnras, 200, 115

\bibitem[{{Steidel} {et~al.}(2003){Steidel}, {Adelberger}, {Shapley},
  {Pettini}, {Dickinson}, \& {Giavalisco}}]{Steidel03}
{Steidel}, C.~C., {Adelberger}, K.~L., {Shapley}, A.~E., {Pettini}, M.,
  {Dickinson}, M., \& {Giavalisco}, M. 2003, \apj, 592, 728

\bibitem[{{Stern} {et~al.}(2012){Stern}, {Assef}, {Benford}, {Blain}, {Cutri},
  {Dey}, {Eisenhardt}, {Griffith}, {Jarrett}, {Lake}, {Masci}, {Petty},
  {Stanford}, {Tsai}, {Wright}, {Yan}, {Harrison}, \& {Madsen}}]{Stern12}
{Stern}, D., {et~al.} 2012, \apj, 753, 30

\bibitem[{{Tozzi} {et~al.}(2006){Tozzi}, {Gilli}, {Mainieri}, {Norman},
  {Risaliti}, {Rosati}, {Bergeron}, {Borgani}, {Giacconi}, {Hasinger},
  {Nonino}, {Streblyanska}, {Szokoly}, {Wang}, \& {Zheng}}]{Tozzi06}
{Tozzi}, P., {et~al.} 2006, \aap, 451, 457

\bibitem[{{Treister} {et~al.}(2009){Treister}, {Urry}, \&
  {Virani}}]{Treister09}
{Treister}, E., {Urry}, C.~M., \& {Virani}, S. 2009, \apj, 696, 110

\bibitem[{{Trouille} {et~al.}(2008){Trouille}, {Barger}, {Cowie}, {Yang}, \&
  {Mushotzky}}]{Trouille08}
{Trouille}, L., {Barger}, A.~J., {Cowie}, L.~L., {Yang}, Y., \& {Mushotzky},
  R.~F. 2008, \apjs, 179, 1

\bibitem[{{Tueller} {et~al.}(2008){Tueller}, {Mushotzky}, {Barthelmy},
  {Cannizzo}, {Gehrels}, {Markwardt}, {Skinner}, \& {Winter}}]{Tueller08}
{Tueller}, J., {Mushotzky}, R.~F., {Barthelmy}, S., {Cannizzo}, J.~K.,
  {Gehrels}, N., {Markwardt}, C.~B., {Skinner}, G.~K., \& {Winter}, L.~M. 2008,
  \apj, 681, 113

\bibitem[{{Tzanavaris} \& {Georgantopoulos}(2008)}]{Tzanavaris08}
{Tzanavaris}, P., \& {Georgantopoulos}, I. 2008, \aap, 480, 663

\bibitem[{{Ueda} {et~al.}(2014){Ueda}, {Akiyama}, {Hasinger}, {Miyaji}, \&
  {Watson}}]{Ueda14}
{Ueda}, Y., {Akiyama}, M., {Hasinger}, G., {Miyaji}, T., \& {Watson}, M.~G.
  2014, \apj, 786, 104

\bibitem[{{Ueda} {et~al.}(2003){Ueda}, {Akiyama}, {Ohta}, \& {Miyaji}}]{Ueda03}
{Ueda}, Y., {Akiyama}, M., {Ohta}, K., \& {Miyaji}, T. 2003, \apj, 598, 886

\bibitem[{{Ueda} {et~al.}(2011){Ueda}, {Hiroi}, {Isobe}, {Hayashida}, {Eguchi},
  {Sugizaki}, {Kawai}, {Tsunemi}, {Mihara}, {Matsuoka}, {Ishikawa}, {Kimura},
  {Kitayama}, {Kohama}, {Matsumura}, {Morii}, {Nakagawa}, {Nakahira},
  {Nakajima}, {Negoro}, {Serino}, {Shidatsu}, {Sootome}, {Sugimori}, {Suwa},
  {Toizumi}, {Tomida}, {Tsuboi}, {Ueno}, {Usui}, {Yamamoto}, {Yamaoka},
  {Yamazaki}, \& {Yoshida}}]{Ueda11}
{Ueda}, Y., {et~al.} 2011, \pasj, 63, 937

\bibitem[{{Ueda} {et~al.}(2001){Ueda}, {Ishisaki}, {Takahashi}, {Makishima}, \&
  {Ohashi}}]{Ueda01}
{Ueda}, Y., {Ishisaki}, Y., {Takahashi}, T., {Makishima}, K., \& {Ohashi}, T.
  2001, \apjs, 133, 1

\bibitem[{{Ueda} {et~al.}(1999){Ueda}, {Takahashi}, {Inoue}, {Tsuru}, {Sakano},
  {Ishisaki}, {Ogasaka}, {Makishima}, {Yamada}, {Akiyama}, \& {Ohta}}]{Ueda99}
{Ueda}, Y., {et~al.} 1999, \apj, 518, 656

\bibitem[{{Vasudevan} {et~al.}(2013){Vasudevan}, {Mushotzky}, \&
  {Gandhi}}]{Vasudevan13b}
{Vasudevan}, R.~V., {Mushotzky}, R.~F., \& {Gandhi}, P. 2013, \apjl, 770, L37

\bibitem[{{Vito} {et~al.}(2014){Vito}, {Gilli}, {Vignali}, {Comastri}, {Brusa},
  {Cappelluti}, \& {Iwasawa}}]{Vito14b}
{Vito}, F., {Gilli}, R., {Vignali}, C., {Comastri}, A., {Brusa}, M.,
  {Cappelluti}, N., \& {Iwasawa}, K. 2014, \mnras, 445, 3557

\bibitem[{{Vito} {et~al.}(2013){Vito}, {Vignali}, {Gilli}, {Comastri},
  {Iwasawa}, {Brandt}, {Alexander}, {Brusa}, {Lehmer}, {Bauer}, {Schneider},
  {Xue}, \& {Luo}}]{Vito13}
{Vito}, F., {et~al.} 2013, \mnras, 428, 354

\bibitem[{{Winter} {et~al.}(2009){Winter}, {Mushotzky}, {Reynolds}, \&
  {Tueller}}]{Winter09}
{Winter}, L.~M., {Mushotzky}, R.~F., {Reynolds}, C.~S., \& {Tueller}, J. 2009,
  \apj, 690, 1322

\bibitem[{{Wirth} {et~al.}(2004){Wirth}, {Willmer}, {Amico}, {Chaffee},
  {Goodrich}, {Kwok}, {Lyke}, {Mader}, {Tran}, {Barger}, {Cowie}, {Capak},
  {Coil}, {Cooper}, {Conrad}, {Davis}, {Faber}, {Hu}, {Koo}, {Le Mignant},
  {Newman}, \& {Songaila}}]{Wirth04}
{Wirth}, G.~D., {et~al.} 2004, \aj, 127, 3121

\bibitem[{{Worsley} {et~al.}(2006){Worsley}, {Fabian}, {Bauer}, {Alexander},
  {Brandt}, \& {Lehmer}}]{Worsley06}
{Worsley}, M.~A., {Fabian}, A.~C., {Bauer}, F.~E., {Alexander}, D.~M.,
  {Brandt}, W.~N., \& {Lehmer}, B.~D. 2006, \mnras, 368, 1735

\bibitem[{{Xue} {et~al.}(2010){Xue}, {Brandt}, {Luo}, {Rafferty}, {Alexander},
  {Bauer}, {Lehmer}, {Schneider}, \& {Silverman}}]{Xue10}
{Xue}, Y.~Q., {et~al.} 2010, \apj, 720, 368

\bibitem[{{Xue} {et~al.}(2011){Xue}, {Luo}, {Brandt}, {Bauer}, {Lehmer},
  {Broos}, {Schneider}, {Alexander}, {Brusa}, {Comastri}, {Fabian}, {Gilli},
  {Hasinger}, {Hornschemeier}, {Koekemoer}, {Liu}, {Mainieri}, {Paolillo},
  {Rafferty}, {Rosati}, {Shemmer}, {Silverman}, {Smail}, {Tozzi}, \&
  {Vignali}}]{Xue11}
---. 2011, \apjs, 195, 10

\bibitem[{{Xue} {et~al.}(2012){Xue}, {Wang}, {Brandt}, {Luo}, {Alexander},
  {Bauer}, {Comastri}, {Fabian}, {Gilli}, {Lehmer}, {Schneider}, {Vignali}, \&
  {Young}}]{Xue12}
---. 2012, \apj, 758, 129

\bibitem[{{Yencho} {et~al.}(2009){Yencho}, {Barger}, {Trouille}, \&
  {Winter}}]{Yencho09}
{Yencho}, B., {Barger}, A.~J., {Trouille}, L., \& {Winter}, L.~M. 2009, \apj,
  698, 380

\bibitem[{{York} {et~al.}(2000){York}, {Adelman}, {Anderson}, {Anderson},
  {Annis}, {Bahcall}, {Bakken}, {Barkhouser}, {Bastian}, {Berman}, {Boroski},
  {Bracker}, {Briegel}, {Briggs}, {Brinkmann}, {Brunner}, {Burles}, {Carey},
  {Carr}, {Castander}, {Chen}, {Colestock}, {Connolly}, {Crocker}, {Csabai},
  {Czarapata}, {Davis}, {Doi}, {Dombeck}, {Eisenstein}, {Ellman}, {Elms},
  {Evans}, {Fan}, {Federwitz}, {Fiscelli}, {Friedman}, {Frieman}, {Fukugita},
  {Gillespie}, {Gunn}, {Gurbani}, {de Haas}, {Haldeman}, {Harris}, {Hayes},
  {Heckman}, {Hennessy}, {Hindsley}, {Holm}, {Holmgren}, {Huang}, {Hull},
  {Husby}, {Ichikawa}, {Ichikawa}, {Ivezi\'{c}}, {Kent}, {Kim}, {Kinney},
  {Klaene}, {Kleinman}, {Kleinman}, {Knapp}, {Korienek}, {Kron}, {Kunszt},
  {Lamb}, {Lee}, {Leger}, {Limmongkol}, {Lindenmeyer}, {Long}, {Loomis},
  {Loveday}, {Lucinio}, {Lupton}, {MacKinnon}, {Mannery}, {Mantsch}, {Margon},
  {McGehee}, {McKay}, {Meiksin}, {Merelli}, {Monet}, {Munn}, {Narayanan},
  {Nash}, {Neilsen}, {Neswold}, {Newberg}, {Nichol}, {Nicinski}, {Nonino},
  {Okada}, {Okamura}, {Ostriker}, {Owen}, {Pauls}, {Peoples}, {Peterson},
  {Petravick}, {Pier}, {Pope}, {Pordes}, {Prosapio}, {Rechenmacher}, {Quinn},
  {Richards}, {Richmond}, {Rivetta}, {Rockosi}, {Ruthmansdorfer}, {Sandford},
  {Schlegel}, {Schneider}, {Sekiguchi}, {Sergey}, {Shimasaku}, {Siegmund},
  {Smee}, {Smith}, {Snedden}, {Stone}, {Stoughton}, {Strauss}, {Stubbs},
  {SubbaRao}, {Szalay}, {Szapudi}, {Szokoly}, {Thakar}, {Tremonti}, {Tucker},
  {Uomoto}, {Vanden Berk}, {Vogeley}, {Waddell}, {Wang}, {Watanabe},
  {Weinberg}, {Yanny}, \& {Yasuda}}]{York00}
{York}, D.~G., {et~al.} 2000, \aj, 120, 1579

\bibitem[{{Young} {et~al.}(2012){Young}, {Brandt}, {Xue}, {Paolillo},
  {Alexander}, {Bauer}, {Lehmer}, {Luo}, {Shemmer}, {Schneider}, \&
  {Vignali}}]{Young12}
{Young}, M., {et~al.} 2012, \apj, 748, 124

\end{thebibliography}

}
\end{document}